\documentclass{iopart} 
\usepackage[english]{babel} 
\usepackage{amssymb,amsmath}
\usepackage[dvipsnames]{xcolor}
\usepackage{graphicx}
\usepackage{mathtools}
\usepackage{verbatim}
\usepackage{enumitem}
\usepackage{caption}
\usepackage{subcaption}
\usepackage{float}
\usepackage{multirow}
\usepackage{bbm}

\usepackage{hyperref}
\hypersetup{
	colorlinks=true,
	linkcolor=WildStrawberry,
	citecolor=ForestGreen, 
	urlcolor=Orange, 
	pdftitle={Eigenvalue spectral tails and localization properties of asymmetric networks},
	pdfauthor={Pietro Valigi, Joseph W. Baron,Izaak Neri, Giulio Biroli, Chiara Cammarota, }
	bookmarks=true,
	bookmarksopen=true,
}

\usepackage[autostyle,italian=guillemets]{csquotes} 
\usepackage[bibstyle=numeric, citestyle=numeric-comp, sorting=none, maxbibnames=10, backend=biber, giveninits=true, hyperref]{biblatex} 
\DeclareNameAlias{author}{family-given}
\DeclareNameAlias{editor}{family-given}
\DeclareNameAlias{translator}{family-given}
\AtEveryBibitem{\clearfield{url}} 
\AtEveryBibitem{\clearfield{month}} 
\renewbibmacro{in:}{ 
	\ifentrytype{article}{}{\bibstring{in}\printunit{\intitlepunct}}}
\DeclareFieldFormat{doi}{ 
  \ifhyperref
    {\href{https://doi.org/#1}{\nolinkurl{#1}}}
    {\nolinkurl{#1}}}

\newcommand{\id}{{\underline{\underline{\mathbbm{1}}}}}
\newcommand{\mmat}{{\underline{\underline{M}}}}

\newcommand{\iu}{\mkern1mu\mathrm{i}\mkern1mu}
\newcommand{\ec}{\mkern1mu\mathrm{e}\mkern1mu}
\newcommand{\dd}{\mathop{}\!\mathrm{d}}
\newcommand{\ipr}{\overline{q^{-1}}}
\newcommand{\ie}{\emph{i.e.}}
\newcommand{\multilinecomment}[1]{} 

\begin{document}
	\title{Eigenvalue spectral tails and localization properties of asymmetric networks}
	\author{Pietro Valigi, Joseph W. Baron, Izaak Neri, Giulio Biroli, Chiara Cammarota}

	\begin{abstract}
		 In contrast to the neatly bounded spectra of densely populated large random matrices, sparse random matrices often exhibit unbounded eigenvalue tails on the real and imaginary axis, called Lifshitz tails.  In the case of  asymmetric matrices, concise mathematical results have proved elusive. In this work, we present an analytical approach to characterising these tails. We exploit the fact that eigenvalues in the tail region have corresponding eigenvectors that are exponentially localised on highly-connected hubs of the network associated to the random matrix. We approximate these eigenvectors using a series expansion in the inverse connectivity of the hub, where successive terms in the series take into account further sets of next-nearest neighbours. By considering the ensemble of such hubs, we are able to characterise the eigenvalue density and the extent of localisation in the tails of the spectrum in a general fashion.   As such, we classify a number of different asymptotic behaviours in the Lifshitz tails, as well as the leading eigenvalue and the inverse participation ratio. We demonstrate how an interplay between matrix asymmetry, network structure, and the edge-weight distribution leads to the variety of observed behaviours.
	\end{abstract}

	\maketitle
	
	\begin{refsection}
	
	\section{Introduction}
	
	Random matrices can be used to model the dynamics of large complex systems, such as, species-rich ecosystems~\cite{may1972will, allesina2015stability}, neural networks~\cite{rajan2006eigenvalue, aljadeff2015transition, thamm2022random}, and financial networks~\cite{laloux2000random}. Such systems are often characterised by a complex interaction network that can be modelled as a sparse, directed, random graph~\cite{newman2018networks}. Although many results are known for random matrices that are either fully connected~\cite{sommers1988spectrum, ginibre1965statistical, girko1985circular, aceituno2019universal, baron2020dispersal, patil2024spectral} or symmetric~\cite{kuhn2008spectra, kim1985density, rodgers1988density, semerjian2002sparse, rogers2008cavity},  understanding the combination of  sparseness  and  directedness is an ongoing effort~\cite{rogers2009cavity,metz2019spectral,neri2020linear, metz2021localization, mambuca2022dynamical,baron2022eigenvalue, valigi2024local}.  
	
\begin{figure}[ht]
		\centering 
		\includegraphics[scale = 0.28]{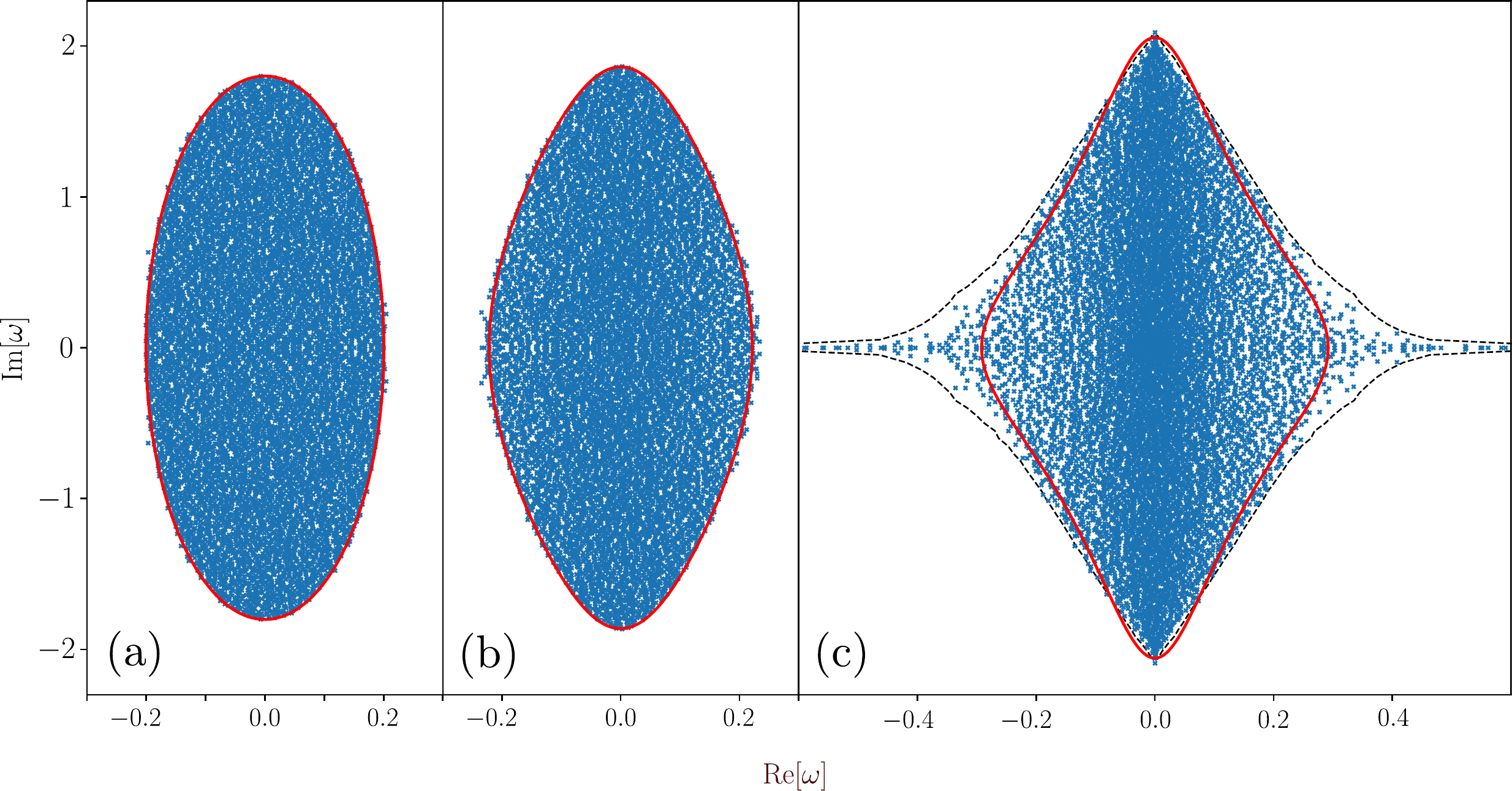}
		\caption{Eigenvalue spectra of Erd\H{o}s-R\'{e}nyi graphs with mean degree $p$, edge weights $J_{ij} = \pm 1/\sqrt{p}$, and $N = 10^4$ nodes. We set $J_{ij} = J_{ji}$ with probability $1-\epsilon = 0.1$ and $J_{ij} = -J_{ji}$ with probability $\epsilon = 0.9$. The solid red lines in all panels use the modified elliptic law from Ref. \cite{baron2025pathintegral}. Panel (a): For fully connected  graphs with $p = N-1$, the spectrum obeys the elliptic law \cite{sommers1988spectrum}. Panel (b): For $p = 20$, deviations from the dense case are apparent. These deviations are well-captured by a $1/p$ expansion about the elliptic law.  Panel (c): For $p = 5$, inaccuracies of the theory [which is accurate only to $\mathcal{O}(1/p)$] begin to show, and we observe that eigenvalue tails emerge. We note that we would also see tails on the imaginary axis for higher values of $N$. The black dashed line represents the non-perturbative estimate of the spectrum boundary obtained using the cavity method, as discussed in Refs.~\cite{metz2019spectral, mambuca2022dynamical}, with parameters $N_{\rm pop}=12500$, $n_{\rm s}^{(1)}=500$ and $n_{\rm s}^{(2)}=1000$.}
		\label{fig:examplespectra}
	\end{figure}

	The eigenvalue spectra of fully-populated matrices and the spectra of sparse random graphs (which have a finite number of entries per row, even in the infinite size limit) display qualitative differences. Perhaps the most evident are the tails of the eigenvalue distribution~\cite{mambuca2022dynamical,valigi2024local}. For fully-populated matrices, the spectrum is usually confined to a compact region of the complex plane (assuming an appropriate scaling of the entries with the matrix size $N$). In contrast, sparse models can exhibit spectral tails on the real and imaginary axis, and in the limit of large sparse matrices the spectrum occupies the entire real and imaginary axis. The emergence of such Lifshitz tails, \cite{khorunzhiy2006lifshitz, kuhn2008spectra, bapst2011lifshitz} as the matrix is made more sparse, can readily be observed in numerically constructed spectra, see Figure~\ref{fig:examplespectra}. Characterising these Lifshitz tails is important, particularly for understanding the stability of complex dynamical systems~\cite{goltsev2012localization, mambuca2022dynamical}, since the tail contains the leading eigenvalue that determines linear stability, and the corresponding eigenvector characterises the dominant mode of disturbances in the system~\cite{may1972will, allesina2015stability}. The purpose of this paper is precisely to develop a quantitative theory for the eigenvalues and the eigenvectors in the tails of the spectra of sparse random matrices.  
	
	For symmetric sparse matrices, the properties of the spectral tails are well understood, see the review~\cite{susca2021cavity}. The first works on these tails are by Rodgers,  Bray and de Dominicis, who characterised the tail distribution of eigenvalues of a nondirected Erd\H{o}s-R\'{e}nyi graph with  binary weights. They have shown that although the distribution covers the whole real axis~\cite{rodgers1988density, rodgers1990density},  it  asymptotically decays to zero faster  than a Gaussian distribution. This behaviour is qualitatively similar  to the decay in the spectrum of random Schr\"{o}dinger operators, as discovered by Lifshitz in the 1960s~\cite{lifshitz1963structure, klopp2002lifshitz}, and hence they were termed  Lifshitz tails~\cite{rodgers1990density, khorunzhiy2006lifshitz, bapst2011lifshitz}. The fast decaying distribution of eigenvalues implies a slow growth of the leading eigenvalue with system sizes, and it was shown that the leading eigenvalue scales as $\sqrt{k_{\rm max}}$~\cite{krivelevich2003largest, chung2003spectra, goltsev2012localization}, where $k_{\rm max}$ is the maximal degree of the graph.  In addition, the Lifshitz tails have  exponentially localised eigenvectors~\cite{semerjian2002sparse, kuhn2008spectra, slanina2012localization}, corresponding with a  spectrum that is pure point-like, yielding Poisson statistics for the eigenvalue density fluctuations~\cite{metz2017level, mirlin2000statistics, alt2023poisson, tarzia2022fully}.

	Particularly since it has been shown that the nature (and even presence) of Lifshitz tails for weighted and directed graphs is  linked with  the  sign patterns of the edge weights \cite{mambuca2022dynamical,valigi2024local}, the extension  to asymmetric matrices is a natural next step.  However, this  requires a non-trivial extension of the analytical techniques, which is one of the main results of our work. Specifically, we develop a systematically improvable analytical method based on  the single-defect approximation~\cite{biroli1999single, semerjian2002sparse}  and the cavity method~\cite{rogers2009cavity, metz2019spectral} to characterize the spectral tails of sparse nonsymmetric graphs.    We demonstrate that the tail eigenvalues correspond to eigenvectors that are exponentially localised around well-connected sites in the weighted network. Exploiting this fact, we perform an expansion in the inverse connectivity of such sites. This allows us to neglect contributions from the full network, and average only over local network configurations. Successively more accurate approximations can be obtained by including the effects of more distant neighbours in a systematic fashion. We thus obtain succinct closed-form approximations for the eigenvalue density, the expected leading eigenvalue, and the inverse participation ratio (IPR) in the tail regions. From these analytical expressions, we see transparently how increasing the number of antagonistic links (asymmetry) leads to more rapidly decaying tails, and can even lead to a crossover in the asymptotic behaviour of the eigenvalue density (from power-law to exponential, for instance). We also find that a balance between antagonistic and reinforcing links minimises the IPR, even eliminating localisation in some special cases.   
	
	We begin by describing the ensemble of sparse random matrices that we will study in Section \ref{section:model}. In Section \ref{section:method}, we introduce the cavity equations, perform the expansion for large eigenvalues,  and derive general expressions for the eigenvalues and IPR in terms of the local network structure around large-connectivity defects. In Section \ref{section:example:constantweight}, we perform the average over the ensemble of possible local configurations in the simple case where the non-zero elements of the random matrix can take the values $\pm a$, deriving expressions for the disorder-averaged eigenvalue density and the IPR for various global network structures. Having understood the simple case, we then generalise our results to the case of arbitrary edge-weight distributions, using a saddle-point procedure in Section \ref{section:saddlepoint}. We then discuss our results and conclude in Section \ref{section:discussion}.

	\section{Weighted and asymmetric network model}\label{section:model}	
	We begin by describing the random matrices of interest. We consider square matrices $\underline{\underline{M}}$ of dimension $N$ with elements that take the form $M_{ij} = J_{ij}  A_{ij}$, where $A_{ij}$ are the entries of an adjacency matrix, $J_{ij}$ are real-valued weights associated with the links of the graph, and $i,j=1,2,\ldots,N$. 
	
	The adjacency matrix $\underline{\underline{A}}$,  describes the underlying (non-directed) network: if a link between sites $i$ and $j$ exists, then $A_{ij} = A_{ji} = 1$, and $A_{ij} = A_{ji} = 0$ otherwise. In the examples, as explained in Section~\ref{appendix:numerics} of the Supplement, we use the algorithm described in Ref.~\cite{delgenio2010efficient} to generate a simple graph with degrees $c_i$ that are drawn independently from a prescribed degree distribution $P_\mathrm{deg}(c)$. Given the mean degree of the graph $p = \sum^{\infty}_{c=1} c P_\mathrm{deg}(c)$, then a link between sites $i$ and $j$ exists with probability $P_{ij} \approx c_i c_j/(pN)$ \cite{carro2016noisy, baron2022eigenvalue, rodgers2005eigenvalue}.  Moreover, we use the notation  $p_2 = \sum^{\infty}_{c=1}  c^2 P_\mathrm{deg}(c)$ for the second moment of the degree distribution.  
    If  $P_\mathrm{deg}(c)$ is a Poisson distribution, then in the large-$N$ limit a random graph with such a prescribed degree distribution   is equivalent to the  Erd\H{o}s-R\'{e}nyi (ER) ensemble~\cite{erdos1960evolution,bollobas1998random, van2017random}.  In the ER ensemble each element $A_{ij}=A_{ji}$ equals one with    a probability $p/N$, independent of the presence of other links in the graph, and is zero otherwise. 
	
	Next we clarify how the weights are associated with the links. The pair of weights $(J_{ij}, J_{ji})$ are drawn independently of all the other pairs, but $J_{ij}$ and $J_{ji}$ may be correlated. For each pair $(i,j)$, the link may be chosen to be either \emph{antagonistic} (sign-antisymmetric) such that $\mathrm{sign}(J_{ij}) = -\mathrm{sign}(J_{ji})$ (with probability $\epsilon$), or \emph{reinforcing} (sign-symmetric) such that $\mathrm{sign}(J_{ij}) = \mathrm{sign}(J_{ji})$ (with probability $1-\epsilon$). Given the relative signs of the two elements, we then draw their precise values. Thus, the edge weight pairs $(J_{ij},J_{ji})$ are drawn independently and identically from the distribution 
	\begin{equation}
		\pi(J_{ij}, J_{ji}) = \epsilon \: \pi_a(J_{ij}, J_{ji}) + (1- \epsilon) \pi_r(J_{ij}, J_{ji}), \label{eq:PiWeight}
	\end{equation}
	where $\pi_a(J_{ij}, J_{ji})$ and $\pi_{r}(J_{ij}, J_{ji})$  are the distributions of the weight pairs $(J_{ij},J_{ji})$ for antagonistic and reinforcing  links, respectively; thus, $\pi_a(J_{ij},J_{ji})=0$ if $J_{ij}J_{ji}>0$ and $\pi_r(J_{ij}, J_{ji}) = 0$ if $J_{ij}J_{ji}<0$.   One notes that in special cases where $\pi(J_{ij}, J_{ji}) \propto \delta(J_{ij}-J_{ji})$, the matrix $\mmat$ is symmetric and the eigenvalues are real, and when $\pi(J_{ij}, J_{ji}) \propto \delta(J_{ij}+J_{ji})$ the matrix is anti-symmetric and the eigenvalues are imaginary.   We note further that if we wish to obtain an eigenvalue spectrum that is confined to a bounded region of the complex plane in the dense limit $p \to \infty$, then the weights must scale as $J_{ij} \sim 1/\sqrt{p}$ \cite{baron2025pathintegral}.
	
	Figure~\ref{fig:examplespectra} illustrates a few examples of the spectra of  Erd\H{o}s-R\'{e}nyi graphs. We draw attention to the right panel that exhibits tails on the real and imaginary axes. These tails appear non-perturbatively, in the sense that a perturbative expansion around the high connectivity limit $p\to \infty$ predicts that the spectrum is confined in a compact region of the complex plane (red, solid line in the figure)~\cite{baron2025pathintegral}.  The cavity method, which is a nonperturbative approach, does capture the tails of the spectra, and the dashed, grey line in Fig.~\ref{fig:examplespectra} shows the boundary as predicted by the cavity method~\cite{mambuca2022dynamical}. However, this boundary is obtained numerically, which has limitations as we show later. Currently an analytical understanding of how network topology and weight statistics determines tails of the spectra is missing. 
	
	\section{Tail eigenvalues due to nodes with large  effective degree  } 
	\label{section:method}
	To quantify the tails of sparse non-Hermitian random matrices, we build on ideas from symmetric sparse random graphs, where similar tails have been observed~\cite{rodgers1988density, kuhn2008spectra}. For symmetric ensembles, the spectral tails  are constructed from  eigenvalues that are associated with nodes of large degree~\cite{krivelevich2003largest, chung2003spectra, benaych2019largest}, and the corresponding eigenvectors are exponentially localised about these nodes~\cite{slanina2012localization, alt2024localized}. Using the single defect approximation that exploits these features, the spectral tails of random symmetric graphs have been characterised in Refs.~\cite{biroli1999single,semerjian2002sparse}. Inspired by these works, we develop here an approach that applies to adjacency matrices of sparse graphs with asymmetric weights. Note that we use the same phrase `single defect' approximation as in Refs.~\cite{biroli1999single,semerjian2002sparse} even though there are differences in the technical implementation of this approximation.  
	
	For sparse (locally tree-like) random graphs, we derive an asymptotic formula for the tail eigenvalues (see Eq.~(\ref{eigenvalueapprox})) in terms of the  local graph architecture. We find that the \textit{effective degree} $|\Phi_i|$,  where 
	\begin{equation}
		\Phi_i =  \sum^{N}_{j=1}A_{ij}J_{ij}J_{ji}  \label{effectivedegree}
	\end{equation}
	is the key quantity that determines whether a node $i$  contributes to the spectral tail. Specifically,  nodes $i$  for which $|\Phi_i|$   is large enough contribute an eigenvalue in the tails of the spectrum. Furthermore, by deriving an analytical expression for the inverse participation ratio (see Eq.~(\ref{iprgeneral})), we  demonstrate that the right eigenvectors associated with the tail eigenvalues are exponentially localised with a localisation length determined by the effective degree. 
		
	\subsection{Quantities of interest}
	We now define other key quantities of interest. We characterise the average spectrum of a large random matrix with the average  eigenvalue spectral density of  $\underline{\underline{M}}$, which is defined as the following distribution of eigenvalues,  
	\begin{equation}
		\mu(z) = \left\langle \frac{1}{N} \sum^N_{\nu=1} \delta\left(z - \lambda_\nu\right)\right\rangle, \label{densitydef2}
	\end{equation}
    where $z\in \mathbb{C}$ is an element of the complex plane and $\delta$ is the Dirac distribution in the complex plane.  The $\left\{ \lambda_\nu\right\}$ are the eigenvalues of a particular instance of the random matrix $\underline{\underline{M}}$, and the angular brackets $\left\langle \cdot \right\rangle$ indicate the disorder average (over realisations of the weights $J_{ij}$ and the network $A_{ij}$). In this paper, we are interested in the tails of the distribution $\mu(z)$ on the real axis. For this reason, we investigate the spectral density of the real parts of the eigenvalues (which in the tail region coincides with the distribution of the real eigenvalues)
	\begin{equation}
		\rho(\omega) = \left\langle \frac{1}{N} \sum^N_{\nu=1} \delta\left(\omega - \Re(\lambda_\nu) \right)\right\rangle, \label{densitydef}
	\end{equation}
    where now $\omega \in \mathbb{R}$ and the Dirac delta is defined on the real axis.
    
	In dynamical systems theory, we are often interested in the leading eigenvalue $\lambda_{\rm max}$~\cite{may1972will,allesina2015stability}, defined as the eigenvalue of  $\underline{\underline{M}}$ that has the largest real part. Numerical evidence shows that the leading eigenvalue in general diverges as a function of $N$~\cite{mambuca2022dynamical, valigi2024local} for asymmetric sparse random matrices, and therefore we will attempt to characterise this growth with the system size $N$ using our theory.  
	
	We also analyse eigenvector statistics. 
     Since the matrix $\underline{\underline{M}}^T$ has the same statistical properties as $\underline{\underline{M}}$, right and left eigenvectors enjoy the same statistical properties. Consequently, in the following, we only focus on right eigenvectors without loss of generality. 
     The main quantity of interest for us is 
   the (disorder-averaged) Inverse Participation Ratio (IPR) of right eigenvectors:
	\begin{align}
		\overline{q^{-1}(\omega)} &= \left\langle \frac{\sum^N_{\nu=1} \delta(\omega - \lambda_\nu)q^{-1}_\nu}{\sum^N_{\nu=1} \delta(\omega - \lambda_\nu)}\right\rangle, \nonumber  \\
		q^{-1}_\nu &=\sum^N_{i=1}\left\vert r_i^{(\nu)} \right \vert^4, \label{iprdef}
	\end{align}
	where $r_i^{(\nu)}$ is the $i$-th component of the normalised right eigenvector of $\underline{\underline{M}}$ corresponding to the eigenvalue $\lambda_\nu$, and $q^{-1}_\nu$ is the IPR corresponding to this eigenvalue. For practical reasons, that will become clear later, we choose to normalise right and left eigenvectors to unity such that $\sum_{i = 1}^N \vert r_i^{(\nu)} \vert^2 = 1$. One notes that for values $\omega$ at which there is a zero eigenvalue density, the $\overline{q^{-1}(\omega)}$ is ill-defined. Since the right eigenvectors have unit norm, one obtains that in the limit of  infinitely large matrices  $\overline{q^{-1}(\omega)}=0$ implies delocalisation, and $\overline{q^{-1}(\omega)}>0$ implies localisation.

	\subsection{Resolvent method}
	To examine the locations of eigenvalues, we  use  the `resolvent' matrix; this is otherwise known as the matrix Green's function and can be related to the Stieltjes transform of the eigenvalue density \cite{potters2020first}.   The elements of the resolvent matrix are given by
	\begin{align}
		G_{ij}(\omega) = \left[\left(\omega\id - \mmat\right)^{-1}\right]_{ij} = \sum^N_{\nu=1} \frac{1}{\vec{l}^{(\nu)}\cdot\vec{r}^{(\nu)}}\frac{(l^{(\nu)}_i)^\star \,\,r_j^{(\nu)}}{\omega - \lambda_\nu } , \label{resdef}
	\end{align}
	where $l_i^{(\nu)}$ is the $i$-th component of the left eigenvector corresponding to the eigenvalue $\lambda_\nu$, and where $\star$ denotes the complex conjugate. For the last equality we have assumed that the matrix  $\mmat$ is diagonalisable.    Hence, we can find  eigenvalues of the random matrix $\underline{\underline{M}}$ by locating the poles of the resolvent matrix elements.

   While for  symmetric  $\mmat$  we can  use the resolvent matrix to find the IPR \cite{economou1972existence, metz2010localization}, in the asymmetric case we must resort to other methods. Therefore, we  define the so-called Hermitised resolvent~\cite{feinberg1997non, metz2019spectral}
	\begin{align}\label{eq:hermitResolv}
		\underline{\underline{\mathcal{H}}}(\eta, \omega)  = \left[\begin{array}{cc}\underline{\underline{\mathcal{H}}}^{11}& \underline{\underline{\mathcal{H}}}^{12} \\\underline{\underline{\mathcal{H}}}^{21}& \underline{\underline{\mathcal{H}}}^{22} \end{array}\right]= \begin{bmatrix}
			\eta \id_{N}&  \omega \id_{N} -\underline{\underline{M}} \\
			( \omega \id_{N}- \underline{\underline{M}} )^\dagger & \eta \id_{N}
		\end{bmatrix}^{-1} , 
	\end{align}
	where $\eta$ is a real regulariser, the $\underline{\underline{\mathcal{H}}}^{\alpha\beta}$ are $N\times N$ matrices, and $(\cdot)^\dagger$ denotes the Hermitian conjugate. For our purposes, we will be mostly interested in the diagonal blocks, which yield the IPR of the right eigenvector corresponding to the eigenvalue $\lambda_\nu$ via \cite{neri2016eigenvalue}
	\begin{align}
		q^{-1}_\nu = \lim_{\eta \to 0} \eta^2 \sum^N_{i=1} \vert\mathcal{H}^{22}_{ii}(\eta, \lambda_\nu) \vert^2, \label{iprfromres}
	\end{align}
    where we note that in order for this formula to be valid, the normalisation $\sum^N_{i=1} \vert l^{(\nu)}_i\vert^2 = 1$ is required (see \ref{app:participNH}).   
    
    Note that  in this paper we focus on the fourth moment through the inverse participation ratio.   Nevertheless, it is  possible to use a cavity method  approach to  evaluate the generalised inverse participation ratios $I_q \equiv \sum_i \vert r_i^{(\nu)}\vert^{2q}$.  However, since the right eigenvectors in the tails are exponentially localised, the qualitative behaviour of the $I_q$ are similar to the inverse participation ratio.

	\subsection{Eigenvalues in the spectral tails from an \texorpdfstring{$1/\omega$}{1/omega}-expansion of the cavity equations}\label{section:expansion} 
	To determine the eigenvalues in the tails of the spectra of sparse random graphs (see panel (c) of Fig.~\ref{fig:examplespectra}), we assume that for $|\omega|$ large enough the spectrum is pure point, so that eigenvalues can be identified as poles of the  diagonal elements of the resolvent matrix  (\ref{resdef}).    Analytically, we can obtain the poles by performing an asymptotic  expansion in $\omega$  of the cavity equations for the diagonal elements of the resolvent.

	Using the fact that  large sparse random graphs are locally tree-like~\cite{dembo2010ising, susca2021cavity}, it has been shown that the diagonal elements of the resolvent matrix in Eq.~(\ref{resdef}) solve the equations~\cite{metz2019spectral}
	\begin{align}
		G_{ii}(\omega) &= \frac{1}{\omega - \sum_{j \in \partial_i}J_{ij}G^{(i)}_{jj}(\omega)J_{ji}}, \nonumber \\
		G_{jj}^{(i)}(\omega) &= \frac{1}{\omega -  \sum_{k \in \partial_j^{/i}} J_{jk} G^{(j)}_{kk}(\omega) J_{kj}}, \label{cavityequations}
	\end{align}
	where $\partial_j^{/i} = \partial_j\setminus\left\{j\right\}$. We refer to $G_{jj}^{(i)}$ as the elements of the cavity resolvent, which is the resolvent of the matrix $\mmat^{(i)}$ obtained from $\mmat$ by removing its $i$-th row and column.   The equations~(\ref{cavityequations}) are called the cavity equations and are   similar to those derived in Refs.~\cite{abou1973selfconsistent, cizeau1994theory, rogers2008cavity,  biroli2010anderson, bordenave2010resolvent, metz2010localization}, albeit with the difference that here the matrix is nonsymmetric. The most straightforward way to derive the cavity equations~(\ref{cavityequations}) is from a recursive application of the Schur-complement formula~\cite{bordenave2010resolvent, metz2019spectral}.   
	
	To determine the spectral properties of  $\mmat$ in the  spectral tails, we use that eigenvectors that are localised in the index $i$ give rise to poles of the local resolvent $G_{ii}(\omega)$ [as can be seen from Eq.~(\ref{resdef})];  a similar approach was used for symmetric matrices~\cite{facoetti2016non, biroli2010anderson}. Note that the poles of the local resolvents do not in general coincide with those of the local cavity resolvent elements $G_{jj}^{(i)}(\omega)$ [see Eqs.~(\ref{cavityequations})]. 
    With this in mind, when focusing on a large eigenvalues on site $j$ (in absence of $i$) we can express the cavity resolvents $G_{kk}^{(j)}(\omega)$ as a series in $1/\omega$, i.e.
	\begin{align}
		G_{kk}^{(j)}(\omega) = \frac{G_{kk}^{(j),0}}{\omega} + \frac{G_{kk}^{(j),1}}{\omega^2} + \frac{G_{kk}^{(j),2}}{\omega^3} + \cdots. \label{cavresseries}
	\end{align}
	As we show in Supplemental Material (SM) Sec.~\ref{appendix:expansion}, substituting the series  Eq.~(\ref{cavresseries}) into the second of Eqs.~(\ref{cavityequations}) and solving order-by-order in $1/\omega$, we  find a series for each cavity resolvent element in terms of  the edge weights $J_{ij}$ and $\omega$ (i.e., Eq.~(\ref{cavresseries2})).  Upon substituting these expressions into the first of Eqs.~(\ref{cavityequations}), we find an expression for the locations of the large eigenvalues.   Specifically, we find that if the effective degree $|\Phi_i|$ is large [see Eq.~(\ref{effectivedegree})], there is an eigenvalue pair at $\omega = \pm\lambda_i$, where 
	\begin{align}
		\lambda^2_i=  \sum_{j \in \partial_i} J_{ij}J_{ji} + \frac{\sum_{j \in \partial_i} J_{ij}J_{ji}\sum_{k \in \partial_j^{/i}} J_{jk}J_{kj} }{\sum_{j \in \partial_i} J_{ij}J_{ji}} + \mathcal{O}\left(|\Phi_i|^{-1} \right)  . \label{eigenvalueapprox}
	\end{align}
    For the  derivation of (\ref{eigenvalueapprox}) we
also require  that the effective degrees  $|\Phi_{i'}|$ of nodes $i'$ in the near vicinity of $i$ are comparatively smaller than $|\Phi_i|$.  The expression in Eq.~(\ref{eigenvalueapprox}) clarifies  why eigenvalues can appear on both the real and imaginary axes, as $J_{ij}J_{ji}$ can be a positive  or negative real number.   For directed graphs for which $J_{ij}J_{ji} = 0$ for all $(i,j)$  there are no spectral tails, as was observed in Ref. \cite{neri2020linear}.
    
There are two kinds of `defect' that occur on or around the site $i$ that would give rise to  a large value of the effective degree $|\Phi_i|=\vert\sum_{j \in \partial_i} J_{ij}J_{ji}\vert$: a large degree of node $i$, or a large value in the link weights connecting to node $i$. For most of the examples that we consider here, highly-connected nodes will be responsible for eigenvalues in the tail regions of the eigenvalue spectrum. We discuss the alternative possibility briefly in Sec.~\ref{appendix:breakdowngeneral} of the SM. The aforementioned approach is similar in philosophy to that used by Feinberg and Zee \cite{feinberg1999non} to describe spectral tails induced by  single `impurities' (on the diagonal of the random matrix) in a non-hermitian random matrix model for asymmetric hopping on a 1D chain. We emphasise however that the `defects' in our case emerge due to the random network topology.
	
	Using the expression in Eq.~(\ref{eigenvalueapprox}), we can thus compute the density of eigenvalues $\rho(\omega)$ in the tails regions. This is accomplished by finding the probability that a local arrangement of connectivities and edge weights gives rise to an eigenvalue at a given $\omega$. We assume (and check later) that these defects occur independently and isolated from one another (hence the term `single-defect' approximation). Once we have the asymptotic eigenvalue density, we can find the cumulative distribution function of the eigenvalue with the largest real part, $\lambda_{\rm max}$, which takes the form   \cite{majumdar2020extreme, vivo2015large}
	\begin{equation}
		Q_N(\lambda_{\rm max}) =  \left[1-\int_{\lambda_{\rm max}}^\infty \dd \omega \,\rho (\omega)\right]^{f_\mathrm{tail}N} \, , \label{eq:lambda_max_cumulative}
	\end{equation}
    where $f_\mathrm{tail}$ is the fraction of all the eigenvalues that are in the positive real tail (we assume that this does not scale with $N$). In the last equality, we have also assumed that there are no eigenvalue correlations, as is known to be the case for localised states~\cite{evers2008anderson}. In order to ensure that the cumulative distribution of $\lambda_{\rm max}$ has a finite value between $0$ and $1$, \ie, $0<Q_N(\lambda_{\rm max})<1$, it is necessary that the integral scales as $1/N$; accordingly, we can find the scaling of the typical leading eigenvalue via the useful relation 
	\begin{equation} \label{eq:max}
		\int_{\lambda_{\rm max}}^\infty \dd \omega \,\rho (\omega) \sim \frac{1}{N} \, .
	\end{equation}

	\subsection{IPR associated with eigenvalues in the spectral tails}

	In order to extract the IPR of right eigenvectors associated with eigenvalues in the spectral tails, we require the elements of the hermitised resolvent matrix in Eq.~(\ref{eq:hermitResolv}). To this end, we use a  hermitised set of cavity equations (given in Sec.~\ref{appendix:hermexpansion} of the Supplement), which are similar in spirit to Eqs.~(\ref{cavityequations}) \cite{rogers2009cavity,metz2019spectral}. 
	
	By performing a similar expansion of the hermitised cavity equations to the one that was discussed in the above Sec.~\ref{section:expansion}, we can evaluated the IPR using the formula in Eq.~(\ref{iprfromres}). We find the following approximate expression for the IPR corresponding to the eigenvalue in Eq.~(\ref{eigenvalueapprox})
	\begin{align}
		q^{-1}_i &= \left(\frac{\vert\sum_{j\in\partial_i} J_{ij} J_{ji}\vert}{\vert\sum_{j\in\partial_i} J_{ij} J_{ji} \vert + \sum_{j\in\partial_i} J_{ji}^2} \right)^2 \left(1 +  \mathcal{O}\left(|\Phi_i|^{-1}\right) \right), \label{iprgeneral}
	\end{align}
	where with some abuse of notation we use $q_i$  for the inverse participation ratio associated with the right eigenvector corresponding to $\lambda_i$. The next-to-leading order term in Eq.~(\ref{iprgeneral}) is given in Sec.~\ref{appendix:hermexpansion} of the Suppelement. Importantly, we find that unless the quantity $\vert\sum_{j\in\partial_i} J_{ij} J_{ji}\vert$ is nil, the states corresponding to highly-connected hubs are localised. 
    
	\subsection{Graphical interpretation of the results}\label{sec:graphical}
    We show here that the series in Eqs.~(\ref{eigenvalueapprox}) and (\ref{iprgeneral}) in the parameter $|\Phi_i|$ can also be obtained by considering the leading eigenvalue of tree graphs centred around the node~$i$.   
	
\begin{figure}[H]
		\centering 
		\includegraphics[scale = 0.28]{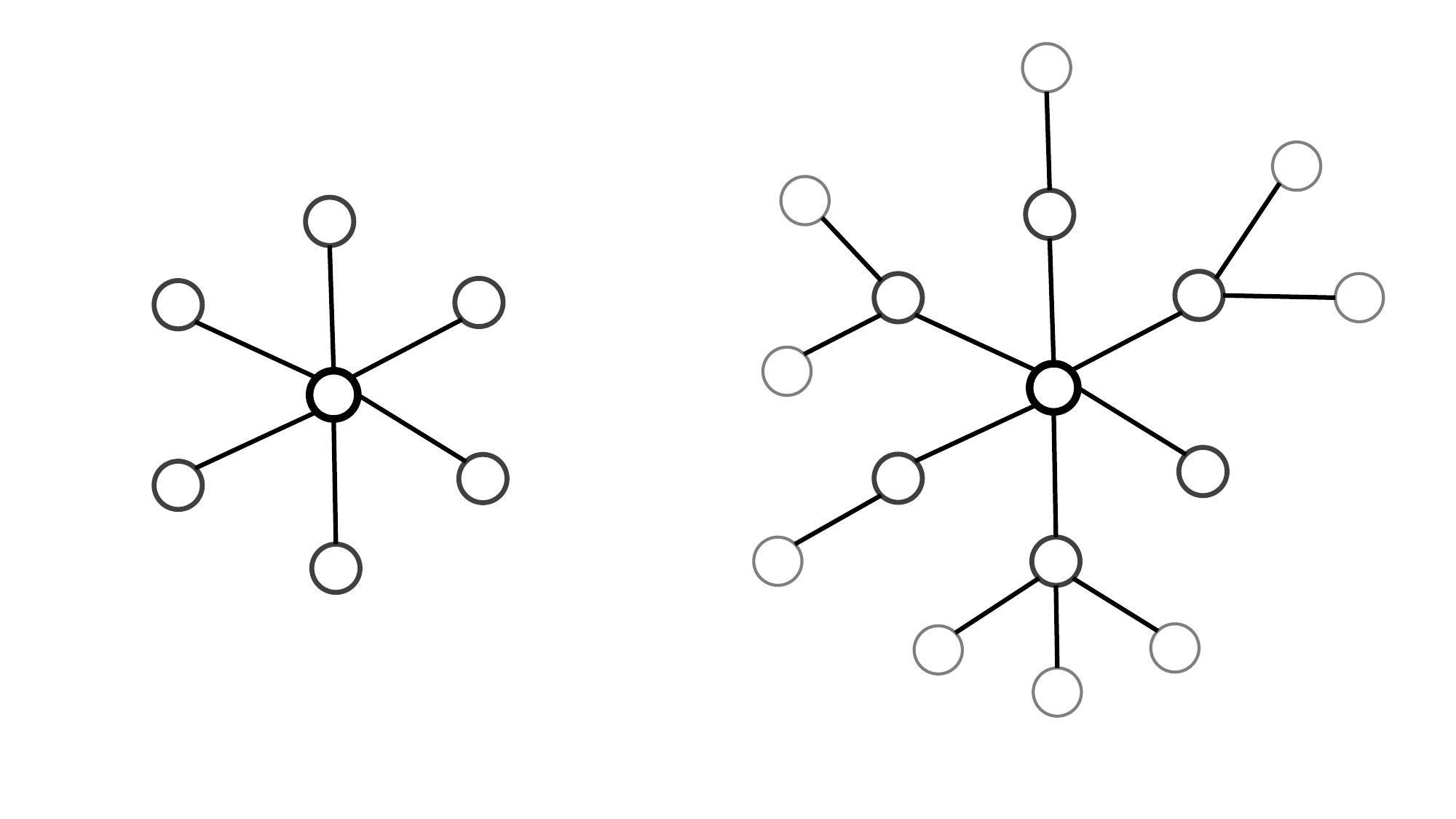}\put(-220,75){\large $i$}
        \put(-270,105){\large $j\in \partial_i$}
        \put(-88,76){\large $i$}
        \put(-35,60){\large $j\in \partial_i$}
          \put(-35,130){\large $k\in \partial^{/i}_j$}
		\caption{The tail eigenvalues are well approximated by the  eigenvalues of star graphs for which the central node (the hub) has a large effective degree $\Phi_i = \sum_{j\in \partial_i}J_{ij}J_{ji}$.      Subleading order contributions are obtained by considering the contribution of neighbouring nodes (the wider network).      Left: example of a star graph  with one generation of descendants.   Right: example of a   tree graph with two generations of descendants.         }\label{fig:stars}
	\end{figure}
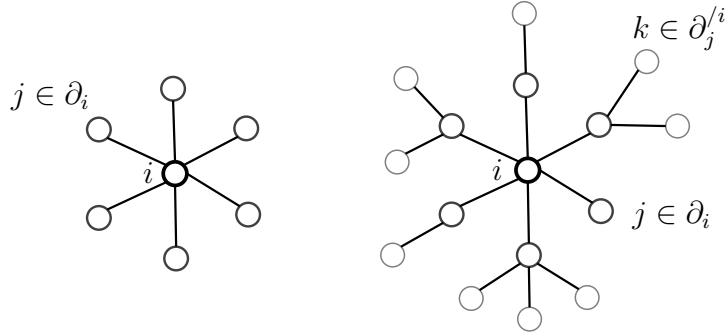

	The leading order terms in Eqs.~(\ref{eigenvalueapprox}) and (\ref{iprgeneral}) are obtained from a star graph, a tree consisting of a central node $i$ (the "hub") and its direct neighbours, as shown in the left panel of Fig.~\ref{fig:stars}.  Indeed, the leading eigenvalue $\lambda_i^{(1)}$ of a star graph centred on a node $i$ takes the form $\lambda_i^{(1)} = \sqrt{\sum_{j\in\partial_i}J_{ij}J_{ji}}$, which equals the leading order term in Eq.~(\ref{eigenvalueapprox}). Computing   the inverse participation ratio of the right eigenvectors corresponding to $\lambda_1$, we also recover the leading term in Eq.~(\ref{iprgeneral})  (see \ref{appendix:starplusneighbours}).          
	
The next leading order terms in Eqs.~(\ref{eigenvalueapprox}) and (\ref{iprgeneral}) follow from the leading eigenvalue (and right eigenvector) of a tree graph centered around node $i$ that has two generations, as shown in the right panel of Fig.~\ref{fig:stars}.  We refer to the generations of nodes beyond the direct neighbours of the hub as the ``wider network''.  Using standard perturbation theory techniques, we compute in ~\ref{appendix:starplusneighbours} the correction to the eigenvalue of the star graph. In so doing, we recover the next order correction terms in Eqs.~(\ref{eigenvalueapprox}) and (\ref{iprgeneral}), which are small compared to the leading term when $|\Phi_i|$ is large, as required. Considering tree graphs of three or more generations, we can in principle obtain the next terms in the series in Eqs.~(\ref{eigenvalueapprox}) and (\ref{iprgeneral}).   
	
	The present graphical argument provides a simple explanation for the  localisation of the right eigenvectors and  the relevance of sign patterns in random directed graphs. Indeed, as we show in  Sec.~\ref{appendix:exploc} of the Supplement, the right eigenvectors corresponding to  well-connected hubs have components whose magnitudes decay exponentially with distance from the hub; similar observations have been made in the context of the (symmetric) adjacency matrix of Erd\H{o}s-R\'{e}nyi graphs \cite{alt2023poisson, tarzia2022fully, alt2024localized}, and are typical in the context of Anderson localisation \cite{frohlich1984rigorous, kramer1993localization, rizzo2024localized}.  The relevance of the sign pattern of the weights is evident from the fact that a well-connected hub with  $\Phi_i \leq 0$ does not produce real eigenvalues, in correspondence with the principle of `local sign stability'~\cite{valigi2024local}.   
	
	\section{Example: Complex networks with asymmetry and constant magnitude of edge weight }\label{section:example:constantweight}
	In Section \ref{section:method}, we showed how rare events in a network, such as well-connected hubs, can give rise to eigenvalues that lie outside the bulk region of the spectrum. Such eigenvalues correspond to localised states and, collectively, form a Lifshitz tail. In this section, we derive the eigenvalue density and disorder-averaged IPR [defined in Eqs.~(\ref{densitydef}) and (\ref{iprdef}) respectively] at large values of $\omega$ for an example ensemble of random networks.

	Presently, we consider networks with a distribution of weights $\pi$ in Eq.~(\ref{eq:PiWeight}) that takes the form 
	\begin{align}
		\pi(u,v) =& \frac{\epsilon}{2} \left[\delta(u - a) \delta(v +a) +  \delta(u + a) \delta(v -a)\right] \nonumber \\
		&+ \frac{(1-\epsilon)}{2} \left[\delta(u - a) \delta(v -a) +  \delta(u + a) \delta(v +a)\right] ,  \label{piconstantmag}
	\end{align}
	so that the magnitude of the weights is fixed to $a$ but there is a finite probability $\epsilon$ to have sign-antisymmetric weights and a probability $1-\epsilon$ to have sign-symmetric weights;  we include the possibility of fluctuating edge weights in Section \ref{section:example:general}.    The distribution (\ref{piconstantmag}) has the benefit of being analytically tractable, and at the same time it 
	demonstrates how sign-asymmetry in the link weights, network structure, and the influence of neighbouring nodes conspire to affect the asymptotic behaviour.   As shown in Ref.~\cite{mambuca2022dynamical,valigi2024local}, for the sign-antisymmetric model with $\epsilon= 1$, Lifshitz tails are absent on the real axis, while the tails on the real axis are present whenever there exist a finite fraction of sign-symmetric weights, $\epsilon\neq 1$. Here, by analytically solving this model we  quantify how  Lifshitz tails vanish when approaching $\epsilon\approx 1$, and we also determine the corresponding changes in the IPR of the right eigenvectors.

    For the degree distribution of the networks, we consider three examples:  (1) geometric, (2) power-law, and (3) Poisson degree distributions (corresponding with Erd\H{o}s-R\'{e}nyi graphs).

    To determine the asymptotic expansion in $\omega$ of the eigenvalue density  and the IPR, we further build on the `{\it single-defect}' approximation \cite{biroli1999single, semerjian2002sparse} that considers each defect independently from the others (see Sec.~\ref{section:expansion} of the Supplement).  
	The single-defect approximation for the tail region is supported by the fact that eigenvectors in the localised phase obey Poisson statistics (i.e., there are no eigenvalue correlations) \cite{metz2017level, mirlin2000statistics, tarzia2022fully}. Further, as we show in Sec.~\ref{appendix:exploc}, the eigenvectors of tail-region eigenvalues are exponentially localised, and so it is unlikely for two defects to occur close enough in the network to interact appreciably. We discuss the limits of the single-defect approximation in Section \ref{section:breakdown}.

    \subsection{Asymptotic eigenvalue spectral density}\label{sec:eigenavlue41}
    When $\pi(u,v)$ is given by Eq.~(\ref{piconstantmag}), it follows from Eq.~(\ref{eigenvalueapprox}) that, to leading order, the eigenvalues in the tail are given by $\lambda=a^2 \Delta$, where $\Delta = \sum_{j\in\partial_i }{\rm sign}(J_{ij}J_{ji})$ is the difference between the number of reinforcing [$(+,+)$ or $(-,-)$] links and the number of antagonistic [$(-,+)$ or $(+,-)$] links that connect the hub at site $i$ with its neighbours. Additionally, we show in Sec.~\ref{appendix:selfaveraging} of the Supplement that the second term in Eq.~(\ref{eigenvalueapprox}), which corresponds to the contribution from the next-nearest neighbours, can be replaced with its average $a^2 (p_2-p)(1-2\epsilon)/p$ since its relative fluctuations go as $1/\lambda^2$.  That is, the eigenvalue associated with a hub that has a particular $\Delta$ is given by  $\lambda^2 = a^2[\Delta + (p_2-p)(1-2\epsilon)/p ]+ \mathcal{O}\left(1/\Delta\right)$.  Recall that  we use the notation $p = \sum^{\infty}_{c=1} c P_\mathrm{deg}(c)$ for the mean degree and  $p_2 = \sum^{\infty}_{c=1}  c^2 P_\mathrm{deg}(c)$ for   the second moment of the degree distribution.
	
	We thus see that many different local network configurations give rise to an eigenvalue in the same location. In order to find the eigenvalue density, we therefore need to determine the  probability that  local network configurations produce an eigenvalue at a location $\omega$.   Because only the quantity $\Delta$ matters in the present model, the eigenvalue spectral density in the tails of the spectrum (in this example we consider $\Re(\omega)\gg1$) is given by the following weighted sum of Dirac delta peaks
	\begin{align}
		\rho(\omega) &\approx\sum_{\Delta= 0}^\infty \sum_{c \in S(\Delta)} P(\Delta, c) \delta\left(\omega-\sqrt{a^2[\Delta + (p_2-p)  (1-2\epsilon)/p ] }\right), \label{configdensitysum} 
	\end{align}
	where, if $\Delta$ is even/odd, $c$ takes values in the set, $S(\Delta)$, of even/odd numbers from $\Delta$ to $\infty$, and 
    \begin{align}
    P(\Delta, c) &= P_\mathrm{deg}(c) \binom{c}{\frac{c+\Delta}{2}} (1-\epsilon)^{\frac{c+\Delta}{2}} \epsilon^{\frac{c-\Delta}{2}}. \label{eq:PDeltaC}
    \end{align}

	We are now tasked with extracting  from Eq.~(\ref{configdensitysum}) the asymptotic behaviour of the density. The crucial observation here is that for a given large $\Delta\gg 1$ there is a connectivity $c$ that hubs are most likely to possess. That is, the distribution $P(c, \Delta)$ is sharply peaked as a function of $c$ for fixed large $\Delta$. It is this observation that is at the heart of most of the subsequent analyses in this work. 
	
	More specifically, by expanding the distribution $P(c \vert \Delta) = P(\Delta,c)/P(\Delta)$  to the second order around its maximum value of  $c$, we are able to approximate the distribution $P(c \vert \Delta)$ with a Gaussian distribution, which allows us to   carry out the sum over $c$ in Eq.~(\ref{configdensitysum}). We then employ a similar coarse graining scheme as in Ref. \cite{semerjian2002sparse} to represent the remaining sum of delta peaks as a continuous eigenvalue density. This procedure is detailed in Secs.~\ref{appendix:configmodelconditional} and~\ref{appendix:eigenvaluedensityconfig} of the Supplement.  
	
\begin{figure}[ht]
	\centering
	\begin{subfigure}[b]{0.6\textwidth}
		\centering
		\includegraphics[width=\textwidth]{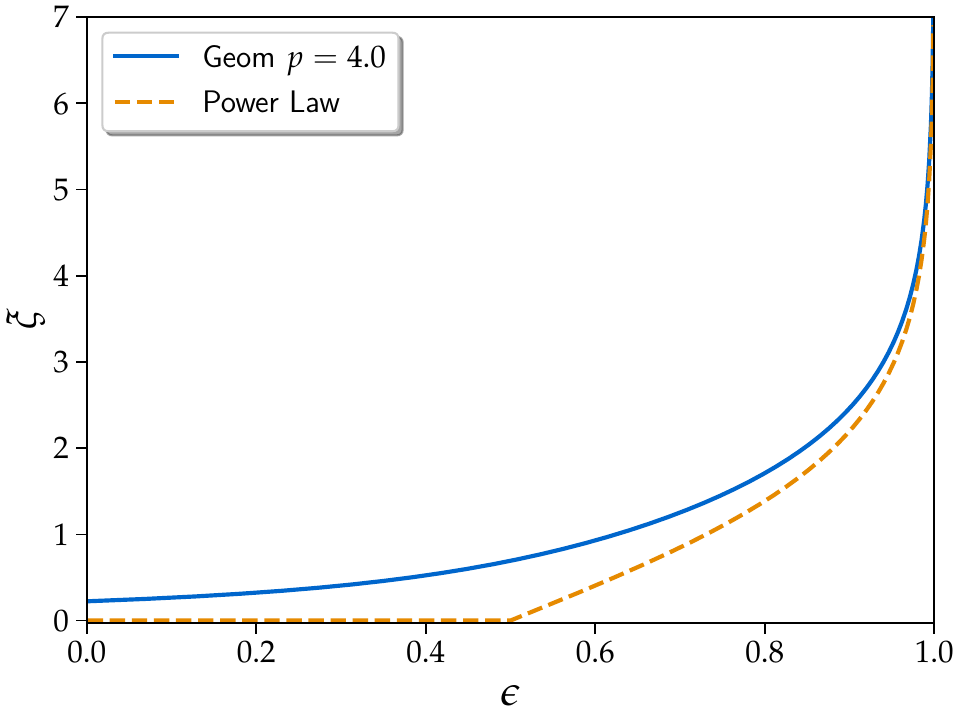}
	\end{subfigure}
	\caption[]{Plot of  $\xi_\mathrm{geom} = \frac{1}{a^2} \ln \left(\frac{p+1}{2(1-\epsilon) p} \left(1 + \sqrt{1- 4 \frac{\epsilon(1-\epsilon)p^2}{(p+1)^2}} \right) \right)$ and  $	\xi_\mathrm{pow} = \frac{1}{a^2} \ln \left(\frac{ \epsilon}{1-\epsilon} \right)$ as a function of $\epsilon$.   These are the characteristic decay constants for the spectral distribution in the tails, see Eqs.~(\ref{eq:geometric_spectral_density_fixed_weights}) and (\ref{epsghalf}).   Parameters chosen are $p=4$ for $\xi_\mathrm{geom}$   and the results for $\xi_\mathrm{pow}$ hold for any $\gamma>3$.}
	\label{fig:xiepsilon_fixed_edge}
\end{figure}

	For the case where the degree distribution is {\it geometric}, i.e.,
	\begin{equation}
		P_\mathrm{deg}(c) = \frac{1}{p+1} \left(\frac{p+1}{p}\right)^{-c} ,
	\end{equation}
	where $c = 0, 1, 2, \cdots$, we  find  for $\omega\gg 1$,
	\begin{equation} \label{eq:geometric_spectral_density_fixed_weights}
			\rho(\omega) \sim \; \omega\exp\left(-\xi_\mathrm{geo} \omega^2 \right), 
	\end{equation}  
	where $\sim$ indicates that the relation holds up to a proportionality constant   and the formula for $\xi_\mathrm{geo}$ is given in  the caption of Fig.~\ref{fig:xiepsilon_fixed_edge}. An analogous expression can be obtained for tails on the imaginary axis, by considering the imaginary eigenvalues of the hubs with large (negative) effective degree. Combining  Eqs.~(\ref{eq:geometric_spectral_density_fixed_weights}) and \eqref{eq:max} we extract the scaling for the typical leading eigenvalue $\lambda_\mathrm{max} \sim \sqrt{\ln N}$.
	
	For the case of a {\it power-law degree} distribution
	\begin{equation}
		P_\mathrm{deg}(c) = \frac{c^{-\gamma}}{\zeta(\gamma, c_{\rm min})} \, ,
	\end{equation}
	where $c \ge c_{\rm min}$ and $c_{\rm min}$ is the minimum degree, we instead have two different asymptotic behaviours, depending on the value of $\epsilon$. For $\epsilon <1/2$, the spectra develop fat tails  decaying as the power-law
	\begin{align}
		\rho(\omega) &\sim \omega^{-(2\gamma-1)},\label{epslhalf}
	\end{align}
	in correspondence with the result for symmetric case $\epsilon=0$ obtained in Ref.~\cite{dorogovtsev2003spectra,rodgers2005eigenvalue}.  
	In contrast, when the sign-antisymmetric links are in the majority, i.e.,  $\epsilon >1/2$, then the  tails decay exponentially fast as in the geometric ensemble, viz.,
\begin{align}
	\rho(\omega) &\sim  \omega^{-(2\gamma-1)}\exp\left(-\xi_\mathrm{pow} 
	\omega^2\right),\label{epsghalf}
\end{align}	
where $\xi_\mathrm{pow}$ is again given in  the caption of Fig.~\ref{fig:xiepsilon_fixed_edge}.

Hence, in the present case of power-law random graphs,  if the majority of links are sign symmetric ($\epsilon<1/2$), then the tails are scale invariant.  On the other hand, if the majority of the links are sign antisymmetric  ($\epsilon>1/2$), then the tail of the distribution has a characteristic decay length determined by $\xi_{\rm pow}$.  Correspondingly, using Eq.~\eqref{eq:max} we find $\lambda_\mathrm{max} \sim N^{1/[2(\gamma -1)]}$ for $\epsilon <1/2$ and $\lambda_\mathrm{max} \sim \sqrt{\ln N}$ for $\epsilon>1/2$.

Figure~\ref{fig:xiepsilon_fixed_edge}  compares the expressions for the decay constants  $\xi_{\rm geo}$ and $\xi_{\rm pow}$ as a function of $\epsilon$ for these two ensembles. We observe that $\xi$ diverges in the sign-antisymmetric limit  $\epsilon\rightarrow 1$, corresponding with the observation in ~\cite{mambuca2022dynamical, valigi2024local} that random graphs with sign-antisymmetric  weights do not have Lifshitz tails.

To recover the results of the fully connected limit for which tails are absent, see~\cite{baron2022eigenvalues, baron2025pathintegral}, we need to use the scaling $a\sim 1/\sqrt{p}$, so that the continuous component of the spectrum is confined to a compact region in the complex plane in the limit $p \to \infty$. With this scaling,  we find that the decay rates $\xi_{\rm geo},\xi_{\rm pow} \sim p$, which confirms that the tails  vanish in the limit of highly connected graphs.  

We note that in the above two cases (geometric and power-law networks), we have kept all terms in $\ln\rho(\omega)$ that contribute up to order $\mathcal{O}(\ln(\omega))$, but we neglect any constant prefactors.  The contribution from the wider network -- i.e., the second term in Eq.~(\ref{eigenvalueapprox}) or the second term in the delta distribution of (\ref{configdensitysum}) -- contributes  to the constant prefactor of $\rho(\omega)$, and we neglect this factor.   However, for $\gamma<3$, the branching coefficient $p_2-p$ diverges, and thus our approximation breaks down, as the second term in Eq.~(\ref{eigenvalueapprox}) is no longer negligible in comparison to the first. We note that, in this case, the continuous part of spectrum of a power-law random graph is supported on the whole complex plane, as has been shown in Refs. \cite{goltsev2012localization,neri2020linear}.

Let us now address the case of the {\it Erd\H{o}s-R\'{e}nyi} graph, where one has (for large $N$) a Poisson degree distribution
\begin{equation}
	P_\mathrm{deg}(c_j) \approx \frac{e^{-p} p^{c_j}}{c_j!} \, ,
\end{equation}
with $p$ the mean degree. In this case, the Gaussian approximation for the conditional distribution $P(c \vert \Delta)$ no longer applies, and we have to be  more careful. This is because the degree distribution is a more quickly decaying function of $c$ than in the previous two cases.  Instead, for a given large $\Delta$ the most likely value for the connectivity $c$ to take is close to its minimal value, i.e., $c \approx \Delta$, and  the number of antagonistic links $m = (c-\Delta)/2$ is Poisson distributed with mean $\mathrm{E}(m\vert \Delta) \approx \frac{e p^2 \epsilon (1-\epsilon)}{ \Delta}$.   With that being said, the crucial property that $c$ is large still applies, and in Sec.~\ref{appendix:poissonian} of the Supplement we use this property to derive   the asymptotic eigenvalue spectral density  and  the IPR. We obtain 
\begin{align}
	\rho(\omega) \sim \sqrt{\frac{p}{2\pi}} e^{-p} \left(\frac{a^2 pe (1-\epsilon)}{\omega^2} \right)^{\Delta(\omega)}  , \label{rhoer}
\end{align}
where  $\Delta(\omega) = \omega^2/a^2-p(1-2\epsilon)$ is the value of $\Delta$ that produces an eigenvalue at location $\omega$ [see Eq.~(\ref{configdensitysum})].    

We present a few comments regarding the result in Eq.~(\ref{rhoer}).    The factor $\omega^{-2\omega^2/a^2}$ determines the leading order decay of the spectral density, which is faster than for power-law random graphs and random graphs with  geometric degree distributions.  Hence, the faster decay of the  Poisson degree distribution (compared to those with geometric and power law degree distributions) yields, correspondingly, a faster decay of the eigenvalue spectral density.   
In the exponent $\Delta(\omega)$  appears a `shift-correction' due to the contribution  of the wider network (viz., the second term in Eq.~(\ref{eigenvalueapprox}) yields  the shift $-p(1-2\epsilon)$ in $\Delta(\omega)$).  
If we neglect the `shift' correction to the eigenvalues that comes from the wider network [i.e. using $\Delta(\omega) = \omega^2/a^2$ in Eq.~(\ref{rhoer})], the  expression (\ref{rhoer}) agrees with those previously obtained for the symmetric case $\epsilon = 0$ \cite{rodgers1988density, semerjian2002sparse}.   Notice that compared with random graphs with geometric and power-law degree distribution, in this case the  correction   due to the wider-network is not a constant, and therefore including it  yields a different functional dependence of the spectral density on $\omega$,  giving our result slightly greater accuracy.     Using  (\ref{rhoer}) in  Eq.~(\ref{eq:max}) we find that in this case the leading eigenvalue  $\lambda_\mathrm{max} \sim \sqrt{\ln N/\ln \ln N}$, in correspondence with results for nondirected graphs~\cite{krivelevich2003largest}.

\begin{figure}[ht]
		\centering
		\begin{subfigure}[b]{1.0\textwidth}
			\centering
			\includegraphics[width=\textwidth]{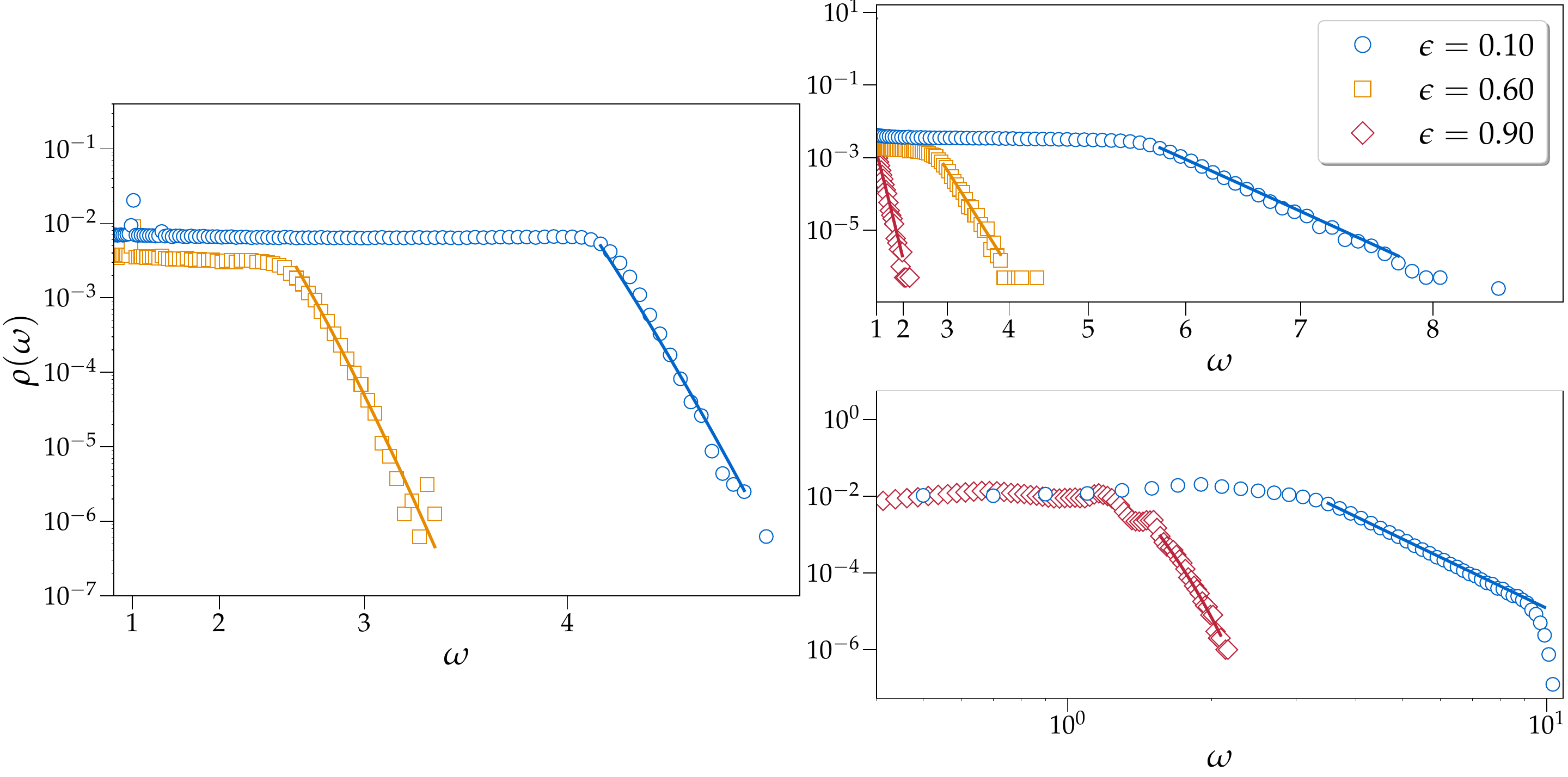}
		\end{subfigure}
		\caption[]{\textit{Tails of the eigenvalue spectral density for random graphs with Poisson (Erd\H{o}s-R\'{e}nyi), geometric and power-law degree distributions with $\pm1$ weights.}  Comparison between  empirical estimates  from  directly diagonalising  $10000$ graphs of size $N=4000$ (markers), generated as discussed in Sec.~\ref{app:generate_graphs_numerics} of the Supplement, and the asymptotic expressions predicted by the theory in Eqs.~\eqref{rhoer}, \eqref{eq:geometric_spectral_density_fixed_weights} and (\ref{epslhalf}-\ref{epsghalf}) with a fitted normalisation constant (solid lines).   The different panels correspond to different degree distributions: Erd\H{o}s-R\'{e}nyi graphs with mean degree  $p=4.0$ (left panel), geometric graphs with mean degree $p=4.0$ (top right panel), and power-law graphs with exponent $\gamma=3.5$ and $c_{\rm min}=2$ (bottom right panel). Each plots shows results for different values of the probability $\epsilon$ that a link is sign-antisymmetric: $\epsilon=0.10$ (blue circles), $\epsilon=0.60$ (orange squares) and $\epsilon=0.90$ (red diamonds). To emphasise the functional different functional forms, $y$-axis is plotted in $\log$ scale and the $x$-axis is plotted in a  quadratic scale for Erd\H{o}s-R\'{e}nyi and geometric degree distributions,  while it is in $\log$ scale for  power-law degree distributions.}
		\label{fig:spectral_denisty_grid_fixed_edge}
	\end{figure}

In Fig.~\ref{fig:spectral_denisty_grid_fixed_edge}, we compare the  expressions Eqs.~(\ref{eq:geometric_spectral_density_fixed_weights}), (\ref{epslhalf}), (\ref{epsghalf}), and (\ref{rhoer}) with the tails of the empirical distribution  of eigenvalues on the real axis estimated from directly diagonalising matrices of finite size $N=4000$. The figures show an excellent agreement between theory and numerical experiments. We observe that the theory captures well the different functional forms of the eigenvalue spectral density, depending on the degree distributions, and the predicted dependency on the fraction $\epsilon$ of sign antisymmetric links.

In order to complement these results, it would be interesting to theoretically investigate the eigenvalue distribution in the tail region graphs using the population dynamics algorithm described in Refs.~\cite{metz2019spectral,mambuca2022dynamical}. However, for the reasons discussed in~\ref{app:popdyn}, the population dynamics algorithm correctly matches the direct diagonalisation results in the bulk region but it does not in the tail region, where it predicts a sharp collapse of the spectral density. This aspect further highlights the significance of the quantitative theory developed in the present paper, which characterises analytically such tails in the spectrum of locally tree-like asymmetric random graphs. 

In the above examples the effect of the terms in Eq.~(\ref{eigenvalueapprox}) coming from the wider network (those that are order $|\Phi_i|^0$)  is small.  For random graphs with geometric or power law degree these correction terms only contribute the constant prefactor, and thus do not affect the functional form of $\rho$ on $\omega$.    However, we show in the following that to capture the variation of $\overline{q^{-1}(\omega)}$ with $\omega$ (beyond the limiting value for $\omega\to \infty$), the effect of the wider network is crucial.	

\subsection{Asymptotic disorder-averaged inverse participation ratio}    

We  turn our attention to the IPR. As we saw in the last section, many different arrangements of the network surrounding a well-connected hub can lead to the same eigenvalue. We therefore wish to find the average IPR \textit{conditioned} on there being an eigenvalue at $\omega$. 

More specifically, from Eq.~(\ref{iprgeneral}), we find that if the edge-weight distribution is given by Eq.~(\ref{piconstantmag}) then the IPR of states with large eigenvalues $\omega$ can be written solely in terms of the values of $c$ and $\Delta$ (see Sec.~\ref{appendix:configurationipr} of the Supplement). Given that $\Delta$ is a proxy for $\omega$ [if $\omega$ is large they are related  by $\Delta = \omega^2/a^2-(p_2-p)(1-2\epsilon)/p$],  we need to  average the expression for the IPR using $P(c \vert \Delta)$, similarly as we have done  in the previous section for  eigenvalue spectral density.

\begin{figure}[ht]
		\centering
		\begin{subfigure}[b]{0.49\textwidth}    \includegraphics[width=\textwidth]{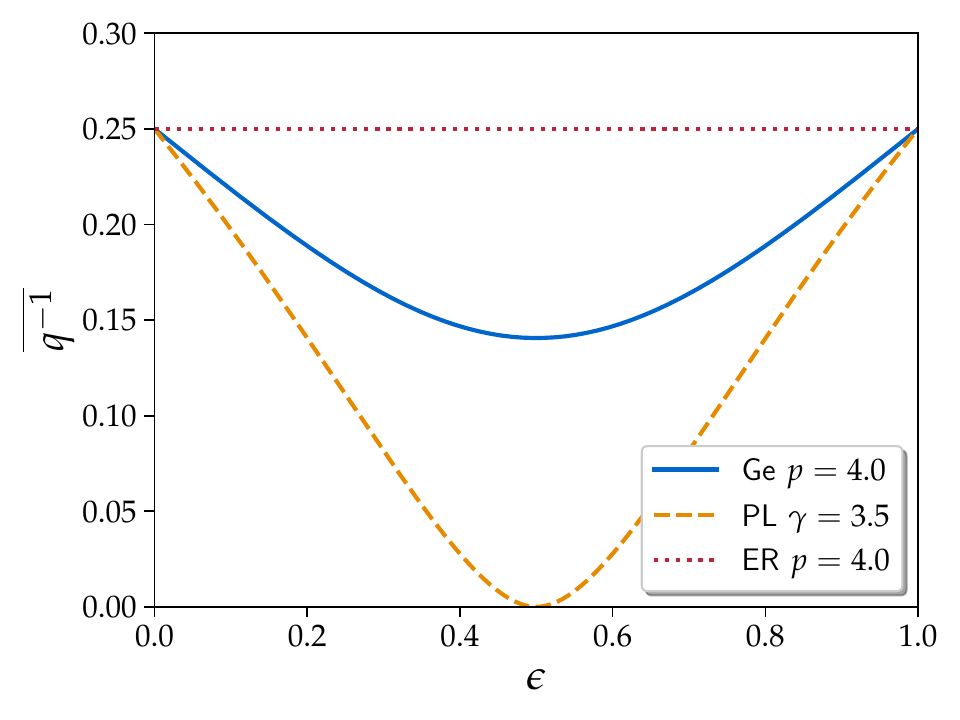}
		\end{subfigure}
		\begin{subfigure}[b]{0.49\textwidth}
			\includegraphics[width=\textwidth]{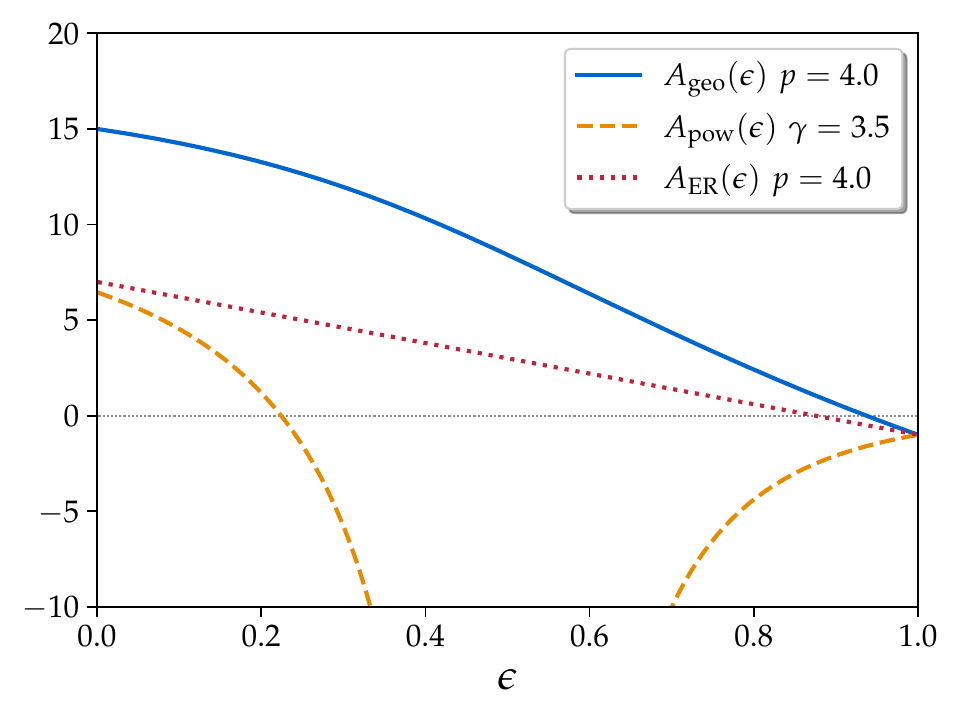}
		\end{subfigure}
		\caption[]{Left Panel: Plot of the asymptotic IPR $\lim_{\omega\rightarrow \infty}\overline{q}^{-1}(\omega)$  as a function of $\epsilon$  for geometric, power-law, and Erd\H{o}s-R\'{e}nyi graphs, as given by Eqs.~(\ref{iprgeo}), (\ref{iprpow}), and (\ref{iprer}), respectively.    Right Panel: the coefficients $A_{\rm geo}$ [Eq.~(\ref{eq:AGeo}) of the Supplement], $A_{\rm pow}$ [Eq.~(\ref{eq:APow}) of the Supplement], $A_{\rm ER}=2p(1-\epsilon)-1$ [Eq.~(\ref{iprer}) of the Supplement] that determine the first order correction as a function of $\epsilon$.    The parameters chosen are $p=4$ for geometric and Erd\H{o}s-R\'{e}nyi, and   $\gamma=3.5$ and $c_{\rm min}=2$ for power-law graphs.}
		\label{fig:ipr_epsilon_fixed_edge}
	\end{figure}

As we derive in Sec.~\ref{appendix:configurationipr} of the Supplement, for graphs with a   geometric degree distribution
\begin{align}
	\overline{q^{-1}(\omega)} = \left( \frac{1}{1 + \frac{\alpha}{\sqrt{\alpha^2 - 4 \epsilon(1-\epsilon)}}}\right)^2 \left[ 1 - \frac{a^2 A_\mathrm{geo}(\epsilon, p)}{\omega^2} \right] + \mathcal{O}\left(\frac{1}{\omega^4}\right), \label{iprgeo}
\end{align}
where $\alpha=1+1/p$.
For graphs with power-law degree distributions, the average IPR takes the form  
\begin{align}
	\overline{q^{-1}(\omega)} = \left( \frac{1}{1 + \frac{1}{\vert1-2\epsilon\vert}}\right)^2 \left[ 1 - \frac{a^2 A_\mathrm{pow}(\epsilon, \gamma)}{\omega^2} \right] + \mathcal{O}\left(\frac{1}{\omega^4}\right). \label{iprpow}
\end{align} 
Note that $\lim_{\omega\rightarrow \infty}\overline{q^{-1}}(\omega)$ is invariant under  $\epsilon\rightarrow 1-\epsilon$ and reaches its minimum value at $\epsilon=1/2$, as shown in the Left Panel  of Figure~\ref{fig:ipr_epsilon_fixed_edge}.   In general, $\lim_{\omega\rightarrow \infty}\overline{q^{-1}}(\omega)>0$, implying that right eigenvectors are localised, except for   power-law random graphs with  $\epsilon=1/2$, for which the IPR equals zero, and thus the right eigenvector is delocalised.      The single-defect approximation, which is based on non-overlapping eigenstates, therefore breaks down in this special case. This matter is discussed further in Sec.~\ref{appendix:crossover} of the Supplement. Together, Eqs.~(\ref{iprgeo}) and (\ref{iprpow}) demonstrate that localisation is minimised when there is a balance of reinforcing and antagonistic links.  

A prescription for the coefficients $A_\mathrm{geo}$ and $A_\mathrm{pow}$ is given in Sec.~\ref{appendix:configurationipr} of the Supplement, and Fig.~\ref{fig:ipr_epsilon_fixed_edge} shows a plot of these coefficients as a function of $\epsilon$.     
  The coefficient $A_\mathrm{geo}(\epsilon, p)$  is usually positive, except when there is an abundance of antagonistic links ($\epsilon\approx 1$), and  $A_\mathrm{pow}(\epsilon, \gamma)$ is usually negative except if there are an abundance of sign-symmetric links ($\epsilon \approx0$).   So,  both the degree distribution and the asymmetry parameter $\epsilon>0$ affect how the amount of localisation varies with $\omega$.

\begin{figure}[ht]
	\centering
	\begin{subfigure}[b]{1.0\textwidth}
		\centering
		\includegraphics[width=\textwidth]{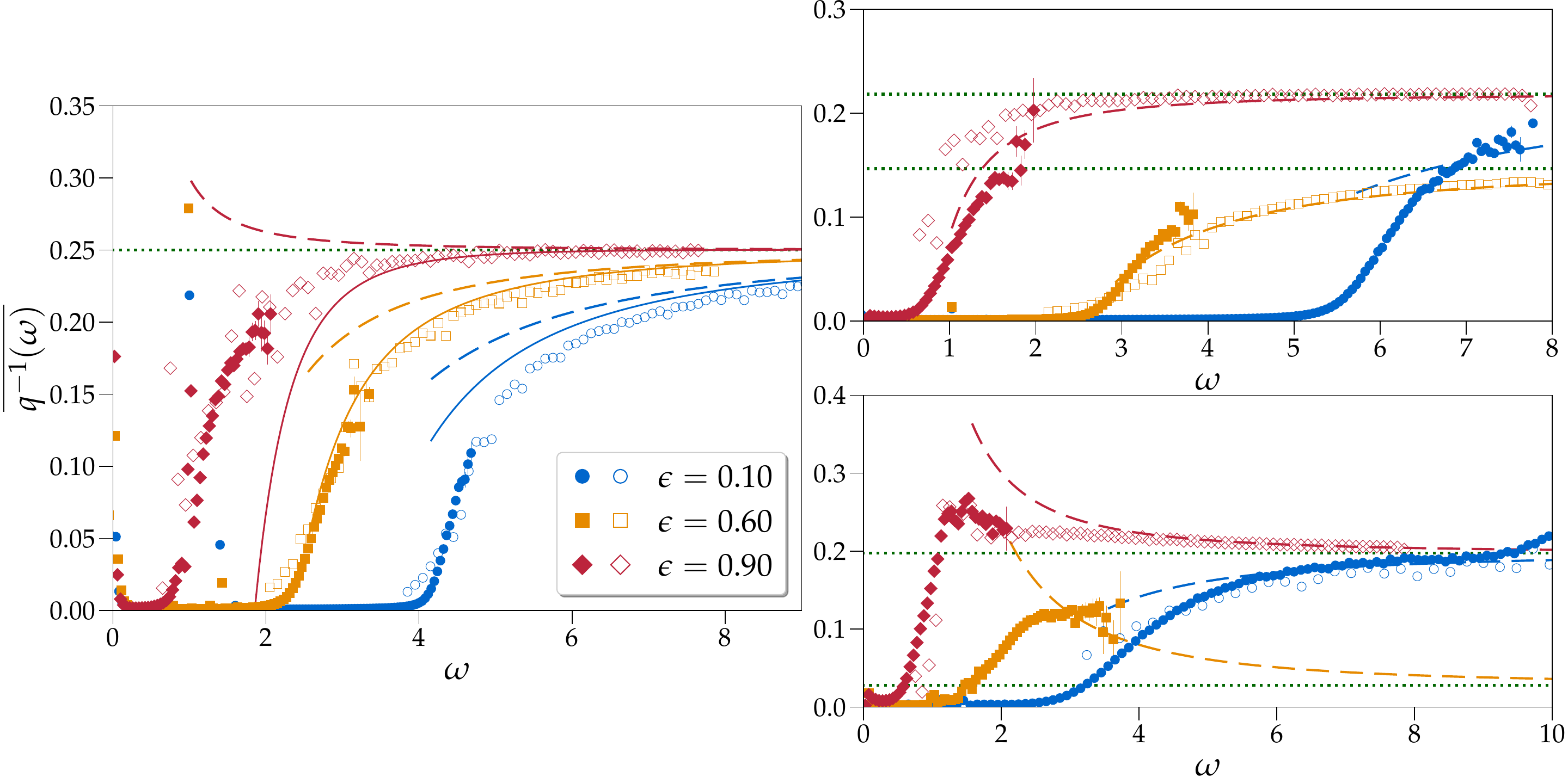}
	\end{subfigure}
	\caption[]{\textit{The disorder-averaged inverse participation ratio for Erd\H{o}s-R\'{e}nyi, geometric, and power-law graphs with $\pm1$ weights.}   Analytical predictions (lines) are compared with numerical diagonalisations results (filled markers) and  numerical results obtained through the single defect numerical approach described in Sec.~\ref{app:single_defect_numerics} of the Supplement (unfilled markers).   The  three panels show  results for, respectively,   Erd\H{o}s-R\'{e}nyi graphs with mean degree $p=4.0$ (left panel), geometric graphs with mean degree $p=4.0$ (top right panel), and  power-law graphs with exponent $\gamma=3.5$ and $c_{\rm min}=2$ (bottom right panel), and for three values of the sign-antisymmetric probability $\epsilon$ as indicated in the legend of the left panel. The geometric and power-law graphs are generated as discussed in Sec.~\ref{app:generate_graphs_numerics} of the Supplement.
    The analytical predictions are plots of the Eqs.~(\ref{iprer}), (\ref{iprgeo}) and (\ref{iprpow}), with dotted lines denoting the leading order contribution  constant in $\omega$, dashed lines includes the $1/\omega^2$ term, and the solid line contains all terms up to order $1/\omega^4$ (only for Erd\H{o}s-R\'{e}nyi graphs). Numerical results are obtained through direct diagonalisation of $10000$ graphs of size $N=4000$ by averaging all the single eigenvalues IPRs in a given interval. The error bars are the standard error of the mean. In Sec.~\ref{appendix:finite_size_effects_ipr} of the Supplement  one can find a finite size  analysis of the disorder-averaged IPR.}
	\label{fig:ipr_grid_fixed_edge}
\end{figure}

The expressions (\ref{iprgeo}) and (\ref{iprpow}) are corroborated in Fig.~\ref{fig:ipr_grid_fixed_edge} with computer generated random matrices.  We observe that greater accuracy is achieved with the inclusion of successive terms in the series for $\overline{q^{-1}(\omega)}$, which correspond to the inclusion of successive layers of neighbouring nodes (see Fig. \ref{fig:stars}). In Fig.~\ref{fig:ipr_grid_fixed_edge}, the dotted line shows the constant leading order contribution, and the dashed lines also contain the subleading $\omega^{-2}$ terms.

For the IPR in the Erd\H{o}s-R\'{e}nyi graph case, we arrive at the  following asymptotic expression for the IPR (see Sec.~\ref{appendix:poissonian} of the Supplement)
\begin{equation} \label{iprer}
	\begin{split}
		\overline{q^{-1}(\omega)} =& \frac{1}{4} \biggr( 1 - \frac{ a^2[2p(1-\epsilon) -1]}{\omega^2} -  \\
		&- \frac{ a^4p^2}{4 \omega^4} \left[ 8 \epsilon(1-\epsilon) + 1 +6(1-2\epsilon) + 5(1-2\epsilon)^2+12/p \right] \biggr) + \mathcal{O}\left(\frac{1}{\omega^6}\right).
	\end{split}
\end{equation}
In this case,  the asymptotic value of the IPR is unaffected by the  fraction of sign asymmetric links (see also Left Panel of Fig.~\ref{fig:ipr_epsilon_fixed_edge}), as $\epsilon$ only appears in the subleading term. This is because the most likely hub configuration at given $\omega$ is one with $\mathcal{O}\left(1/\Delta\right) = \mathcal{O}\left(1/\omega^2\right)$ antagonistic links. For this reason, we were able to include the effects of a second layer of nodes in Eq.~(\ref{iprer}), thus improving  the accuracy of the approximation up to order $\mathcal{O}(\omega^{-4})$.  

We note that some previous attempts at finding the IPR in the tail regions of the ER graph spectrum using the so-called Effective-Medium and Shell approximations have failed to reproduce numerical results \cite{slanina2012localization}, incorrectly predicting an asymptotic value of $\lim_{\omega\to\infty}\overline{q^{-1}(\omega)} = 1$. Our prediction appears a better fit for the numerical results in Ref.~\cite{slanina2012localization} for example, which studies the symmetric case $\epsilon = 0$. We also test the results for the ER graph numerically for $\epsilon \neq 0$ in Fig. \ref{fig:ipr_grid_fixed_edge}, well corroborating the formula (\ref{iprer}).

	\section{Asymptotic eigenvalue density and IPR on weighted graphs: general theory}\label{section:saddlepoint}
	
	In Section \ref{section:example:constantweight}, we have extracted the eigenvalue spectral density and disordered-averaged IPR for large values of $\omega$ by averaging over local network configurations with high effective degree $|\Phi_i|$. This was done in the special case where the edge weights $J_{ij}$ have a fixed absolute value. We now extend the asymptotic analysis to the case of continuously varying edge weights. We also discuss briefly how our approach can break down when the edge-weight distribution is heavy-tailed, favouring states localised on single edges.

    In the interest of space, we focus on the leading order term in the asymptotic expansion of the eigenvalue density and the IPR. Consequently, we ignore the subleading contributions that come from either the wider network or from fluctuations of the hub connectivity $c$ about its most likely value for a given $\omega$.  This approach is equivalent with  the consideration that the eigenvalues in the tails are those of   `star' graphs (see the left-hand panel of Fig. \ref{fig:stars}), and we thus seek to describe the `typical' kinds of star graphs that yield a large given value of $\omega$.  

    In Secs.~\ref{section:saddledensity} and \ref{sec:saddlePointIPR}  we develop  the general approach for obtaining the  eigenvalue density and the IPR, respectively, in the spectral tails of sparse non-Hermitian random matrices.      In these sections, we focus on the method, and do not yet specify the distribution of the weights $J_{ij}$.  
    In Sec.~\ref{section:example:general}, we derive explicit expressions for the eigenvalue density for an example, namely, with weights $J_{ij}$ that are drawn from an exponential distribution.     In Sec.~\ref{sec:discsec4}, we discuss how  the different qualitative results   for the eigenvalue spectral density and the IPR are related to the differences in the properties of network. In addition, in  Sec.~\ref{sec:discsec4} we discuss the assumptions on  which the method we develop in this section relies and how they can break down.

	\subsection{Eigenvalue spectral density}\label{section:saddledensity}

    According to the formula (\ref{eigenvalueapprox}) sparse non-Hermitian matrices have tail eigenvalues of the form (up to leading order) $\lambda^2=  \Phi_i$
    where  $\Phi_i$ is the effective degree of a 'hub' $i$ with large $|\Phi_i|$.     Therefore, for   large enough values of $\omega$ we approximate  the eigenvalue density as follows
	\begin{align}
		\rho(\omega) \approx \omega \int \dd\Phi \sum^{\infty}_{c=0} \delta(\omega^2 - \Phi ) P(\Phi \vert c)P_\mathrm{deg}(c) , \label{densitygeneral}
	\end{align}
    and thus 
	\begin{align}
		P(\Phi \vert c ) &=  \int \prod^c_{j=1}(\dd J_{ij} \dd J_{ji} \pi(J_{ij}, J_{j,i}) )   \delta\left(\Phi - \sum_{j= 1}^c J_{ij}J_{ji}\right) \nonumber \\
		&= \int \frac{\dd \hat\Phi}{2\pi} \exp\left({\rm i} \hat \Phi \Phi + c \ln \left\langle e^{-{\rm i}\hat\Phi uv}\right\rangle_{\pi(u,v)} \right) .\label{pdeltagivenk}
	\end{align}
   Note that here  $P(\Phi \vert c)P_\mathrm{deg}(c)$ in (\ref{densitygeneral})  is the analogue of $P(\Delta, c)$ in Eq.~(\ref{configdensitysum}) for the case of continuously distributed edge weights.

    We analyse the  distribution $P(\Phi|c)$ of the effective degree $\Phi$  conditioned on the degree $c$ in the limit when $\Phi$ is large.   
    For now, we  presume that if   $\Phi$ is large then also  $c$  is  large.    We will re-examine this assumption \textit{a posteriori}.   Using that in the tails of the spectrum $\Phi$ is large, we 
    perform a saddle-point approximation for the integral over $\hat\Phi$ in (\ref{pdeltagivenk}), and we obtain (after relabelling $\hat\Phi_s=\iu\xi$, where $\hat\Phi_s$ is the value of $\hat\Phi$ at the saddle point)
		\begin{align}
			\ln P(\Phi \vert c) \approx -\Phi\xi  + c \beta , \label{lnpdgivek}
		\end{align}
		where $\beta$ is  defined through 
		\begin{align}
			\beta = \ln \left\langle e^{uv\xi} \right\rangle_{\pi(u,v)}, \label{betadef}
		\end{align}
		and  $\xi$ is the solution of the  saddle point equation
		\begin{align}
			\frac{\left\langle uv \: e^{uv\xi } \right\rangle_{\pi(u,v)}}{\left\langle e^{uv\xi } \right\rangle_{\pi(u,v)}} = \frac{\Phi}{c} . \label{saddlexi}
		\end{align}
        Note that both $\xi$ and $\beta$ are functions of $\Phi/c$.

        Next, we need to evaluate the sum over $c$ in Eq.~(\ref{densitygeneral}).     For this purpose, we will replace the sum over the degrees $c$, i.e.,   $\sum_c \to f(\omega)\int \dd\kappa$,   through a transformation of the form 
		\begin{equation}
			 \kappa  = \frac{c}{f(\omega)}\,  \label{eq:continuousAssum}
		\end{equation}
       where $f(\omega)$ is a function that captures how typical values of $c$  scale with $\omega$.

	To find the function $f(\omega)$, we  assume that  $P(\Phi \vert c)P_\mathrm{deg}(c)$ is peaked around some typical value of $c$ for fixed large $\Phi$ (similarly as for the quantity $P(\Delta ,c)$ in  Sec.~\ref{section:example:constantweight}), which we call $c_{\rm max}(\omega)$.   If this condition applies, which examples show is often the case, then there is a trade-off between $P_\mathrm{deg}(c)$ (which will always be a decreasing function of $c$ for large $c$) and $P(\Phi \vert c)$ (which is  peaked around some other value $c^{(0)}_{\rm max}(\omega)$ of $c$, as we discuss in Sec.~\ref{app:PHiC}). Finding the value $c_{\rm max}(\omega)$ of $c$ that maximises the product $P(\Phi \vert c)P(c)$ for fixed $\Phi$ tells us the typical connectivity of hubs that contribute  an eigenvalue $\omega = \sqrt{\Phi}$. This typical value $c_{\rm max}$ dominates the sum over $c$, and thus provides us with the leading order term in the large $\omega$-expansion of the  eigenvalue density.    We capture the asymptotic   scaling of $c_{\rm max}(\omega)$ with $\omega$ through a function $f(\omega)$  so that 
    \begin{equation}
    \lim_{\omega \rightarrow \infty}  \frac{c_{\rm max}(\omega)}{f(\omega)}  \label{eq:fomegaDef}
    \end{equation}
    converges to a nonzero constant.     One notes that although $f(\omega) \sim \omega^2$ for the cases studied in Section \ref{section:example:constantweight}, this is no longer necessarily the case here.

    Assuming that $f(\omega)$ is large for large values of $\omega$, we can use 
  (\ref{eq:continuousAssum})  to replace the sum in (\ref{densitygeneral}) by an integral.  Using again that $\Phi$ is large, we   perform a further saddle-point approximation of the integral over $\kappa$ to arrive at 
		\begin{align}
			\ln \rho(\omega) \sim -\omega^2 \xi+ f(\omega)\kappa \ln \langle e^{uv \xi} \rangle_{\pi(u,v)}  + \ln P_\mathrm{deg}\left(f(\omega) \kappa\right).  \label{logrho}
		\end{align} 
		Here, $\kappa$ is a constant that solves the saddle point equations
		\begin{align}
			\ln \langle e^{uv \xi} \rangle_{\pi(u,v)} + \frac{P_\mathrm{deg}'\left(f(\omega) \kappa\right)}{P_\mathrm{deg}\left(f(\omega) \kappa\right)} = 0, \label{saddlekappa}
		\end{align}
         $P'_{\rm deg}$ is the derivative of $P_{\rm deg}$, 
       and  $\xi$ solves 
      \begin{align}
			\frac{\left\langle uv \: e^{uv\xi } \right\rangle_{\pi(u,v)}}{\left\langle e^{uv\xi } \right\rangle_{\pi(u,v)}} = \frac{\omega^2}{\kappa f(\omega)}. \label{saddlexi2}
		\end{align}
    To derive the above formulae   we have also used the delta distribution in  (\ref{densitygeneral}) to  equate $\Phi = \omega^2$.

        In conclusion, the asymptotic expression for $\rho(\omega)$ is given by  Eq.~(\ref{logrho}) with $\xi(\omega)$, $\kappa(\omega)$, and $f(\omega)$ functions of $\omega$ that are     are determined from solving  the coupled set  of saddle point equations  (\ref{saddlekappa}) and (\ref{saddlexi2}) for large, but arbitrary values of $\omega$.   The function $f(\omega)$ is defined through  the scaling in Eq.~(\ref{eq:fomegaDef}).   However, in practice we determine $f(\omega)$  by requiring that   the function  $\kappa(\omega)$ solving (\ref{saddlekappa}) converges for large values of $\omega$  to a constant value   (note that  $\xi$ may grow indefinitely with $\omega$).   
       We need to \textit{a posteriori} verify that $f(\omega)\to \infty$ as $\omega\to \infty$, as otherwise  the second saddle-point approximation made is not valid (we discuss this in more detail in  Sec.~\ref{section:breakdown}).

		\subsection{Disorder-averaged IPR at fixed \texorpdfstring{$\omega$}{omega}} \label{sec:saddlePointIPR}
		
		We can also compute the asymptotic value of the IPR as $\omega\to\infty$ by proceeding along similar lines. For  positive real $\omega$ we get 
		\begin{align}
			\overline{q^{-1}(\omega ) }&\approx [\rho(\omega)]^{-1}\sum_{c}  P_\mathrm{deg}(c) \int  \prod_{j = 1}^{c} \dd J_{ij} \dd J_{ji} \pi\left( J_{ij}, J_{ji} \right)    \nonumber \\
			&\hspace{2cm}\times\delta\left( \omega - \sqrt{\sum_{j = 1}^{c} J_{ij}J_{ji}  }  \right) \left(\frac{\sum^c_{j=1} J_{ij}J_{ji}}{ \sum^c_{j=1} J_{ij}J_{ji} + \sum^c_{j=1} J_{ji}^2} \right)^2, \label{generaliprintegral}
		\end{align}
		where we have kept only the leading term from Eq.~(\ref{iprgeneral}), and thus will obtain only the asymptotic value of the IPR for $\omega\to \infty$. To obtain  the subleading  correction term  of order  $\mathcal{O}(1/\omega^2)$ to the IPR, we must take into account corrections to the saddle-point approximation and the higher-order terms from Eq.~(\ref{iprgeneral}). 
		
		Assuming that the tail is dominated by contributions from highly connected hubs, we find that the asymptotic value of the IPR is given by (as we show in Sec.~\ref{appendix:saddleipr} of the Supplement)
		\begin{align}
			\lim_{\omega\to \infty}\overline{q^{-1}(\omega ) }&= \lim_{\omega \to \infty}\left(\frac{\omega^2}{\omega^2 + \Psi(\omega)}\right)^2, \label{iprpsiphi}
		\end{align}
		where $\Psi(\omega)$ solves the following set of simultaneous equations (eliminating $\kappa$ and $\xi$)
		\begin{equation} \label{saddlepsi}
			\begin{split}
				\Psi(\omega) &= f(\omega)\kappa \frac{\langle v^2 e^{uv \xi}\rangle_{\pi(u,v)}}{\langle e^{uv \xi}\rangle_{\pi(u,v)}},
			\end{split}
		\end{equation}
        and where $\xi$ and $\kappa$ solve the Eqs.~(\ref{saddlexi2}) and (\ref{saddlekappa}), respectively.  
		Note that the formula for the IPR applies under the same circumstances for which the saddle-point approximation of the eigenvalue density given in Section \ref{section:saddledensity}  holds (we come back on this assumption  in more detail in  Sec.~\ref{section:breakdown}).
		
		\subsection{Example: Exponential distribution of the product \texorpdfstring{$\vert J_{ij}J_{ji} \vert$}{}} \label{section:example:general}
		
		We  examine the spectral tails for the three canonical random graphs (with geometric, power-law and Poisson degree distributions) and with weights drawn from an example distribution $\pi(J_{ij}, J_{ji})$ of the form given by Eq.~(\ref{eq:PiWeight}), where now $J_{ij}$ take values on the full real axis.   

        We construct $\pi(J_{ij}, J_{ji})$ according to the following procedure.
       We first randomly set the signs of $J_{ij}$ and $J_{ji}$ such that the product $J_{ij}J_{ji}$ is positive with probability $1-\epsilon$, and negative with probability $\epsilon$. We subsequently draw the absolute values of the product $y=\vert J_{ij}J_{ji} \vert$ and of the ratio $z=\vert J_{ij}/J_{ji} \vert$ from continuous distributions $P(y)$ and $R(z|y)$.   Hence, the distributions  $\pi_{\rm r}$ and $\pi_{\rm a}$  in Eq.~(\ref{eq:PiWeight}) can be expressed as 
           \begin{eqnarray}
    \fl \pi_{\rm r}(J_{ij},J_{ji})  = \frac{1}{2}\int^{\infty}_0 \dd y  P(y) \int^{\infty}_0 \dd z R(z|y)
      \nonumber\\ 
     \times \left(\delta(J_{ij}-\sqrt{yz})\delta(J_{ji}-\sqrt{y/z}) +\delta(J_{ij}+\sqrt{yz})\delta(J_{ji}+\sqrt{y/z}) \right) \label{eq:pir}
      \end{eqnarray}
      and 
        \begin{eqnarray}
    \fl \pi_{\rm a}(J_{ij},J_{ji})  = \frac{1}{2} \int^{\infty}_0 \dd y  P(y) \int^{\infty}_0 \dd z R(z|y)
      \nonumber\\ 
     \times \left(\delta(J_{ij}-\sqrt{yz})\delta(J_{ji}+\sqrt{y/z}) +\delta(J_{ij}+\sqrt{yz})\delta(J_{ji}-\sqrt{y/z}) \right). \label{eq:pia}
      \end{eqnarray} 
      
    For exponentially distributed weights we set 
    \begin{equation} \label{expedgedist}
    			\begin{split}
    				P(y) = \frac{\ec^{-y/d}}{d} \quad , \quad y \in [0, +\infty],
    			\end{split}
    \end{equation}
    and 
    \begin{equation}
    R(z|y) = \frac{\delta(z  + \nu) + \delta(z  + \nu^{-1}) }{2}.
    \end{equation}
    The parameter $d$ is a scaling parameter for the weights, and the  $\nu$ tells us about the magnitude of the asymmetry between $J_{ij}$ and $J_{ji}$.

        In Sec.~\ref{appendix:expproduct}, we use the  saddle point procedure of Sec.~\ref{section:saddledensity} and \ref{sec:saddlePointIPR} to  calculate the asymptotic eigenvalue densities and the IPR in the cases  of the geometric, power-law, and Poisson degree distributions.   We report on the final results below.  
		
\begin{figure}[ht]
			\centering
			\begin{subfigure}[b]{0.49\textwidth}
				\centering
				\includegraphics[width=\textwidth]{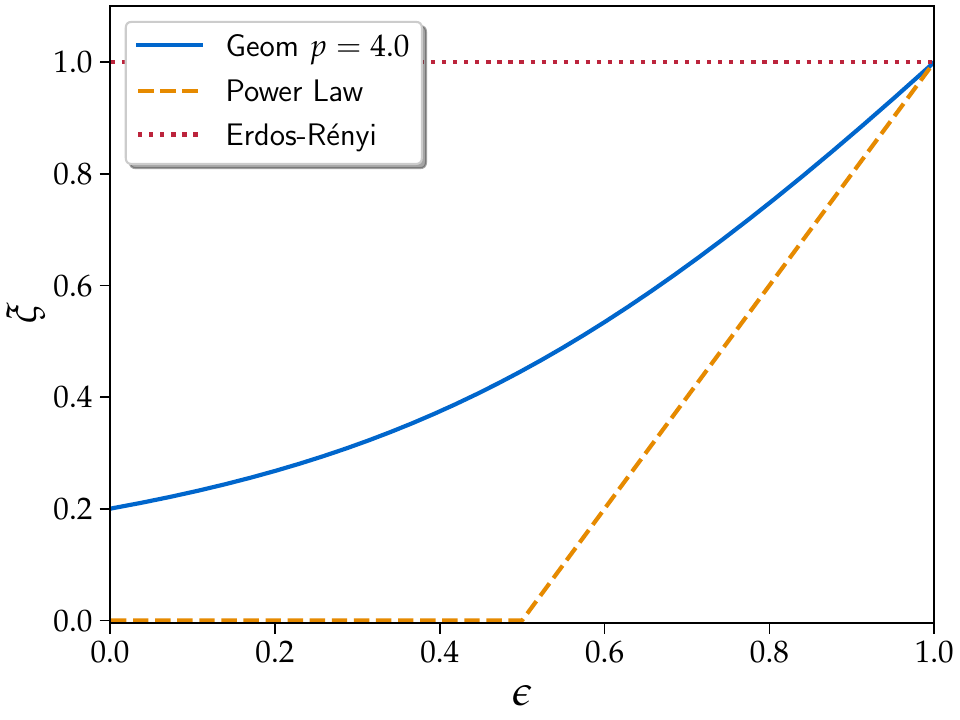}
			\end{subfigure}
			\begin{subfigure}[b]{0.49\textwidth}
				\centering
				\includegraphics[width=\textwidth]{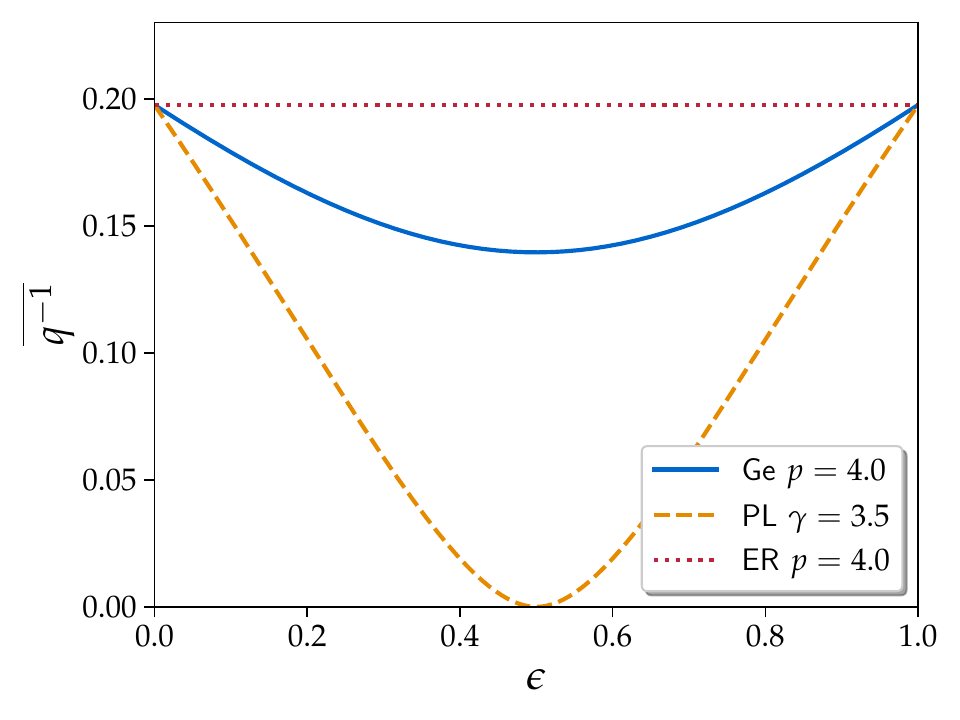}
			\end{subfigure}
			\caption[]{On the left: plots of $
				\xi$ as a function of $\epsilon$ for geometric and  power-law random graphs given by, respectively,  $\xi_{\rm geo} = -\frac{(1-2\epsilon)p}{2d(p+1)} + \frac{1}{d}\sqrt{\left[\frac{(1-2\epsilon)p}{2(p+1)}\right]^2 + \frac{1}{(p+1)}}$, $\xi_{\rm pow} = \Theta(\epsilon-1/2)\frac{2\epsilon-1}{d}$, and $\xi_{\rm ER} = 1/d$, which are independent of $\nu$. On the right: plots of the asymptotic prediction for the inverse participation ratio $\lim_{\lambda\to\infty}\ipr(\lambda)$ as a function of $\epsilon$.  For power-law degree distributions this is Eq.~\eqref{eq:ipr_saddle_point_exp_weights_powerlaw}, for Erd\H{o}s-R\'{e}nyi graphs this is Eq.~\eqref{eq:IPR_exp_weights_prod_er}, and for geometric degree distributions the plot shows the function obtained from combining Eq.~\eqref{eq:xi_saddle_point_exp_weights_geometric_app} with Eq.~\eqref{eq:ipr_example_exp_weights} of the Supplement.  The parameters are set to  $d=1$ (for both plots) and $\nu=2$ (only relevant for the right panel).   }
			\label{fig:xi&ipr_epsilon_expprod_weights}
		\end{figure}

		In the case of the geometric degree distribution with $\ln P_\mathrm{deg}(c) \sim -c\ln ((p+1)/p)$, we obtain for the leading order asymptotics of the spectral density the expression
		\begin{align} \label{eq:spectral_density_saddle_point_exp_weights_geometric}
			\rho(\omega) \sim \omega e^{-\xi_{\rm geo} \omega^2}, 
		\end{align}
		with $\xi_{\rm geo}$ a decay constant independent of $\omega$ whose explicit expression as a function of the model parameters is given in Fig.~\ref{fig:xi&ipr_epsilon_expprod_weights}.     As with the case of constant edge weights,  the typical connectivity of a hub associated with an eigenvalue $\omega$ scales as  $c \sim f(\omega) = \omega^2$.  We therefore find that the density has a similar exponential decay with $\omega^2$ to the case with fixed edge weights (see Eq.~\eqref{eq:geometric_spectral_density_fixed_weights}), albeit with a different decay constant $\xi_{\rm geo}$, and the same scaling of the typical leading eigenvalue $\lambda_{\rm max} \sim \sqrt{\ln N}$. The explicit expression for $\ipr(\omega)$ is obtained in Sec.~\ref{app:geom5} of the Supplement, but  is too long to be presented here (see the right Panel of Fig.~\ref{fig:xi&ipr_epsilon_expprod_weights}).   Therefore, we examine three informative limiting cases:
		\begin{equation} \label{eq:ipr_example_exp_weights_geom}
			\lim_{\omega \to \infty}\ipr(\omega) = \left\{
			\begin{aligned}
				&\; \left(\frac{1}{1+\frac{1}{4}\left(\nu+\frac{1}{\nu}\right)\frac{2+p}{\sqrt{1+p}}}\right)^2 &, \; {\rm if}& \; \epsilon=0.5 \; ,\\
				\\
				&\; \left(\frac{2\nu}{(1+\nu)^2}\right)^2 &, \; {\rm if}& \; \epsilon=0 \quad{\rm or}\quad \epsilon=1 \; .
			\end{aligned} \right. 
		\end{equation}
    Notice that states are less localised  if there is an imbalance between the magnitudes of $J_{ij}$ and $J_{ji}$, as the IPR is decreased by moving away from the value $\nu = 1$.

		In the case of the power-law degree distribution with $P_\mathrm{deg}(c) \sim -\gamma\ln c$ we obtain instead
		\begin{equation} \label{eq:spectral_density_saddle_point_exp_weights_powerlaw}
			\rho(\omega) \sim \left\{
			\begin{aligned}
				&\; \omega^{-2\gamma+1} &, \; {\rm if}& \; \epsilon<0.5 \; ,\\
				\\
				&\; \omega^{-2\gamma+1} \; e^{-\xi_{\rm pow} \omega^2}\negthickspace\negthickspace &, \; {\rm if}& \; \epsilon>0.5 \; ,
			\end{aligned} \right.  
		\end{equation}
		where $\xi_{\rm pow}$ is given in Fig. \ref{fig:xi&ipr_epsilon_expprod_weights}, and the formula should again be understood as describing the leading order asymptotics for $\omega\gg 1$.   Just as for the geometric case, the quality of the tail behaviour has not changed compared to the fixed-weight case [c.f. Eqs.~(\ref{epsghalf}) and (\ref{epslhalf})], resulting in the same scaling of the leading eigenvalue $\lambda_{\rm max}$ with the size $N$. The scaling of the typical connectivity is once again $c \sim f(\omega) = \omega^2$, and the  IPR in this case is given by 
		\begin{equation} \label{eq:ipr_saddle_point_exp_weights_powerlaw}
			\begin{split}
				&\lim_{\omega \to \infty} \ipr(\omega)  = \left[\frac{1}{1+\frac{\left(\nu + \frac{1}{\nu}\right)}{2\lvert 1-2\epsilon \rvert}}\right]^2 \; .
			\end{split}
		\end{equation}
		
		Lastly, for Erd\H{o}s-R\'{e}nyi graphs with $P_\mathrm{deg}(c) \sim c - c\ln(c/p)$ we find the asymptotic expression
		\begin{equation} \label{eq:spectral_density_saddle_point_exp_weights_er}
			\rho(\omega) \sim \omega \exp\left( -\left(\omega - \sqrt{(1-\epsilon) pd} \; \right)^2 \Big/ d\, \right) \; ,
		\end{equation}
		which is, in this case as well, a Gaussian decay with a constant decay parameter $\xi_{\rm ER}= 1/d$, yielding, once again, the asymptotic scaling of the leading eigenvalue $\lambda_{\rm max} \sim \sqrt{\ln N}$. However, in this case there is  shift depending on the various model  parameters.  Thus, for the Erd\H{o}s-R\'{e}nyi ensemble  the edge-weight distribution changes the quality of the decay of the tail [viz., Eq.~(\ref{rhoer}) and (\ref{eq:spectral_density_saddle_point_exp_weights_er}) have a different functional form], due to the fact that the edge-weight distribution decays slower than the degree distribution in this case.  Indeed,  the typical connectivity of hubs is also modified to $c \sim f(\omega) = \omega$. For the IPR, we get
		\begin{align} \label{eq:IPR_exp_weights_prod_er}
			\lim_{\omega \to \infty}\ipr(\omega) =\left[\frac{2\nu}{(1+\nu)^2} \right]^2.
		\end{align}
		which does not depend on the sign-antisymmetric probability $\epsilon$, as was also found previously for the case with fixed weights.   

In the Left Panel of Fig.~\ref{fig:xi&ipr_epsilon_expprod_weights} we compare the expressions for the decay constant $\xi$ as a function of the probability $\epsilon$ across the different models.  Note the similarity with the behaviour of $\xi$ in  Fig.~\ref{fig:xiepsilon_fixed_edge} for the case with fixed edge weights.    In particular, in both cases $\xi$ is an increasing function with $\epsilon$.   However,  a notable difference can be observed  in the functional behaviour of $\xi$ when $\epsilon$ goes to $1$.  For bounded weights,  $\xi$  diverges  when $\epsilon\rightarrow 1$, whereas for unbounded weights $\xi$ converges to a finite value in this limit. Since  there are no spectral tails for random graphs with sign-antisymmetric weights~\cite{mambuca2022dynamical},  we can set $\xi=\infty$  for $\epsilon=1$, and thus for exponentially distributed weights there is a discontinuous  jump in the decay length of the spectral tail at $\epsilon = 1^{-}$ (see Fig.~\ref{fig:xi&ipr_epsilon_expprod_weights}).   On the other hand, the decay length vanishes continuously when the weights are bounded (see the divergence of $\xi$ for $\epsilon\rightarrow 1^-$ in Fig.~\ref{fig:xiepsilon_fixed_edge}).    Taken together, the qualitative appearance of the tails  as  a function of the fraction $\epsilon$ of sign-antisymmetric links depends on the distribution of weights.  

The Right Panel of Fig.~\ref{fig:xi&ipr_epsilon_expprod_weights}  compares the IPR as a function of $\epsilon$ across the different models.    The qualitative behaviour is similar to the one observed in the fixed-weight case of Fig.~\ref{fig:xiepsilon_fixed_edge}.  However, states are less localised if there is an 
 imbalance between the magnitudes of $J_{ij}$ and $J_{ji}$, as the IPR is decreased by moving away from the values $\nu = 1$.
		
\begin{figure}[ht]
			\centering
			\begin{subfigure}[b]{1.0\textwidth}
				\centering
				\includegraphics[width=\textwidth]{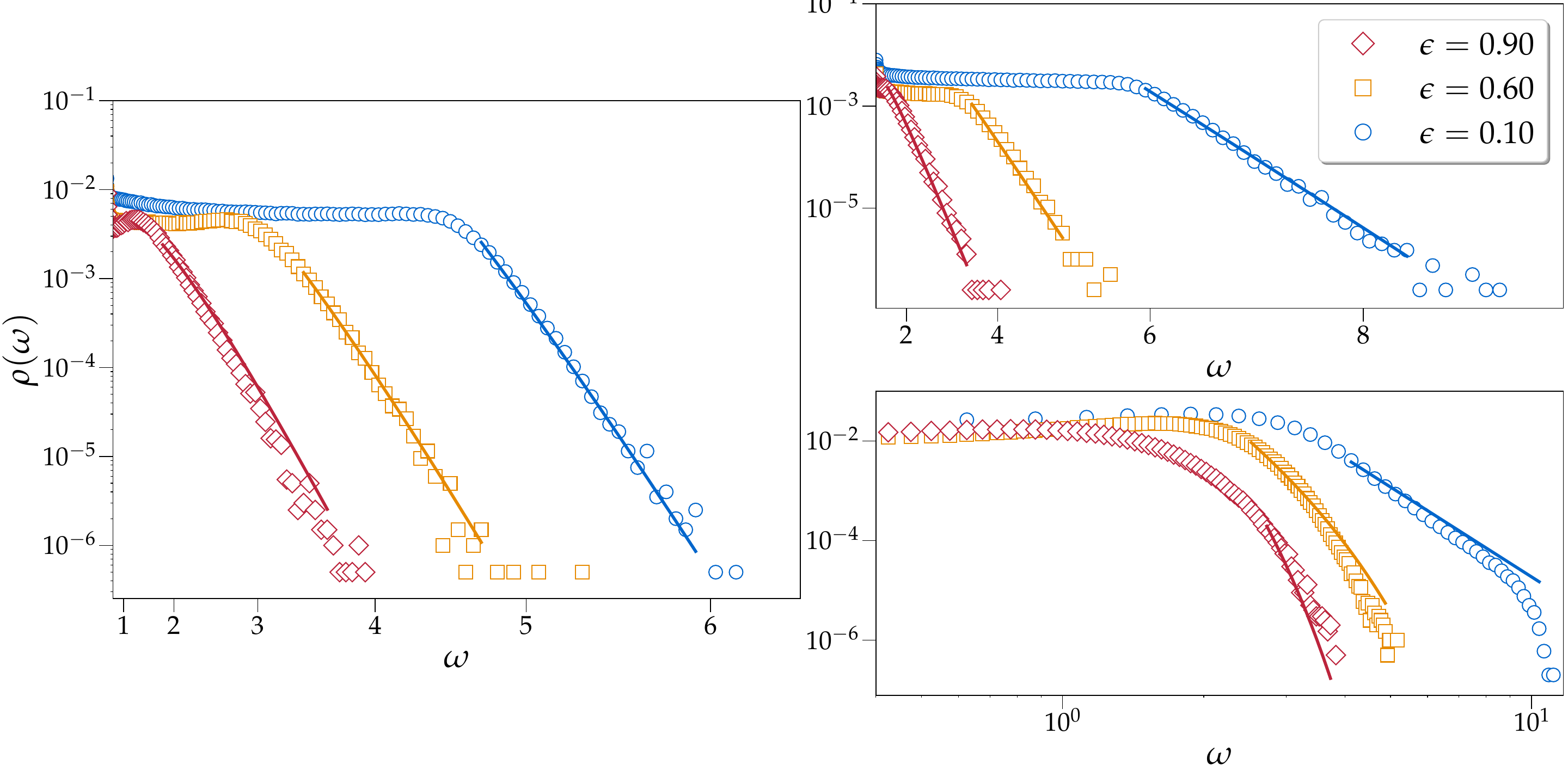}
			\end{subfigure}
			\caption[]{\textit{Tails of the eigenvalue density for random graphs with weights that are exponentially distributed.} Comparison between the asymptotic expression for $\rho(\omega)$   predicted by the theory  in Eqs.~\eqref{eq:spectral_density_saddle_point_exp_weights_er}, \eqref{eq:spectral_density_saddle_point_exp_weights_geometric} and \eqref{eq:spectral_density_saddle_point_exp_weights_powerlaw} (solid lines, obtained by simply fitting the normalisation constant) and numerical results obtained through direct diagonalisation (markers).  The edge weights are drawn from a distribution $\pi(J_{ij},J_{ji})$  as described at the beginning of Sec.~\ref{section:example:general} with parameters $d=1.0$ and $\nu=2$ and $\epsilon$ as given in the legend [$\epsilon=0.10$ (blue circles), $\epsilon=0.60$ (orange squares) and $\epsilon=0.90$ (red diamonds)].   The three panels correspond with  
            Erd\H{o}s-R\'{e}nyi graphs with connectivity $p=4.0$ (Left Panel), geometric graphs with connectivity $p=4.0$ (Top Right Panel) and power-law graphs with exponent $\gamma=3.5$ and $c_{\rm min}=2$ (Bottom Right Panel).  To emphasise the asymptotic scaling of $\rho(\omega)$ with $\omega$, the plots use logarithmic scaling on the  $y$-axis,  quadratic  scaling  on the  $x$-axis for Erd\H{o}s-R\'{e}nyi and geometric graphs, and logarithmic scaling on the $x$-axis for power-law random graphs.   Each  of the numerical estimates of  $\rho(\omega)$ are based on numerically diagonalising $10000$ graphs of size $N=4000$.}
			\label{fig:spectral_denisty_grid_expprod_weights}
		\end{figure}

\begin{figure}[ht]
			\centering
			\begin{subfigure}[b]{1.0\textwidth}
				\centering
				\includegraphics[width=\textwidth]{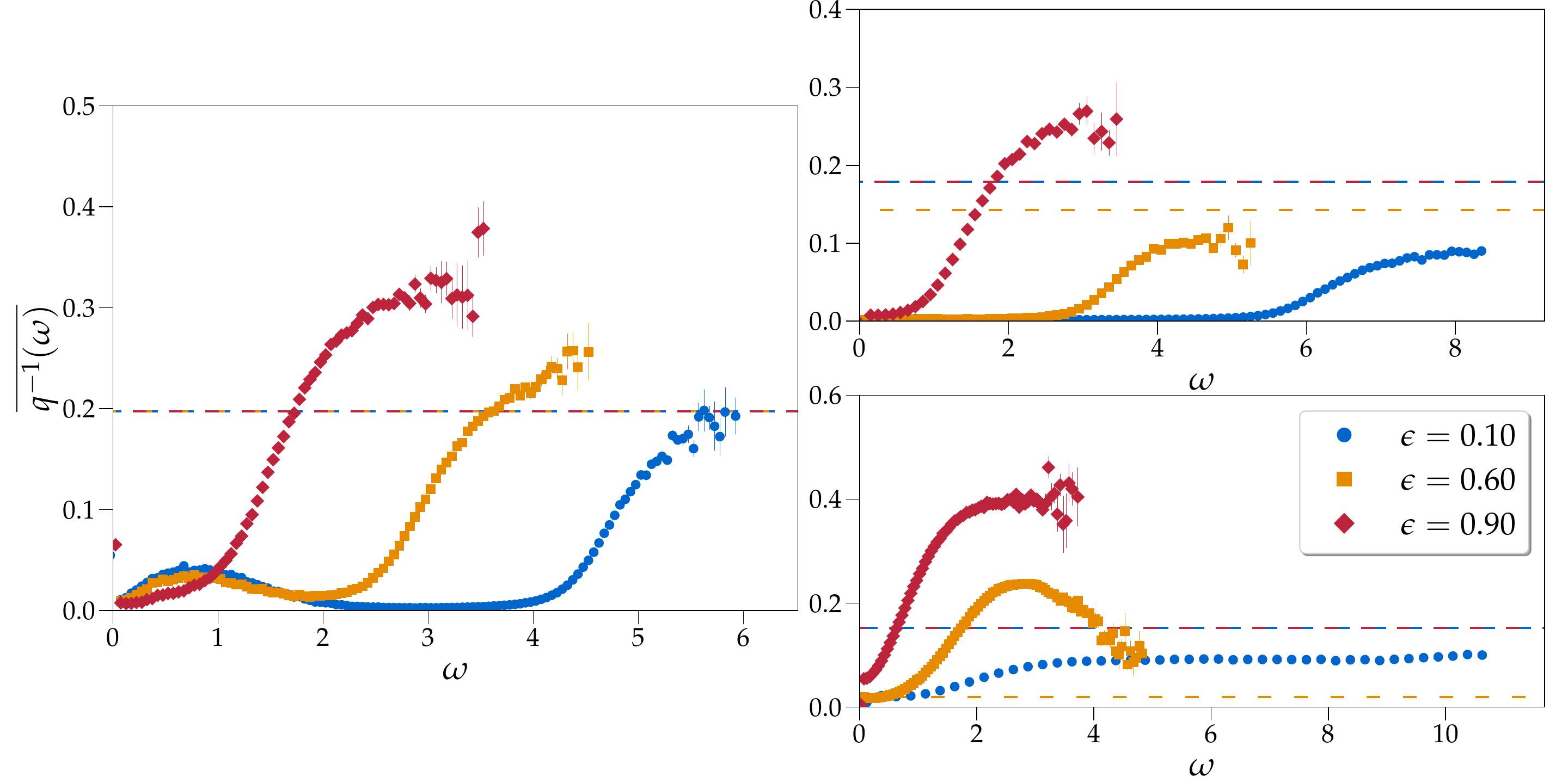}
			\end{subfigure}
			\caption[]{\textit{Inverse participation ratio for random graphs with  weights that are exponentially distributed.} Comparison between the analytical  prediction for the asymptotic value of the disordered-average IPR as given by Eqs.~\eqref{eq:IPR_exp_weights_prod_er}, \eqref{eq:ipr_example_exp_weights_geom} and \eqref{eq:ipr_saddle_point_exp_weights_powerlaw} (dashed lines) with  numerical results obtained through direct diagonalisation (markers). The edge weights are drawn from a distribution $\pi(J_{ij},J_{ji})$  as described at the beginning of Sec.~\ref{section:example:general} with parameters $d=1.0$ and $\nu=2$ and $\epsilon$ as given in the legend [$\epsilon=0.10$ (blue circles), $\epsilon=0.60$ (orange squares) and $\epsilon=0.90$ (red diamonds)].   The three panels correspond with  
            Erd\H{o}s-R\'{e}nyi graphs with connectivity $p=4.0$ (Left Panel), geometric graphs with connectivity $p=4.0$ (Top Right Panel) and power-law graphs with exponent $\gamma=3.5$ and $c_{\rm min}=2$ (Bottom Right Panel).    Numerical estimates for each ensemble are  based on $10000$ graphs of $N=4000$ nodes, and the disordered-average IPR is extracted from this data by averaging all the single eigenvalues IPRs in a given bin and the error bars are the standard deviation of the mean. The analysis of finite size effects for the disorder-averaged IPR is provided in Sec.~\ref{appendix:finite_size_effects_ipr} of the Supplement.}
			\label{fig:ipr_grid_expprod_weights}
		\end{figure}

To test numerically  the asymptotic expressions for the  eigenvalue density and the inverse participation ratio, in Fig.~\ref{fig:spectral_denisty_grid_expprod_weights} and Fig.~\ref{fig:ipr_grid_expprod_weights} respectively, we compare the analytical predictions with direct diagonalisation of matrices of size $N=4000$.    For Erd\H{o}s-R\'{e}nyi graphs and for graphs with geometric degree distributions, the agreement between the theoretical expressions for the tails of  $\rho(\omega)$ are in very good agreement with direct diagonalisation results, while for random graphs with power-law degree distributions we find good qualitative correspondence [the power law behaviour is recovered for $\epsilon<0.5$, while for $\epsilon>0.5$ the decay is exponentially fast, in agreement with Eq.~(\ref{eq:spectral_density_saddle_point_exp_weights_powerlaw})].   The  numerical tests for the IPR in Fig.~\ref{fig:ipr_grid_expprod_weights} are overall not conclusive, since the values of $\omega$ we are able to access through direct diagonalisation are not sufficiently large  to numerically verify our analytical predictions.   
However, the results are consistent with the results in Sec.~\ref{section:example:constantweight} and the numerics  in Fig.~\ref{fig:ipr_epsilon_fixed_edge}, which demonstrate that     the  asymptotic value for the IPR is not sufficient to corroborate the theory with numerical results based on the diagonalisation of $10000$ matrices of size $N=4000$ (in order to find a good agreement between theory and numerics  in Fig.~\ref{fig:ipr_epsilon_fixed_edge}, we required the subleading terms to the asymptotic  value of order $\mathcal{O}(1/\omega^2)$ in the case of power-law and geometric random graphs,  and we required the $\mathcal{O}(1/\omega^4)$ terms to get a good correspondence  for the Erd\H{o}s-R\'{e}nyi case).     This stands in contrast to the  eigenvalue density, which is modified only sightly by the inclusion of the effects from the wider network.  
		
        \subsection{Discussion}\label{sec:discsec4}

        \subsubsection{Relating \texorpdfstring{$\rho$}{rho} and \texorpdfstring{$\overline{q}^{-1}$}{IPR} to the typical local network configuration that makes up  a hub. } \label{sec:RhoQ}

The different qualitative  behaviours of $\rho(\omega)$ (and   $\overline{q^{-1}(\omega)}$)  for $|\omega|\gg 1$ that we obtained in Sec.~\ref{section:example:general} can be understood  as resulting from the combination of the three model aspects (the degree distribution, sign-asymmetry of the edge weights, and the tails of the distribution of edge weights), with different factors playing the dominant role depending on the circumstances.   To understand the different qualitative behaviours, we now examine the typical local network configuration that makes up a hub contributing a tail eigenvalue.  

From Eq.~(\ref{densitygeneral}) we recognise that the degree  $c_i$ of a hub at an eigenvalue $\lambda^2_i = \Phi$ is determined by the degree  value  $c_{\rm max}$ at which  $P(\Phi|c)P_{\rm deg}(c)$ attains its maximum value.  Let us therefore analyse $c_{\rm max}$ as a function of $\Phi$ for the different  degree distributions that we have considered.

Let us first discuss the cases for which  the degree distribution is a more slowly varying function of $c$ than exponential (so that $\frac{d}{dc} \ln P_\mathrm{deg}(c)$ vanishes for $c\to \infty$); the power-law degree distribution is an example of a degree distribution satisfying this property.   In this case, the value  $c_{\rm max}$ for which the maximum of  $P(\Phi|c)P_{\rm deg}(c)$  is attained equals the value $c^{(0)}_{\rm max}$ for which  $P(\Phi|c)$ attains its maximum.  This is because  $P(\Phi|c)\sim c$ [see Eq.~(\ref{lnpdgivek})], and thus the  degree distribution plays a sub-leading role in the sum of Eq.~(\ref{densitygeneral}).        
  
 We show in Sec.~\ref{app:PHiC} of the Supplement that $P(\Phi \vert c)$ has a unique maximum.    
If $\epsilon<1/2$ (when the majority of the links in the network are sign-symmetric), then the degree of the hub is
\begin{equation}
c^{(0)}_\mathrm{max} = \frac{\Phi}{ \langle uv \rangle_{\pi(u,v)}} =  \frac{\Phi}{d (1-2\epsilon)} \, .
\end{equation}
As  for tail eigenvalues $\Phi = \sum_{j\in \partial_i} J_{ij}J_{ji}$ (in leading order),   this means that the most likely arrangement of  the weights   is  the one in which $\sum_{j\in \partial_i} J_{ij}J_{ji}/c_{i}$ takes its average value $d (1-2\epsilon)$.    

On the other hand, if $\epsilon>1/2$ (when the majority of the links are sign-antisymmetric), then  $\sum_{j\in \partial_i} J_{ij}J_{ji}/c_{i}$ has a negative value.   In this case, there is a trade-off between having a sufficiently many sign-symmetric links to produce an eigenvalue pair at $\pm\sqrt{\Phi}$ and the fact that such links are unlikely compared to sign-antisymmetric ones. Indeed, in this case  
\begin{equation}
c^{(0)}_\mathrm{max}=\Phi  \frac{\langle e^{uv \xi}  \rangle_{\pi(u,v)}}{\left\langle uv \:  e^{uv \xi}    \right\rangle_{\pi(u,v)}} . 
\end{equation}
with $\xi = \xi_{\rm pow} = (2\epsilon-1)/d$ governing the trade-off between sign-symmetric and sign-antisymmetric links (see Sec.~\ref{app:PHiC} of the Supplement).

This simple argument clarifies the result in Eq.~(\ref{eq:spectral_density_saddle_point_exp_weights_powerlaw}).  For $\epsilon < 1/2$,  the probability of seeing a hub with particular eigenvalue $\lambda_i = \sqrt{\Phi}$ is simply the probability of seeing a hub with connectivity $c_\mathrm{max} = \Phi/[d(1-2\epsilon)]$.  This means that the asymptotic eigenvalue density is for $\epsilon<1/2$ dictated by the asymptotic form of $P_\mathrm{deg}(c)$,  and hence $\rho(\omega)\sim \omega^{-2\gamma +1}$ for the case of a  power-law degree distribution.  On the other hand for   $\epsilon>1/2$, it is impossible for hubs for which the $J_{ij}J_{ij}$ take their average values of $d(1-2\epsilon)$ to produce a positive value of $\Phi_i = \sum_{j\in \partial_i} J_{ij}J_{ji}$. Instead, for $\epsilon>1/2$ hubs should have an atypical configuration of weights consisting of a  large number of sign-symmetric links.  Hence, the leading order term, $\exp(-(2\epsilon-1)\omega^2/d)$, in the asymptotic decay of the spectral density is independent of the degree distribution,  but is   determined by  the fraction $\epsilon$ of sign-antisymmetric links.

		As the degree distribution becomes more quickly varying with $c$, this same reasoning no longer applies, as $c_{\rm max}\neq c^{(0)}_{\rm max}$.   
         Since $\ln P(\Phi|c)\sim c$,  the degree distribution starts playing a more significant role in determining the hub as soon as $\ln P_{\rm deg}(c)$ decays as fast as a geometric distribution, or faster.    In the geometric case, since we have that $\ln P(\Phi \vert c_\mathrm{max}) \sim \ln P_\mathrm{deg}(c_\mathrm{max}) \sim c_\mathrm{max}$, the asymptotic behaviour of the eigenvalue density is always  of the form $\exp(-\xi \omega^2)$.    The value of $c_\mathrm{max}$ is now determined by a non-trivial trade-off between minimising the total number of links, and minimising the number of atypically weighted links that are required to produce a value $\Phi = \sum_{j\in \partial_i} J_{ij}J_{ji}$ [obtained from solving  Eqs.~(\ref{logrho}-\ref{saddlexi2})].
		
		Finally, the case of the Erd\H{o}s-R\'{e}nyi graph is an example of a degree disribution that decays faster than a geometric distribution. In this case,  the degree distribution is so quickly varying with $c$ that the most likely local arrangement of links that will produce a fixed large value $\Phi = \sum_{j\in \partial_i} J_{ij}J_{ji}$ is one in which the number of links is smaller than the other examples, and the set of products of weights $J_{ij}J_{ji}$ `align' with each other, taking the same sign. We saw this also in the example with fixed link weights in Section \ref{section:example:constantweight}, where the typical fraction of antagonistic links $m/c$ was vanishing and $c_\mathrm{max} \approx \omega^2/a^2$. In the above example where the link weights are allowed to vary however, the most likely arrangement is to have a much smaller number of links such that  $c_\mathrm{max} \sim \sqrt{\Phi} \sim \omega$, where the products $J_{ij}J_{ji}$ are heavily weighted in the positive direction. More precisely, one finds in this case that the typical arrangement of links contributing to eigenvalues at $\omega \sim \sqrt{\Phi}$ has the property $\sum_{j\in \partial_i} J_{ij}J_{ji}/c_\mathrm{max} \sim \omega$, whereas for the other degree distributions we still had $\sum_{j\in \partial_i} J_{ij}J_{ji}/c_\mathrm{max} \sim \omega^0$ when the edge weights were allowed to vary.

        \subsubsection{Breakdown of the saddle-point procedure.}\label{section:breakdown}
		The approximations for the IPR and the eigenvalue density for large $\omega$ that were discussed in this section rely on two assumptions: (1) The states in the tail region are sufficiently localised around network defects so that we can consider such defects as isolated. We termed this the single-defect approximation, and this approximation implies that  the eigenvalues in the tails take the form (\ref{eigenvalueapprox}), where the central node $i$ has a large effective degree $|\Phi_i|$.   
        (2) There is a trade-off between $P(\Phi \vert c)$ and $P_\mathrm{deg}(c)$ such that an eigenvalue $\omega$ is associated with a typical large connectivity $c_\mathrm{max}(\omega)$. That is, the network defects that are responsible for large eigenvalues are well-connected hubs, so that the first term  on the right-hand side of  (\ref{eigenvalueapprox})  has a large number of terms.  
		
		We saw in the example of the power-law degree distribution how assumption (1) can break down when there is a balance between the number of antagonistic and reinforcing links [i.e., $\epsilon \approx 1/2$ -- see the discussion around Eqs.~(\ref{epslhalf}) and (\ref{epsghalf})]. In this case, the eigenvectors in the tail region are poorly localised, and so defects can no longer be considered isolated. A second case for which assumption (1) fails is  when  the degree distribution is sufficiently slowly decaying. Specifically, if the second moment of the degree distribution diverges, a non-negligible number of the neighbours of highly-connected nodes will also be highly connected. This manifests in the fact that the second term in Eq.~(\ref{eigenvalueapprox}) no longer being much smaller than the first, indicating the breakdown of our approximation scheme~\cite{goltsev2012localization}. Strong degree correlations of the type considered in Ref.\cite{rogers2010spectral} could also lead to this same issue.	
		
		It is also possible for assumption (2) to break down.   This can happen in different scenarios.  First, is the case  
        when the distribution of edge weights decays slower than an exponential. Indeed,  we demonstrate in Sec.~\ref{appendix:breakdowngeneral} of the Supplement  that for a power-law distribution of the edge weights, $P(\Phi \vert c)$ no longer has a maximum as a function of $c$. We argue that, in this case, the states corresponding to large values of $\omega$ are localised on single edges. This observation has also previously been made in the context of dense symmetric L\'evy random matrices \cite{cizeau1994theory}.   A second example for which  Assumption (2)  breaks down is in the case of Husimi trees with unbounded weight distribution \cite{valigi2024local}, where the defects responsible for tail eigenvalues are short cycles with an unbalance between the clockwise and counter-clockwise weights.

		\section{Conclusion}\label{section:discussion}
		In this work, we have studied the tail regions of the eigenvalue spectra of sparse networks with arbitrary degree and edge-weight distributions. We showed that the eigenvectors  in these tail regions are localised around network defects with high connectivity or heavily weighted edges. By performing an  expansion of the cavity equations in $1/\omega$, we were able to show that eigenvalues in the tail region (and their corresponding eigenvectors) are well approximated by considering only the local network structure in the vicinity of defects.  This allowed us to derive succinct analytical results for the asymptotic  eigenvalue density and the IPR in the spectral tails of sparse and non-Hermitian random  matrices. With this method, we described how the degree distribution, the edge-weight distribution, and the degree of asymmetry in the network combine to give different asymptotic behaviours for the spectral density and the IPR. The expressions that we derived were simple and closed-form, allowing us to see clearly the influence of the various model aspects.
		
		In particular, our approach enabled us to study analytically the role of asymmetry on the tails of the eigenvalue density and on the disorder-averaged inverse participation ratio. We  demonstrated  that the eigenvalue density  decays more quickly  with $\omega$ for networks with a large number of sign antisymmetric weights, in agreement with the observation that sign-antisymmetry tends to act as a stabilising influence in both sparse \cite{mambuca2022dynamical, valigi2024local} and dense \cite{sommers1988spectrum, baron2022eigenvalue,cure2023antagonistic} disordered dynamical systems.  Furthermore, in some cases, particularly when the degree distribution of the network is a slowly decaying function of connectivity, the functional decay of the spectral tails changes qualitatively as a function of the fraction of sign-antisymmetric weights in the network.   The  fraction of sign-antisymmetric  versus sign-symmetric links also affects the localisation of right eigenvectors, and we have found that  localisation is minimised if there is a balance between sign-antisymmetric and sign-symmetric weights.  These results are compatible with other results in the literature, for which it has been observed that non-Hermiticity suppresses localisation, particularly in  one-dimensional systems with asymmetric couplings \cite{kawabata2021nonunitary, hatano1996localization, hatano1997vortex, hatano1998non}. In systems where non-Hermiticity takes the form of complex on-site energies (or complex diagonal entries, in the random matrix language) however, it has been found that non-Hermiticity tends to expand the range of system parameters for which localisation occurs \cite{deTomasi2023non}. While this might at first seem to be in contrast with our claim that `non-hermiticity suppresses localisation', we emphasise that we do not comment on the comparative prevalence of localised states as a function of asymmetry, only how localised the states are, as measured by the IPR. 
		
		Our analysis opens the door to understanding the behaviour of more complicated nonlinear models with sparse interactions. For example, recent progress has been made in understanding the coexistence \cite{marcus2024local} and the dynamics \cite{marcus2022local} of species in the sparsely interacting Lotka-Volterra model, also in terms of the underlying architecture of the interaction network. Given the observation that non-linearities can curtail Lifshitz tails \cite{rodgers1988density, hertz1979marginal, bray1982eigenvalue}, it will be interesting to see if and how the observations of the present work carry over into non-linear models.

		\section{Acknowledgements}
		The authors would like to thank Ivan Khaymovich for comments and fruitful discussions. JWB acknowledges grants from the Simons Foundation (\#454935 Giulio Biroli), and he also thanks the Leverhulme Trust for support through the Leverhulme Early Career Fellowship scheme. PV has received funding  under the 'Avvio alla Ricerca 2024' grant (\#AR22419078ACB3CE SToRAGE), PV and CC acknowledge grants from "Progetti di Ricerca Grandi 2023"  (\#RG123188B449C3DE) both from Sapienza University of Rome. This project has been supported by the FIS 1 funding scheme (SMaC - Statistical Mechanics and Complexity) from Italian MUR (Ministry of University and Research).
		\pagebreak
		\appendix

		\section{Eigenvalues and right eigenvectors of tree graphs} \label{appendix:starplusneighbours}

We first analyse in \ref{app:tree1} the eigenvalues and eigenvectors of a tree graph with one generation of descendants (as in  the left panel of Fig.~\ref{fig:stars}) and then in \ref{app:tree2} we use a perturbation approach to  derive formulae for the leading eigenvalue and right eigenvector of a tree graph with two generations of descendants (as in  the right panel of Fig.~\ref{fig:stars}) .

        \subsection{Tree graph with one generation of descendants}\label{app:tree1}
     Let $A^{\rm star}_{jk} = \delta_{i,j}(1-\delta_{k,i}) + \delta_{k,i}(1-\delta_{j,i})$ be the adjacency matrix of a tree graph, where $\delta_{i,j}$ is the Kronecker delta function, and $i$ is the index labelling the hub of the graph.   Let $M^{\rm star}_{jk} = A^{\rm star}_{jk}J_{jk}$ be the adjacency matrix of a weighted star graph.    
     
     The leading eigenvalue $\lambda_1(\underline{\underline{M}}^{\rm star})$, i.e., the eigenvalue with the largest real part, equals 
     \begin{equation}
     \lambda_1 = \sqrt{\sum_{j\in \partial_i}J_{ij}J_{ji}},  
     \end{equation}
which corresponds with the leading order contribution of (\ref{eigenvalueapprox}).
     
     The right  and left eigenvectors corresponding with the eigenvalue $\lambda_1$ have the components  
		\begin{equation}
			r^{(1)}_\ell =   \frac{\lambda_{1}}{\sqrt{\lambda_{{1}}^2 + \sum_{\color{red}{\ell'\in \partial_i}}J^2_{\color{red}{\ell'} i} }}\delta_{\ell,i}  + \frac{J_{\ell i}}{\sqrt{\lambda_{1}^2 + \sum_{\color{red}{\ell'\in \partial_i}}J^2_{\color{red}{\ell'} i} }} (1-\delta_{\ell,i})  \label{eq:rGraph}
		\end{equation}
		and 
		\begin{equation}
			l^{(1)}_\ell =   \frac{\lambda_{1}}{\sqrt{\lambda_{1}^2 + \sum_{\color{red}{\ell'\in \partial_i}}J^2_{ i \color{red}{\ell'}} }}\delta_{\ell,i}  + \frac{J_{i \ell}}{\sqrt{\lambda_{1}^2 + \sum_{\color{red}{\ell'\in \partial_i}}J^2_{i\color{red}{\ell'}} }} (1-\delta_{\ell,i})  ,
		\end{equation}
		where we have used the same normalisation $\sum^N_{j=1}|r^{(1)}_j|^2 = \sum^N_{j=1}|l^{(1)}_j|^2 =1$ as considered throughout.  The corresponding IPR is thus 
        \begin{eqnarray}
        q^{-1}_1 &=\sum_{j\in \partial_i}\left\vert r_j^{(1)} \right \vert^4 + \left\vert r_i^{(1)} \right \vert^4 =  \frac{\sum_{j\in \partial_i} J^4_{j i} + \left(\sum_{j\in \partial_i}J_{ij}J_{ji}\right)^2}{\left(|\sum_{j\in \partial_i}J_{ij}J_{ji}| + \sum_{j\in\partial_j}J^2_{j i}\right)^2}, 
        \end{eqnarray}
        which corresponds with the result in (\ref{iprgeneral}).

        \subsection{Perturbative expansion for tree graphs with two generations of descendants}\label{app:tree2} 
		To obtain the next leading order term in Eqs.~(\ref{eigenvalueapprox}) and (\ref{iprgeneral}), we consider a tree graph centered around node $i$ that has two generations,   as shown in the right panel of Fig.~\ref{fig:stars}.   We may represent the adjacency matrix of the second generation graph as
		\begin{equation}
        {\underline{\underline{A}}}_2 =      {\underline{\underline{M}}}^{\rm star} +     {\underline{\underline{B}}} \label{eq:Perturb}
		\end{equation}
		where  ${\underline{\underline{B}}}$ is a matrix that contains all the links between  the first and second generation nodes (next nearest neighbour links).

     Although we were not able explicitly to evaluate the  leading eigenvalue of the second generation star graph $\color{red}{\underline{\underline{A}}}_2 $,  here we are particularly interested in the case that the central node $i$ is a  hub with a large effective degree $|\Phi_i|$.  In this scenario, the contribution ${\underline{\underline{B}}}$ can be considered a perturbation of ${\underline{\underline{M}}}^{\rm star}$, which becomes evident if we write    (\ref{eq:Perturb}) as 
     \begin{equation}
        {\underline{\underline{A}}}_2 =    \sqrt{\sum_{j\in \partial_i}J_{ij}J_{ji}}\left(  \frac{1}{\sqrt{\sum_{j\in \partial_i}J_{ij}J_{ji}}}{\underline{\underline{M}}}^{\rm star}  +     \frac{1}{\sqrt{\sum_{j\in \partial_i}J_{ij}J_{ji}}}{\underline{\underline{B}}}\right),  \label{eq:Perturb2}
		\end{equation}
        where the norm of the matrix $\underline{\underline{M}}^{\rm star}/\sqrt{\sum_{j\in \partial_i}J_{ij}J_{ji}}$ remains finite in the limit of $\sqrt{\sum_{j\in \partial_i}J_{ij}J_{ji}}\rightarrow \infty$, while the norm of    ${\underline{\underline{B}}}/\sqrt{\sum_{j\in \partial_i}J_{ij}J_{ji}}$ vanishes in this limit.  Hence, if the effective degree $\color{red}{\sqrt{\sum_{j\in \partial_i}J_{ij}J_{ji}}}$ of the hub is large, then we can use perturbation theory with $1/\sqrt{\sum_{j\in \partial_i}J_{ij}J_{ji}}$ as small parameter.   
        
        Using standard pertrubation theory, see~Ref.~\cite{wilkinson},  we find that the   leading eigenvalue is  perturbed as 
		\begin{equation}
			\lambda_1(        {\underline{\underline{A}}}_2)  =  \sqrt{\sum_{j\in \partial_i}J_{ij}J_{ji}}\left(1  +  \frac{\lambda^{(1)}_1}{\sqrt{\sum_{j\in \partial_i}J_{ij}J_{ji}}} + \color{red}{\mathcal{O}\left(\frac{1}{\sum_{j\in \partial_i}J_{ij}J_{ji}}\right)}\right)\color{red}{,}  
		\end{equation} 
		where 
		\begin{equation}
			\lambda^{(1)}_1 = \frac{\vec{l}^{(1)}\cdot {\underline{\underline{B}}}\vec{r}^{(1)}}{\vec{l}^{(1)}\cdot \vec{r}^{(1)}} \color{red}{.}\label{eq:correction}
		\end{equation}
        Note that in (\ref{eq:correction}) we recognise  the  next-leading correction term   of $\lambda_1$ in Eq.~(\ref{eigenvalueapprox}).   
        Considering tree graphs of three or more generations, we can in principle obtain the next terms in the series (\ref{eigenvalueapprox}) and (\ref{iprgeneral}).

          \section{Derivation of Eq.~(\ref{iprfromres}) for the IPR}\label{app:participNH}
        Following similar arguments as in ~\cite{neri2016eigenvalue}, let us consider the matrix 
            \begin{equation}
        \begin{bmatrix}
			\eta \id_{N}&  \omega \id_{N} -\underline{\underline{M}} \\
			( \omega \id_{N}- \underline{\underline{M}} )^\dagger & \eta \id_{N} 
		\end{bmatrix} \color{red}{.}  \label{eq:origMatrix}
        \end{equation}
        Using the expression for the determinant of a block matrix, we find for $\eta=0$  the characteristic polynomial
        \begin{equation}
        {\rm det}\left(\lambda^2\id_{N} - (  \omega \id_{N} -\underline{\underline{M}})( \omega \id_{N}- \underline{\underline{M}} )^\dagger\right) =0,
        \end{equation}
        and hence the eigenvalues of (\ref{eq:origMatrix}) are  given by   $\eta \pm s_\nu(\omega)$,  with $s_\nu(\omega)$ the singular values of $\omega \id_{N}- \underline{\underline{M}} $.    

        Next, let us set $\omega = \color{red}{\lambda_\nu}+\eta$, for a certain $\color{red}{\nu}\in \left\{1,2,\ldots,N\right\}$.     In that case, the matrix (\ref{eq:origMatrix}) has the eigenvalue $\eta$ with two  eigenvectors, one that has the entries  
\begin{equation}
e^{+,{(\nu)}}_j = \left\{\begin{array}{ccc} l^{(\nu)}_j &{\rm if}& \color{red}{1 \leq j\leq N}, \\ r^{(\nu)}_j  \ &{\rm if}& \color{red}{N+1 \leq j\leq 2N,} \end{array}\right.
\end{equation}
and a second one with the entries 
 		\begin{equation}
e^{-,{(\nu)}}_j = \left\{\begin{array}{ccc} -l^{(\nu)}_j&{\rm if}& \color{red}{1 \leq j\leq N}, \\ r^{(\nu)}_j\ &{\rm if}& \color{red}{N+1 \leq j\leq 2N.}   \end{array}\right.
\end{equation} 

Therefore, in the limit of $\eta\rightarrow 0$, we can express 
\begin{equation}
\mathcal{H}^{22}_{ii}(\eta,\eta+\lambda_\nu) = \frac{2}{\eta} \frac{ |r^{(\nu)}_i|^2}{\sum^N_{j=1}|r^{(\nu)}_j|^2 +\sum^N_{j=1}|l^{(\nu)}_j|^2 } + \mathcal{O}(\eta^0).\label{eq:B5}
\end{equation}
If we choose $\sum^N_{j=1}|r^{(\nu)}_j|^2 = \sum^N_{j=1}|l^{(\nu)}_j|^2  = 1$, we recover Eq.~\eqref{iprfromres} in the main text.

    \section{Population Dynamics prediction for the eigenvalues density} \label{app:popdyn}

\begin{figure}[H]
		\centering
		\begin{subfigure}[b]{0.7\textwidth}
			\centering
			\includegraphics[width=\textwidth]{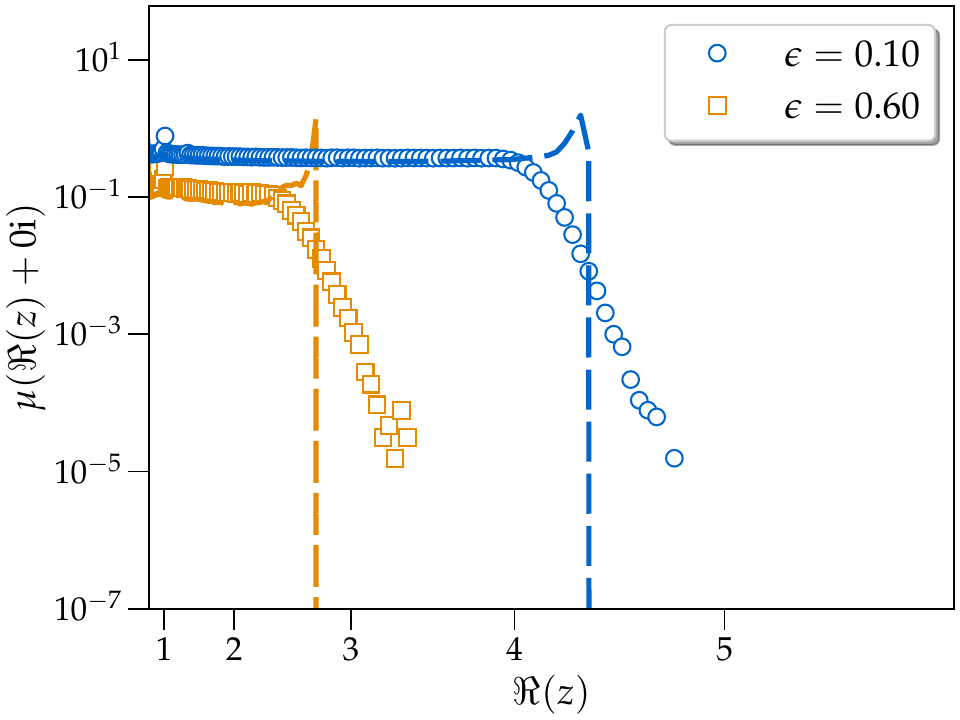}
		\end{subfigure}
		\caption[]{Comparison between empirical estimates of the complex spectral density on the real axis from  directly diagonalising $10000$ Erd\H{o}s-R\'{e}nyi graphs of size $N=4000$ with mean degree $p=4.0$ and $\pm1$ weights (markers) and the estimates from population dynamics (dashed lines), for different values of the sign-antisymmetric probability $\epsilon$: $\epsilon=0.10$ (blue circles) and $\epsilon=0.60$ (orange squares). The $y$-axis is plotted in $\log$ scale and the $x$-axis is plotted in a  quadratic scale. The theoretical predictions from the population dynamics algorithm have been obtained as discussed in Appendix E  of  Ref.~\cite{mambuca2022dynamical}, with the following parameters: a population size $N_{\rm p}=125000$,  a number of equilibration sweeps $N_{\rm eq}=1500$, a number of measurement sweeps $n_{\rho}=1000$, and regulariser $\eta=10^{-100}$. }
	   \label{fig:popdyn_spectral_denisty_grid_fixed_edge}
	\end{figure}

    A popular approach for computing the eigenvalue density in the complex plane of a sparse asymmetric random graph is to solve the cavity equations for the Hermitised resolvent [see Sec.~\ref{app:HermitisedCavPre} and Eq.~\eqref{cavityherm}] using the population dynamics algorithm as described in Refs.~\cite{rogers2009cavity,metz2019spectral,mambuca2022dynamical}. In Figure~\ref{fig:popdyn_spectral_denisty_grid_fixed_edge}, we compare the results obtained using population dynamics with the eigenvalue density in the complex plane estimated from directly diagonalising matrices of size $N=4000$. Notice that Fig.~\ref{fig:popdyn_spectral_denisty_grid_fixed_edge} displays the spectral distribution of eigenvalues in the complex plane, while Fig.~\ref{fig:spectral_denisty_grid_fixed_edge} plots the distribution of real eigenvalues. Interestingly, the population dynamics algorithm is correctly matching direct diagonalisation in the bulk region, while it is not able to correctly predict the spectral tails, where it yields an abrupt collapse of the eigenvalue density. In fact, it is generally difficult to obtain the pure point spectrum (\ie, the eigenvalues corresponding with localised eigenvectors, such as those in the spectral tails) with population dynamics, as it requires a fine-tuning of the regulariser~\cite{kuhn2008spectra, susca2021cavity}. In addition, the population dynamics prediction in the tail region is known to display a dependence on the size of the population~\cite{mambuca2022dynamical}.

\printbibliography[heading=subbibliography]
\end{refsection}

\pagebreak

\vspace*{-0.3cm}
\begin{center}
	\Large{\textbf{--- Supplemental Material ---}}
\end{center}
\vspace*{-0.3cm}

\setcounter{page}{1}
\setcounter{figure}{0}
\setcounter{equation}{0}
\setcounter{section}{0}
\setcounter{table}{0}

\newcommand\numberthis{\addtocounter{equation}{1}\tag{\theequation}}
\renewcommand\numberthis{\addtocounter{equation}{1}\tag{\theequation}}
\renewcommand\numberthis{\addtocounter{figure}{1}\tag{\thefigure}}
\renewcommand\numberthis{\addtocounter{section}{1}\tag{\thesection}}

\counterwithout{equation}{section} 
\counterwithout{figure}{section} 

\renewcommand{\thesection}{S\arabic{section}} 		
\renewcommand{\thepage}{S\arabic{page}} 			
\renewcommand{\theequation}{S\arabic{equation}}  	
\renewcommand{\thefigure}{S\arabic{figure}}  		
\renewcommand{\thetable}{S\arabic{table}}  		

    \begin{refsection}
    \newrefcontext[labelprefix=S] 

    \section{General structure of the eigenvalue spectrum and properties of the eigenvectors}
   
    The properties of the eigenvalue spectrum of matrices $\mmat$ of the type described in Section \ref{section:model} have some common features. We describe these features in the case of symmetric matrices, and then explain how these features generalise in the case of asymmetric random matrices.
    
    In the case of symmetric sparse random matrices, it is generally observed that the eigenvalue spectrum has two contributions: a continuous `bulk' region and a point-like Lifshitz tail. 
    
    As with dense random matrices, there is a `bulk' part of the eigenvalue spectrum in which the eigenvalues form a continuum on the real axis in the limit $N\to\infty$, and the eigenvalue density is given by the Wigner semi-circle law \cite{wigner1958distribution, wigner1967random}. For finite connectivity $p \sim \mathcal{O}(N^{0})$ (also in the thermodynamic limit $N\to \infty$), this continuous part of the spectrum remains, but deviates from the Wigner semi-circle. Such corrections to the semi-circle law due to sparsity and network structure can be characterised using perturbative methods \cite{rodgers1988density, baron2025pathintegral} and other approximation schemes \cite{semerjian2002sparse}. 
    
    In contrast to the dense case however, the spectrum of a sparse matrix exhibits a so-called Lifshitz tail of eigenvalues. This part of the spectrum is point-like (i.e. the eigenvalues do not form a continuum in the limit $N \to \infty$). As has been well understood in context of Anderson localisation \cite{kramer1993localization, frohlich1984rigorous, rizzo2024localized}, it can be reasoned generically that the states in the point-like tail region correspond to localised states, whereas those in the continuous part of the spectrum correspond to extended states. Localised eigenvectors have an $\mathcal{O}(N^{0})$ number of components that have a magnitude $\vert r_i^{(\nu)} \vert^2 \sim \mathcal{O}(N^{0})$, whereas a number $\mathcal{O}(N)$ of the components of the extended eigenvectors have a magnitude $\mathcal{O}(1/N)$. Thus, the IPR defined in Eq.~(\ref{iprdef}) takes values $q^{1} \sim \mathcal{O}(1)$ in the tail of the eigenvalue spectrum, but vanishes in the thermodynamic limit in the continuous part. In the case where there is no diagonal disorder (unlike for Anderson localisation), it has been suggested that the states in the tail regions of the spectra of sparse random matrices correspond to well-connected hubs of the network \cite{biroli1999single, semerjian2002sparse}. 
    
    In the case of asymmetric matrices, an analogous phenomenology is expected, but is less well-understood. Indeed, it has been shown that as sparsity/network structure is introduced, deviations of the continuous bulk part of the eigenvalue spectrum from the dense elliptic law are observed, which can be characterised perturbatively \cite{baron2025pathintegral, baron2022eigenvalue, poley2024eigenvalue} (see also Fig. \ref{fig:examplespectra}). Additionally, using tools such as the population dynamics algorithm along with the cavity equations, infinitely extending tails have been observed along both the imaginary and real axes \cite{mambuca2022dynamical, valigi2024local}. 
    
    We show in the following Supplementary Sections how one can find closed form expressions for the eigenvalues in the tail regions of the eigenvalue spectrum in terms of the local properties of the network. This is accomplished by performing a $1/\omega$ expansion of the cavity equations for the resolvent matrix. We thus see directly that eigenvalues in the tail region correspond to highly-connected `hubs' of the network (or, in some circumstances, to highly-weighted edges). We emphasise that this result emerges organically from our analysis. We also show explicitly how the eigenvectors of eigenvalues in the tail can be shown to be exponentially localised. The remaining sections deal with the evaluating the eigenvalue density and the disorder-averaged IPR for various realisations of the network and edge weights.
    
    \section{Tail eigenvalues from an asymptotic expansion of the resolvent for large values of \texorpdfstring{$\omega$}{omega}}\label{appendix:expansion}
		
	We derive the expression (\ref{eigenvalueapprox}) for eigenvalues  in the spectral tails, which are associated with nodes $i$ that have a large effective degree $|\Phi_i|$.   We identify the tail eigenvalues  as poles of the diagonal elements of the resolvent.     Since we aim to find  eigenvalues with large absolute value, we  perform an asymptotic expansion in $1/\omega$ of the resolvent elements.   We develop this asymptotic expansion with the    cavity Eqs.~(\ref{cavityequations}).    From this asymptotic expansion, we  find explicit expressions for the poles of the diagonal elements of the resolvent in terms of the local network structure surrounding node $i$.  Consistency of the argument requires that the poles occur at large enough values of $\omega$, which yields the condition that the hub node $i$  has a large  effective degree $|\Phi_i|$.

    \subsection{Asymptotic expansion of the cavity equations}
    From the first line of Eq.~(\ref{cavresseries}), we observe that the poles of $G_{ii}$ are determined by the diagonal elements of the cavity resolvent $G^{(i)}_{jj}$ (as the poles are defined as the values of $\omega$ for which the denominator equals zero).   Therefore, we derive an asymptotic expression for the cavity resolvents $G^{(i)}_{jj}$.   In what follows,  we use the same labeling of nodes as in the main text: the central hub is labeled by $i$,  the nodes neighbouring the central hub are labeled by $j\in \partial_i$, and the next nearest neighbours by $k\in \partial^{/i}_j$ (as illustrated in the right-panel of Fig.~\ref{fig:stars}).  
    
    Substituting the series expansion  Eq.~(\ref{cavresseries}) for the cavity resolvent in Eqs.~(\ref{cavityequations}), we obtain  for the series coefficients (up to the fifth order): 
		\begin{align} \label{eq:D1}
			G_{jj}^{(i), 0} &= 1, \nonumber \\
			G_{jj}^{(i), 1} &= 0, \nonumber \\
			G_{jj}^{(i), 2} &= \sum_{k \in \partial_j^{/i}} J_{jk} G^{(j),0}_{kk} J_{kj} = \sum_{k \in \partial_j^{/i}} J_{jk} J_{kj} , \nonumber \\
			G_{jj}^{(i), 3} &= \sum_{k \in \partial_j^{/i}} J_{jk} G^{(j),1}_{kk} J_{kj} = 0, \nonumber \\
			G_{jj}^{(i), 4} &= \sum_{k \in \partial_j^{/i}} J_{jk} G^{(j),2}_{kk} J_{kj} + \left( \sum_{k \in \partial_j^{/i}} J_{jk} G^{(j),0}_{kk} J_{kj}\right)^2 \nonumber \\
			&= \sum_{k \in \partial_j^{/i}} J_{jk} J_{kj}\sum_{l \in \partial_k^{/j}} J_{kl} J_{lk}  + \left( \sum_{k \in \partial_j^{/i}} J_{jk} J_{kj}\right)^2 , \nonumber  \\
			G_{jj}^{(i), 5} &= \sum_{k \in \partial_j^{/i}} J_{jk} G^{(j),3}_{kk} J_{kj} +  2\sum_{k \in \partial_j^{/i}} J_{jk} G^{(j),0}_{kk} J_{kj} \sum_{k \in \partial_j^{/i}} J_{jk} G^{(j),1}_{kk} J_{kj} = 0. \nonumber \\
		\end{align}
        From the above relations, we observe that to obtain an approximation for $G_{jj}^{(i)}$ that is valid up to order $\mathcal{O}(1/|\omega|^{2r-1})$, we need to take into account contributions from the cavity resolvents from $r-1$ sets of nearest neighbours. We thus obtain from Eq.~(\ref{cavresseries})  a series expression for the cavity resolvent  in terms of successive sets of edge weights 
		\begin{align}
			G_{jj}^{(i)}(\omega) =& \frac{1}{\omega} + \frac{\sum_{k \in \partial_j^{/i}} J_{jk}J_{kj} }{\omega^3} \nonumber \\
			&+ \frac{\sum_{k \in \partial_j^{/i}}  J_{jk} J_{kj}\sum_{l \in \partial_k^{/j}} J_{kl} J_{lk}  + \left( \sum_{k \in \partial_j^{/i}} J_{jk} J_{kj} \right)^2}{\omega^5}+ \mathcal{O}\left(\omega^{-7}\right). \label{cavresseries2}
		\end{align}

        \subsection{Tail eigenvalues from the poles of the resolvent}
		Now that we have explicit expressions for the cavity resolvent elements in terms of the random matrix elements, we can determine the poles of $G_{ii}$, and thus also the graph  features that lead to the presence of eigenvalues in the tail region.    Indeed,  upon substituting the expression  Eq.~(\ref{cavresseries2}) into the first of Eqs.~(\ref{cavityequations}), we are able to extract the locations of the poles of the local resolvent $G_{ii}(\omega)$ (i.e. the eigenvalues that make up the tail). Poles occur at values of $\omega=\lambda_i$ with 
		\begin{align}
			\lambda_i^2 =& \sum_{j \in \partial_i} J_{ij}J_{ji} + \frac{\sum_{j \in \partial_i} J_{ij}J_{ji}\sum_{k \in \partial_j^{/i}} J_{jk}J_{kj} }{\lambda_i^2} \nonumber \\
			&+ \frac{\sum_{j \in \partial_i} J_{ij}J_{ji}\left[\sum_{k \in \partial_j^{/i}}  J_{jk} J_{kj}\sum_{l \in \partial_k^{/j}} J_{kl} J_{lk}  + \left( \sum_{k \in \partial_j^{/i}} J_{jk} J_{kj} \right)^2 \right]}{\lambda_i^4}+ \cdots. \label{implicitomi}
		\end{align}
In order for our assumption of large $\lambda_i$ to be valid, we thus require that the effective degree $|\Phi_i| = \vert \sum_{j \in \partial_i} J_{ij}J_{ji}\vert$ be large. By inserting the expression for $\lambda_i^2$ into the right-hand side of Eq.~(\ref{implicitomi}) iteratively and collecting terms of the same order in $\sum_{j \in \partial_i} J_{ij}J_{ji}$, we obtain Eq.~(\ref{eigenvalueapprox}) of the main text.

\subsection{Poles in the resolvent of neighbouring nodes}

We show now that if $G_{ii}$ has a pole at $\omega = \lambda_i$, then $G_{jj}$ must also have a pole at this same location. This can be seen via the following useful corollary  from the cavity equations (\ref{cavityequations}),
		\begin{align}
			G_{jj} = G_{jj}^{(i)}G_{ii} \frac{1}{G_{ii}^{(j)}} . \label{messagepassing}
		\end{align}
 It states that  the resolvent element on one node can be related to that of a neighbouring node via the cavity resolvent.   In particular, we see that if $G_{ii}$ has a pole, then  $G_{jj}$ must have the same pole, as long as the cavity resolvents does not have that pole.

		\section{Series expansion of the hermitised resolvent and asymptotic expression for the IPR}\label{appendix:hermexpansion}

		We derive the expression Eq.~(\ref{iprgeneral}) for the IPR of the tail eigenvalues $\lambda_i$ of nodes with high effective degree $|\Phi_i|$ (see Eq.~(\ref{eigenvalueapprox})).  As will be justified \textit{a posteriori}, we will see that neglecting contributions from the next-to-nearest neighbours to the IPR is equivalent to neglecting a contribution $\mathcal{O}(1/\vert\Phi_i\vert)$. That is, the sum on the right-hand side of the IPR in Eq.~(\ref{iprfromres}) can be truncated as follows   
\begin{align}
		q^{-1}_i(\omega) = \lim_{\eta \to 0} \eta^2  \vert\mathcal{H}^{22}_{ii}(\eta, \omega) \vert^2 +  \lim_{\eta \to 0} \eta^2 \sum_{j\in \partial_i }\vert\mathcal{H}^{22}_{jj}(\eta, \omega) \vert^2 + \mathcal{O}(1/|\Phi_i|^{-1}), \label{iprfromres2}
	\end{align}
    if the IPR is evaluated at a tail eigenvalue $\omega=\lambda_{i}$; here, as in the main text, we use the notation  $q^{-1}_i $  to highlight that $\lambda_{i}$ is the pole associated with node $i$.     Since the tail eigenvalues are large, we  develop in Sec.~\ref{app:HermitisedCav} an asymptotic expansion of the quantities $\mathcal{H}_{ii}(\omega)$ and  $\mathcal{H}^{22}_{jj}(\omega)$ ($j\in \partial_i$) for large values of $\omega$; this expansion relies on the  Hermitised cavity equations, which we present in Sec.~\ref{app:HermitisedCavPre}.    Then in Sec.~~\ref{app:hermitisedCav2}  we evaluate the asymptotic expressions   at the tail eigenvalue and expand in the effective degree $|\Phi_i|$ to obtain the final result Eq.~(\ref{iprgeneral}) for the IPR. 
    
\subsection{Hermitised cavity equations}\label{app:HermitisedCavPre}

The Hermitised cavity equations take the form \cite{metz2019spectral}
		\begin{align}
			\mathcal{H}_{ii} &= \left[\begin{array}{cc}\mathcal{H}_{ii}^{11}& \mathcal{H}_{ii}^{12} \\ \mathcal{H}_{ii}^{21}& \mathcal{H}_{ii}^{22} \end{array}\right] = \left(\mathcal{W} - \sum_{j\in\partial_i} \mathcal{J}_{ij} \mathcal{H}^{(i)}_{jj} \mathcal{J}_{ji} \right)^{-1}\label{cavityhermP}
            		\end{align}
                    and
		\begin{align}
			\mathcal{H}^{(i)}_{jj} &= \left(\mathcal{W} - \sum_{k\in\partial_j^{/i}} \mathcal{J}_{jk} \mathcal{H}^{(j)}_{kk} \mathcal{J}_{kj} \right)^{-1}. \label{cavityherm}
		\end{align}
		Here calligraphic capital letters without underlines are used to denote $2\times2$ matrices and we have defined
		\begin{align}
			\mathcal{W} =\begin{bmatrix}
				\eta &  \omega \\
				\omega^\star & \eta
			\end{bmatrix}, \hspace{2cm} \mathcal{J}_{ij} =\begin{bmatrix}
				0&  J_{ij} \\
				J_{ji} & 0
			\end{bmatrix},
		\end{align}
        where $\omega^\star$ is the complex conjugate of $\omega$.   The Hermitised cavity equations  are analogous to the simple cavity equations in Eq.~(\ref{cavityequations}), but applied to the Hermitised resolvent. 
        In fact,  we recognise in the off-diagonal entries of $\mathcal{H}_{ii}$ the diagonal entries of the resolvent,  viz.,
        \begin{align}
			\mathcal{H}_{ii}   &= \left[\begin{array}{cc}A_{ii}& G_{ii}^{\star} \\ G_{ii} & D_{ii} \end{array}\right] 
            \quad {\rm and}\quad \mathcal{H}^{(j)}_{ii} = \left[\begin{array}{cc}A^{(j)}_{ii}& G_{ii}^{(j)\star} \\ G_{ii}^{(j)}& D^{(j)}_{ii} \end{array}\right].
        \end{align}
        Where we have used the shorthands $A_{ii}$ and $D_{ii}$ for the diagonal entries of the local Hermitised resolvent $\mathcal{H}_{ii}$.

        \subsection{Asymptotic expansion of the Hermitised cavity equations}\label{app:HermitisedCav}	
We develop here an asymptotic expansion in $1/\omega$ of the Hermitised cavity equations that provide us with the quantities $\mathcal{H}^{22}_{ii}$ and  $\mathcal{H}^{22}_{jj}(\omega)$ ($j\in \partial_i$) that we need to evaluate the disordered average IPR at a tail eigenvalue.    This asymptotic expansion is  complementary to the one we have developed in Sec.~\ref{appendix:expansion}  for the standard  cavity equations of the resolvent.   

As in Sec.~\ref{appendix:expansion},  we label the central hub of large effective degree by $i$, its neighbours by $j$, and the next-nearest neighbours by $k$ (see right Panel of Fig.~\ref{fig:stars}).   Approximation signs in this apppendix signify that    terms of higher order in $1/\omega$ and $\eta$ are neglected.

\subsubsection{Hermitised resolvent \texorpdfstring{$\mathcal{H}_{ii}$}{Hii} associated with the hub \texorpdfstring{$i$}{i}.}

	We have seen in Sec.~\ref{appendix:expansion} that in the asymptotic limit of large values of $\omega$,  the influence of node $i_1$ on the resolvent at node $i_2$  decays  as $\omega^{-2d(i_1,i_2)-1}$, where $d(i_1,i_2)$ is the graph distance between $i_1$ and $i_2$, when calculating the location of large eigenvalues.
	An analogous principle applies to the entries of the hermitised resolvent.   To  obtain an expression that is accurate up to terms of order $\mathcal{O}(1/\omega^4)$ (i.e., it neglects the precise form of terms of order $1/\omega^4$ or higher), we need  to consider the Hermitised cavity equations at the hub location $i$ (i.e., the node with large effective degree $|\Phi_i|$), at its neighbours $j \in \partial_i$, and at its  next-nearest neighbours $k \in \partial_j^{/i}$, while nodes that are located at  a larger distance of $i$ contribute to higher order terms in $\omega$ and can thus be neglected.

        Hence, we start with the next-nearest neighbours $k \in \partial_j^{/i}$
        Hence, we start by setting the Hermitised resolvent of the next-nearest neighbours, $k \in \partial_j^{/i}$,  equal to their asymptotic value for large $|\omega|$,  viz., 
		\begin{align}
			\mathcal{H}_{kk}^{(j)} &\approx \begin{bmatrix}
				-\frac{\eta}{\vert \omega\vert^2} &  \frac{1}{\omega^\star} \\
				\frac{1}{\omega} & -\frac{\eta}{\vert \omega\vert^2}
			\end{bmatrix} ,\label{eq:Hkkj}
        \end{align}   
        where the approximation sign indicates we have neglected higher order contributions to $\mathcal{H}_{kk}^{(j)}$  in $\omega$ and in the regulator $\eta$ (anticipating that we et $\eta\rightarrow 0$ at the end of the calculation).     

        Next, we use (\ref{eq:Hkkj}) in the Eqs.~(\ref{cavityherm}) to find the Hermitised resolvent at the neighbours $j\in \partial_i$.     In particular, we find that 
            \begin{align}
			(\mathcal{H}_{jj}^{(i)})^{-1} &= \begin{bmatrix}
				\eta - \sum_{k\in\partial_j^{/i}} J_{kj}^2 D_{kk}^{(j)}  &  \omega - \sum_{k\in\partial_j^{/i}} J_{jk} G_{kk}^{(j)} J_{kj} \\
				\omega^\star -\sum_{k\in\partial_j^{/i}} J_{jk} G_{kk}^{(j)\star} J_{kj} & \eta - \sum_{k\in\partial_j^{/i}} J_{kj}^2 A_{kk}^{(j)}
			\end{bmatrix} \nonumber \\
			&\approx \begin{bmatrix}
				\eta\left(1 + \frac{1}{\vert\omega\vert^2} \sum_{k\in\partial_j^{/i}} J_{jk}^2 \right) &  \omega - \frac{1}{\omega}\sum_{k\in\partial_j^{/i}} J_{jk} J_{kj} \\
				\omega^\star -\frac{1}{\omega^\star}\sum_{k\in\partial_j^{/i}} J_{jk}  J_{kj} & \eta\left(1+ \frac{1}{\vert\omega\vert^2} \sum_{k\in\partial_j^{/i}} J_{kj}^2 \right)
			\end{bmatrix}, \label{truncatedhermcavres}
		\end{align}
		where we anticipate taking the limit $\eta \to 0$ by keeping only leading order terms in $\eta$.    Inverting  Eq.~(\ref{truncatedhermcavres}) we obtain 
		\begin{align}
			\mathcal{H}_{jj}^{(i)}	&\approx \frac{1}{\Delta} \begin{bmatrix}
				-\eta\left(1 + \frac{1}{\vert\omega\vert^2} \sum_{k\in\partial_j^{/i}} J_{kj}^2 \right) &  \omega - \frac{1}{\omega}\sum_{k\in\partial_j^{/i}} J_{jk} J_{kj} \\
				\omega^\star -\frac{1}{\omega^\star}\sum_{k\in\partial_j^{/i}} J_{jk}  J_{kj} & -\eta\left(1+ \frac{1}{\vert\omega\vert^2} \sum_{k\in\partial_j^{/i}} J_{jk}^2 \right)
                \end{bmatrix},  \label{eq:HMatrixEQ}
                		\end{align}
               where        
		\begin{align}
			\Delta &=  \left\vert\omega - \frac{1}{\omega}\sum_{k\in\partial_j^{/i}} J_{jk} J_{kj} \right\vert^2 .
		\end{align}
		Notice that from the $(2,1)$ element of the matrix equation above, we  recover the expression for the cavity resolvent in Eq.~(\ref{cavresseries2}) up to next-to-leading order, viz.,
		\begin{align}
			G_{jj}^{(i)} &\approx \frac{1}{\omega - \frac{1}{\omega} \sum_{k\in\partial_j^{/i}} J_{jk} J_{kj} } \approx  \frac{1}{\omega} + \frac{1}{\omega^3} \sum_{k\in\partial_j^{/i}} J_{jk} J_{kj} .   \label{eq:AppCGjji}
		\end{align}
		For the diagonal entries  of (\ref{eq:HMatrixEQ}) we obtain
		\begin{align}
			A_{jj}^{(i)} &\approx \frac{-\eta\left(1 + \frac{1}{\vert\omega\vert^2} \sum_{k\in\partial_j^{/i}} J_{kj}^2 \right)}{\left\vert\omega - \frac{1}{\omega}\sum_{k\in\partial_j^{/i}} J_{jk} J_{kj} \right\vert^2} \nonumber \\
			&\approx  -\frac{\eta }{\vert\omega\vert^2}\left(1 +  \frac{1}{\vert\omega\vert^2}\sum_{k\in\partial_j^{/i}} J_{kj}^2 + 2\,\Re\left(\omega^{-2}\right)\sum_{k\in\partial_j^{/i}} J_{jk} J_{kj} \right)  , \label{cavityhermexpansionP}
            \end{align}
            and
            \begin{align}
			D_{jj}^{(i)} &\approx \frac{-\eta\left(1 + \frac{1}{\vert\omega\vert^2} \sum_{k\in\partial_j^{/i}} J_{jk}^2 \right)}{\left\vert\omega - \frac{1}{\omega}\sum_{k\in\partial_j^{/i}} J_{jk} J_{kj} \right\vert^2} \nonumber \\
			&\approx  -\frac{\eta}{\vert\omega\vert^2} \left(1 + \frac{1}{\vert\omega\vert^2}\sum_{k\in\partial_j^{/i}} J_{jk}^2 + 2\,\Re\left(\omega^{-2}\right)\sum_{k\in\partial_j^{/i}} J_{jk} J_{kj}  \right) , \label{cavityhermexpansion}
		\end{align}
        respectively.

        Lastly, we obtain an asymptotic expression for the Hermitised resolvent at the hub, $\mathcal{H}_{ii}$.   
		Substituting Eqs.~(\ref{cavityhermexpansionP}-\ref{cavityhermexpansion}) into the Eqs.~(\ref{cavityhermP}) yields
		\begin{align}
			(\mathcal{H}_{ii}^{-1})^{11}&= \eta - \sum_{j\in\partial_i} J_{ij}^2 D_{jj}^{(i)} \approx \eta\left[1 + \sum_{j\in\partial_i} \frac{J_{ij}^2}{\vert\omega\vert^2}  +  \sum_{j\in\partial_i} \frac{J_{ij}^2}{\vert\omega\vert^2} \sum_{k\in\partial_j^{/i}}\left(\frac{J_{jk}^2}{\vert\omega\vert^2} + \frac{2J_{jk} J_{kj} }{\omega^2}  \right) \right], \nonumber \\
			(\mathcal{H}_{ii}^{-1})^{12}&=  \omega - \sum_{j\in\partial_i} J_{ij} G_{jj}^{(i)} J_{ji} \approx \omega - \frac{1}{\omega}\sum_{j\in\partial_i} J_{ij} J_{ji} - \frac{1}{\omega^3} \sum_{j\in\partial_i} J_{ij} J_{ji} \sum_{k\in\partial_j^{/i}} J_{jk} J_{kj},\nonumber \\
			(\mathcal{H}_{ii}^{-1})^{21}&= \left[(\mathcal{H}_{ii}^{-1})^{12}\right]^\star ,\nonumber \\
			(\mathcal{H}_{ii}^{-1})^{22}&= \eta - \sum_{j\in\partial_i} J_{ji}^2 A_{jj}^{(i)} \approx	\eta\left[1 + \sum_{j\in\partial_i} \frac{J_{ji}^2}{\vert\omega\vert^2}  +  \sum_{j\in\partial_i} \frac{J_{ji}^2}{\vert\omega\vert^2} \sum_{k\in\partial_j^{/i}}\left(\frac{J_{kj}^2}{\vert\omega\vert^2} + \frac{2J_{jk} J_{kj} }{\omega^2}  \right) \right] , \label{hermresi}
		\end{align}
		where we have used the fact that the solutions that we obtain in Eq.~(\ref{eigenvalueapprox}) for $\omega^2$ are real.     Inverting these equations we can get  an asymptotic expression for $\mathcal{H}_{ii}$.  
        
        Notice that for the local resolvent, corresponding with the off-diagonal elements of $\mathcal{H}_{ii}$,  we obtain expressions consistent with those previously obtained in Sec.~\ref{appendix:expansion}.  Indeed, after taking the matrix inverse and the limit $\eta \to 0$,  we find 
		\begin{align}\label{Giiii}
			G_{ii} = \left( \omega - \frac{1}{\omega}\sum_{j\in\partial_i} J_{ij} J_{ji} - \frac{1}{\omega^3} \sum_{j\in\partial_i} J_{ij} J_{ji} \sum_{k\in\partial_j^{/i}} J_{jk} J_{kj}\right)^{-1} , 
		\end{align}
		which  has poles that satisfy Eq.~(\ref{implicitomi}) (up to the next-to-leading order term in $1/\omega^2$), and from which we can recover Eq.~(\ref{eigenvalueapprox}). 

\subsubsection{Hermitised resolvent \texorpdfstring{$\mathcal{H}_{jj}$}{Hjj} associated with nodes \texorpdfstring{$j\in \partial_i$}{j}.}

To derive an asymptotic expression for $\mathcal{H}_{jj}$, we can use  a formula similar to  Eq.~(\ref{messagepassing}), viz.,  
		\begin{align}
			\mathcal{H}_{jj}^{-1} &= (\mathcal{H}_{jj}^{(i)})^{-1} \mathcal{H}_{ii}^{-1}\mathcal{H}_{ii}^{(j)}, \label{hermmesspass}
		\end{align} 
        Equation (\ref{hermmesspass}) follows from the Hermitised cavity equations (\ref{cavityherm}) and (\ref{cavityhermP}).   We will use     (\ref{hermmesspass}) specifically at the tail eigenvalue $\lambda_i$, as then the right-hand side simplified as we will see.

\subsection{IPR of a tail eigenvalue with large effective degree \texorpdfstring{$|\Phi_i|$}{|Phi|} }\label{app:hermitisedCav2}
 We evaluate  the expressions for $\mathcal{H}^{22}_{ii}(\omega)$  and $\mathcal{H}^{22}_{jj}(\omega)$, as obtained in Eqs.~(\ref{hermresi}) and (\ref{hermmesspass}), respectively, at the poles $\omega=\lambda_i$ of the resolvent.    The poles of the resolvent take the form 
  [see Eq.~(\ref{eigenvalueapprox}) or (\ref{Giiii})]
        	\begin{align}
			\vert \lambda_i\vert^2 \approx \left\vert \sum_{j \in \partial_i} J_{ij}J_{ji} \right\vert + \frac{\sum_{j \in \partial_i} J_{ij}J_{ji}\sum_{k \in \partial_j^{/i}} J_{jk}J_{kj} }{\vert\sum_{j \in \partial_i} J_{ij}J_{ji} \vert} .
		\end{align}  
Importantly, at the pole 
\begin{equation}
(\mathcal{H}_{ii}^{-1}(\lambda_i))^{12} = (\mathcal{H}_{ii}^{-1}(\lambda_i))^{21} = 0,  \label{eq:prop}
\end{equation} 
which we will have to use to obtain  $\mathcal{H}^{22}_{ii}(\omega)$  and $\mathcal{H}^{22}_{jj}(\omega)$.

First, let us evaluate the   leading order term in the IPR  $\mathcal{H}^{22}_{ii}$.   From Eqs.~(\ref{hermresi}), using (\ref{eq:prop}), we find 
\begin{align}
			&\mathcal{H}^{22}_{ii}(\lambda_i) \approx \frac{1}{\eta} \Bigg[1 + \frac{\sum_{j\in\partial_i} J_{ji}^2}{\vert\sum_{j\in\partial_i} J_{ij} J_{ji}\vert} - \frac{\left(\sum_{j\in\partial_i} J_{ji}^2\right)\left(\sum_{j\in\partial_i} J_{ij} J_{ji} \sum_{k\in\partial_j^{/i}} J_{jk} J_{kj}\right)}{\left\vert\sum_{j\in\partial_i} J_{ij} J_{ji}\right\vert^3}\nonumber \\
			&\hspace{2cm}+\frac{\sum_{j\in\partial_i} J_{ji}^2 \sum_{k\in\partial_j^{/i}}J_{kj}^2 }{\left\vert\sum_{j\in\partial_i} J_{ij} J_{ji}\right\vert^2} + \frac{2\sum_{j\in\partial_i} J_{ji}^2 \sum_{k\in\partial_j^{/i}}  J_{jk} J_{kj}}{\left(\sum_{j\in\partial_i} J_{ij} J_{ji}\right)^2}\Bigg]^{-1}, \nonumber \\
			&\approx \frac{1}{\eta} \frac{\vert\sum_{j\in\partial_i} J_{ij} J_{ji}\vert}{\vert\sum_{j\in\partial_i} J_{ij} J_{ji} \vert + \sum_{j\in\partial_i} J_{ji}^2} \nonumber \\
			&+ \frac{1}{\eta}\left(\frac{\vert\sum_{j\in\partial_i} J_{ij} J_{ji}\vert}{\vert\sum_{j\in\partial_i} J_{ij} J_{ji} \vert + \sum_{j\in\partial_i} J_{ji}^2}\right)^2 \frac{\left(\sum_{j\in\partial_i} J_{ji}^2\right)\left(\sum_{j\in\partial_i} J_{ij} J_{ji} \sum_{k\in\partial_j^{/i}} J_{jk} J_{kj}\right)}{\left\vert\sum_{j\in\partial_i} J_{ij} J_{ji}\right\vert^3}\nonumber \\
			&- \frac{1}{\eta}\left(\frac{\vert\sum_{j\in\partial_i} J_{ij} J_{ji}\vert}{\vert\sum_{j\in\partial_i} J_{ij} J_{ji} \vert + \sum_{j\in\partial_i} J_{ji}^2} \right)^2 \Bigg[\frac{\sum_{j\in\partial_i} J_{ji}^2 \sum_{k\in\partial_j^{/i}}J_{kj}^2 }{\left\vert\sum_{j\in\partial_i} J_{ij} J_{ji}\right\vert^2} \nonumber \\
			&\hspace{7cm}+ \frac{2\sum_{j\in\partial_i} J_{ji}^2 \sum_{k\in\partial_j^{/i}}  J_{jk} J_{kj}}{\left(\sum_{j\in\partial_i} J_{ij} J_{ji}\right)^2} \Bigg]  , \label{dii}
		\end{align}
        where the last approximation sign follows from expanding in the effective degree $|\Phi_i|=\vert\sum_{j\in\partial_i} J_{ij} J_{ji}\vert$.

        Second we evaluate  $\mathcal{H}^{22}_{jj}(\lambda_i)$ from Eq.~(\ref{hermmesspass}).  Beginning with the cavity equations Eqs.~(\ref{cavityherm}), using (\ref{eq:prop}) we have for small $\eta$ that 
		\begin{align} \label{eq:H_ii^{(j)}}
			\mathcal{H}_{ii}^{(j)}(\lambda_i) &= \begin{bmatrix}
				\mathcal{O}(\eta)  &  \left(J_{ij} J_{ji}G_{jj}^{(i)\star} \right)^{-1} \\
				\left(J_{ij} J_{ji}G_{jj}^{(i)} \right)^{-1}& \mathcal{O}(\eta) 
			\end{bmatrix}.
		\end{align}
		Combining Eqs.~(\ref{cavityhermexpansionP}), (\ref{cavityhermexpansion}), (\ref{hermresi}) and (\ref{eq:H_ii^{(j)}}) with (\ref{hermmesspass}), we  find that
		\begin{align}
			\mathcal{H}_{jj}^{-1}  &= \begin{bmatrix}
				\left(J_{ij}J_{ji} D_{ii} \left(G_{jj}^{(i)} \right)^2 \right)^{-1}  &  \mathcal{O}(\eta^2) \\
				\mathcal{O}(\eta^2) & \left(J_{ij}J_{ji} A_{ii} \left(G_{jj}^{(i)\star} \right)^2 \right)^{-1} 
			\end{bmatrix} , \label{hermresopposite}
		\end{align}
		and, by inverting Eq.~\eqref{hermresopposite},
		\begin{align}
			\mathcal{H}^{22}_{jj} &= J_{ij} J_{ji} \mathcal{H}^{11}_{ii} \left( G_{jj}^{(i)\star} \right)^2 + \mathcal{O}(\eta^2) , \nonumber \\
			\mathcal{H}^{11}_{jj} &= J_{ij} J_{ji} \mathcal{H}^{22}_{ii} \left( G_{jj}^{(i)} \right)^2 + \mathcal{O}(\eta^2) . \label{messagepassingasym1}
		\end{align} 
        Using   $(G_{jj}^{(i)})^2 \sim 1/\lambda_i^2$ in  the first line of (\ref{messagepassingasym1}) and the expression of  $\mathcal{H}^{11}_{ii}(\lambda_i)$ obtained from  (\ref{hermresi}),  we obtain the final expression for $\mathcal{H}^{22}_{jj}$.   

        Substitution of the asymptotic expressions for $\mathcal{H}^{22}_{ii}$ [Eq.~(\ref{dii})] and  $\mathcal{H}^{22}_{jj}$  [the first line of (\ref{messagepassingasym1})] in (\ref{iprfromres2}), we find  the  following expression for the IPR     up to next-to-leading order in $1/\lambda_i^2$ (i.e., up to terms of order $|\Phi_i|^{-1}$), 
		\begin{align}
			q^{-1}_i &\approx \left(\frac{\vert\sum_{j\in\partial_i} J_{ij} J_{ji}\vert}{\vert\sum_{j\in\partial_i} J_{ij} J_{ji} \vert + \sum_{j\in\partial_i} J_{ji}^2} \right)^2 \nonumber \\
			&+ 2\left(\frac{\vert\sum_{j\in\partial_i} J_{ij} J_{ji}\vert}{\vert\sum_{j\in\partial_i} J_{ij} J_{ji} \vert + \sum_{j\in\partial_i} J_{ji}^2} \right)^3 \frac{\left(\sum_{j\in\partial_i} J_{ji}^2\right)\left(\sum_{j\in\partial_i} J_{ij} J_{ji} \sum_{k\in\partial_j^{/i}} J_{jk} J_{kj}\right)}{\left\vert\sum_{j\in\partial_i} J_{ij} J_{ji}\right\vert^3} \nonumber \\
			&- 2\left(\frac{\vert\sum_{j\in\partial_i} J_{ij} J_{ji}\vert}{\vert\sum_{j\in\partial_i} J_{ij} J_{ji} \vert + \sum_{j\in\partial_i} J_{ji}^2} \right)^3 \Bigg[\frac{\sum_{j\in\partial_i} J_{ji}^2 \sum_{k\in\partial_j^{/i}}J_{kj}^2 }{\left(\sum_{j\in\partial_i} J_{ij} J_{ji}\right)^2} \nonumber \\
			&\hspace{7cm}+ \frac{2\sum_{j\in\partial_i} J_{ji}^2 \sum_{k\in\partial_j^{/i}}  J_{jk} J_{kj}}{\left(\sum_{j\in\partial_i} J_{ij} J_{ji}\right)^2} \Bigg]  \nonumber \\
			&+ \left(\frac{\vert\sum_{j\in\partial_i} J_{ij} J_{ji}\vert}{\vert\sum_{j\in\partial_i} J_{ij} J_{ji} \vert + \sum_{j\in\partial_i} J_{ji}^2}\right)^2 \frac{\left(\sum_{j\in\partial_i} J_{ij}^2J_{ji}^2\right)}{ \left(\sum_{j\in\partial_i} J_{ij} J_{ji}\right)^2}.\label{iprasymgeneral}
		\end{align} Neglecting the subleading terms in (\ref{iprasymgeneral}), we recover the Eq.~(\ref{iprgeneral}) that we were meant to derive. 
        
        For symmetric random matrices with  $J_{ij} = J_{ji}$, we  obtain the following simpler expression
		\begin{align}
			q^{-1}_i(\lambda_i) &\approx  \frac{1}{4} \left( 1 - 2\frac{\sum_{j\in\partial_i}J_{ij}^2 \sum_{k \in \partial_j^{/i}} J_{jk}^2}{\left(\sum_{j\in \partial_i}J_{ij}^2\right)^2} + \frac{\sum_{j\in\partial_i}J_{ij}^4}{\left(\sum_{j\in \partial_i}J_{ij}^2\right)^2}\right) \label{iprsym}.
		\end{align}
        
        Although the  expression (\ref{iprasymgeneral}) is rather cumbersome,  it can simplify greatly when taking its   disorder-average  [as defined in Eq.~(\ref{iprdef})]. We demonstrated this simplification in  Sec.~\ref{section:example:constantweight} for some special cases.

		\section{Exponential localisation of eigenvectors}\label{appendix:exploc}
		In this appendix, we  give an intuitive argument for why  the right eigenvectors corresponding to large eigenvalues are exponentially localised around well-connected hubs of the network. For this purpose, we evaluate the average of $|r^{(i)}_{j}|$ for a tail eigenvalue $\lambda_i$ as a function of the distance  $d(i,j)$ from the central hub $i$. We note that the superscript $(i)$ in $r^{(i)}_{j}$ refers to the large eigenvalue produced by a hub at site $i$; it is not to indicate the cavity graph produced by removing site $i$.
        
\begin{figure}[H]
			\centering 
			\includegraphics[scale = 0.5]{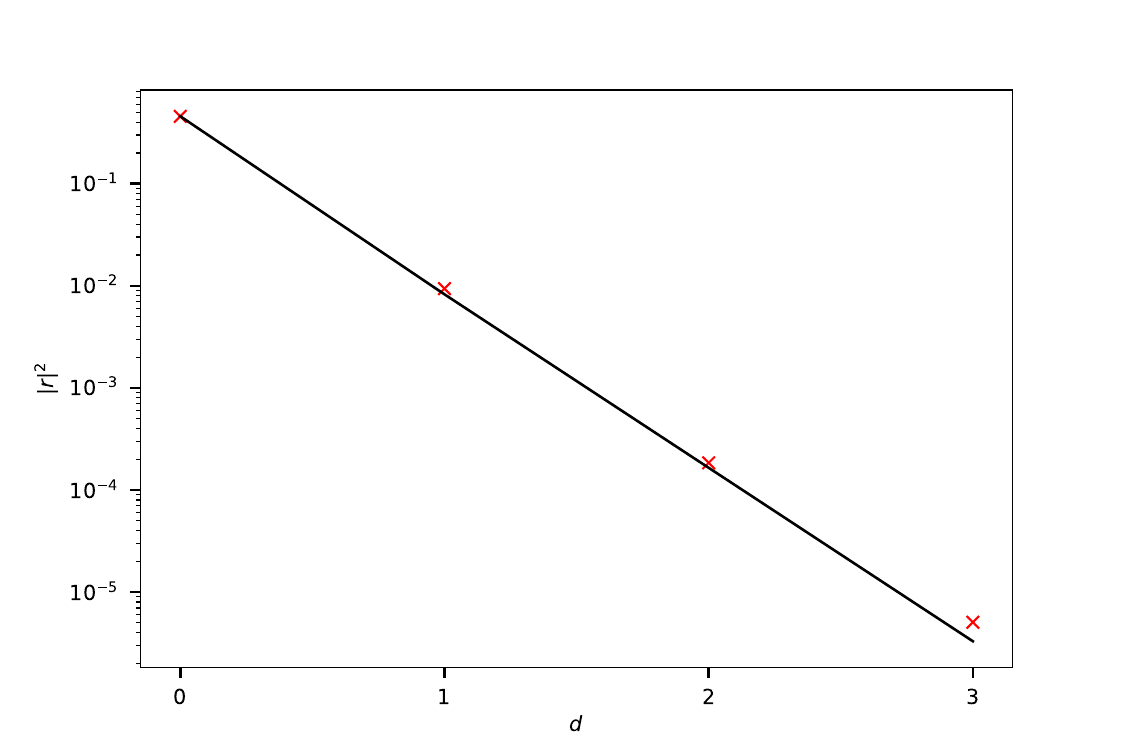}
			\caption{Average square magnitude of the right eigenvector components $\sum_{j:d(i,j)=d}|r^{(i)}_j|^2$ as a function of the distance $d$ from the well-connected hub at site $i$, where $r^{(i)}_j$ are the components of the right eigenvector $r^{(i)}$ localised on the hub.   A single hub with $\Delta = 50$ and $c = \alpha\Delta/\sqrt{\alpha^2 - 4 \epsilon(1-\epsilon)}$ [see the text surrounding Eq.~(\ref{iprgeo})] was added to a network with a geometric degree distribution, constructed according to the prescription in Sec. \ref{section:example:constantweight} of the main text with $N = 4000$, $p = 2$, $\epsilon = 0.25$ and $ a= 1$. The solid black line is the theory prediction discussed in the text above.}\label{fig:exploc}
		\end{figure}

		Given the assumption that there is a  well-connected hub at site $i$, from Eq.~(\ref{dii}) and the fact that $D_{ii} \sim \eta^{-1} \vert r^{(i)}_i\vert^2$  \cite{neri2016eigenvalue}, we  obtain 
		\begin{align}
			\vert r^{(i)}_i\vert^2 \approx \frac{\vert\sum_{j\in\partial_i} J_{ij} J_{ji}\vert}{\vert\sum_{j\in\partial_i} J_{ij} J_{ji} \vert + \sum_{j\in\partial_i} J_{ji}^2},
		\end{align}
        where the approximation sign indicates that we have neglected subleading order terms in (\ref{dii}).
		From the relationship between the Hermitised resolvents at neighbouring sites in Eq.~(\ref{messagepassingasym1}), we can determine  the structure of the right eigenvectors in the vicinity of $i$. Indeed, using (\ref{messagepassingasym1}) we find 
		\begin{align}
			\frac{1}{c}\sum_{j\in\partial_i}\vert r^{(i)}_j\vert^2 \approx \frac{\vert r^{(i)}_i\vert^2}{c} \sum_{j\in\partial_i} \frac{J_{ij}J_{ji}}{\lambda_i^2} \approx \frac{\vert r^{(i)}_i\vert^2}{c} 
		\end{align}
        for the average square magnitude of the eigenvector components of the neighbours. 
		Furthermore, if we consider  the next-nearest neighbours, we obtain
		\begin{align}
			\frac{1}{c}\sum_{j\in\partial_i}\frac{1}{c_j}\sum_{k\in\partial j}\vert r^{(i)}_k\vert^2 \approx \frac{1}{c}\sum_{j\in\partial_i}\frac{\vert r^{(i)}_j\vert^2}{c_j}\sum_{k\in\partial j} \frac{J_{jk}J_{kj}}{\lambda_i^2} \approx  \frac{\vert r^{(i)}_i\vert^2 \langle uv\rangle_{\pi(u,v)}}{c \lambda_i^2} , \label{eq:D3}
		\end{align}
        where we have applied the law of large numbers to the sum over $j$ as $|\partial_i|$ is large.
		Going one step further yields
		\begin{align}
			\frac{1}{c}\sum_{j\in\partial_i}\frac{1}{c_j}\sum_{k\in\partial j}\frac{1}{c_k}\sum_{l\in\partial_k}\vert r^{(i)}_l\vert^2 \approx \frac{\vert r^{(i)}_i\vert^2 \left(\langle uv\rangle_{\pi(u,v)}\right)^2}{c \, \lambda_i^4} ,
		\end{align}
		and this pattern continues as we move further away from the node $i$. Hence, there is an exponential (or geometric) decay of the magnitude of the eigenvector components as we move further from the hub at node $i$, as also shown in Fig.~\ref{fig:exploc} for the example of a random graph with geometric degree distribution.

        \section{Self-averaging contribution from the wider network}\label{appendix:selfaveraging}
		
		One crucial observation for the remaining calculations is that the contribution from the wider network can often be treated as a deterministic quantity. More specifically, we show that the second term in Eq.~(\ref{eigenvalueapprox}) takes the  deterministic value 
         \begin{equation}
    \Phi_2 \approx \frac{(p_2-p)}{p \Phi} \left\langle J_{12}J_{21}\right\rangle_\pi \label{eq:Phi2Det}
    \end{equation}
    if the first term 
    \begin{equation}
    \Phi = \sum_{j\in \partial_i}J_{ij}J_{ji},
    \end{equation}
    is large, and if node $i$  has a large degree.   In other words, we show that  the contribution from the wider network is self-averaging when the hub in question is well-connected. 
    
    Note that also in Eq.~(\ref{configdensitysum}) of the main text we neglect the fluctuations of $\Phi_2$ [i.e. the second term in Eq.~(\ref{eigenvalueapprox})], due to the justification above. Furthermore, when $\Phi_2$ can be treated as a deterministic quantity as stated above in Eq.~(\ref{eq:Phi2Det}),  the wider network contributes for large $\omega$ to the prefactor of the eigenvalue spectral density, as $\Phi_2 \sim \mathcal{O}(\Phi^0)$. In Sec.~\ref{appendix:configurationipr} we  use a similar  reasoning to approximate terms in the IPR by their average values.

    Let us now derive (\ref{eq:Phi2Det}) from a central-limit-like argument.    Within our single-defect approximation, where the highly-connected hubs can be considered as effectively isolated from one another, we can using Eq.~(\ref{eigenvalueapprox}) express the spectral density at large $\omega$ by  
		\begin{align}
			\rho(\omega ) &\approx \sum_{c_i}  P_\mathrm{deg}(c_i) \int  \prod_{j \in \partial_i}\dd J_{ij} \dd J_{ji} \pi\left( J_{ij}, J_{ji} \right)  \Bigg\{ \sum_{c_j}P_\mathrm{deg}(c_j \vert c_i) \int\prod_{k \in \partial_j^{/i}}  \dd J_{jk} \dd J_{kj} \pi\left( J_{jk}, J_{kj} \right) \Bigg\} \nonumber \\
			&\hspace{2cm}\times\delta\left( \omega - \sqrt{\sum_{j \in \partial_i} J_{ij}J_{ji} + \frac{\sum_{j\in \partial_i} J_{ij}J_{ji} \sum_{k\in\partial_j^{/i}} J_{jk}J_{kj}}{\sum_{j \in \partial_i} J_{ij}J_{ji}}   }  \right) , \label{generaldensityintegral}
		\end{align}
        where we note that $c_i = \sum_{j\in \partial_i} 1$ and $c_j = 1+\sum_{k\in \partial_j^{/i}} 1$, where $P_\mathrm{deg}(c_j\vert c_i)$ is the conditional probability that the neighbour of a node with degree $c_i$ has degree $c_j$. For the case where degree correlations are negligible (e.g., in the case of configuration model networks~\cite{bollobas1980probabilistic,van2017random, newman2018networks}), one has the following simple rule:
        \begin{equation}
        P_\mathrm{deg}(c_j \vert c_i) = P_\mathrm{deg}(c_j) \frac{c_j}{p}. \label{eq:PDegD}
        \end{equation}
        
        Let us consider the object 
        \begin{equation}
        \Phi_2 = \frac{1}{\Phi}\sum_{j\in\partial_i}J_{ij}J_{ji} \sum_{k\in \partial_j^{/i}} J_{jk}J_{kj} 
        \end{equation}
        that appears in the argument of the $\delta$ distribution of (\ref{generaldensityintegral}), where $\Phi = \sum_{j\in \partial_i} J_{ij}J_{ji}$ is a signed version of the effective degree of node $i$. The probability distribution of $\Phi_2$, conditioned on the values of $\{J_{ij}, J_{ji}\}_{j\in \partial_i}$, for which we use the notation $P(\Phi_2|\Phi)$, can be written as follows (assuming large $\Phi$) 
		\begin{align}
			&P(\Phi_2 \vert \Phi) = \int \frac{\dd\hat\Phi_2}{2\pi} e^{{\rm i}\hat\Phi_2 \Phi_2} \prod_{j \in \partial_i} \Bigg\{ \nonumber\\
			&\sum_{c_j}P_\mathrm{deg}(c_j \vert c_i) \int\prod_{k \in \partial_j^{/i}} \left( \dd J_{jk} \dd J_{kj} \pi\left( J_{jk}, J_{kj} \right)\right) \exp\left(-{\rm i} \frac{\hat\Phi_2}{\Phi} J_{ij}J_{ji} \sum_{k \in \partial_j^{/i}}J_{jk}J_{kj}\right)\Bigg\} \nonumber \\
			&\approx \int \frac{\dd\hat\Phi_2}{2\pi} e^{{\rm i}\hat\Phi_2 \Phi_2} \prod_{j\in \partial_i} \Bigg\{\sum_{c_j}P_\mathrm{deg}(c_j \vert c_i) \int\prod_{k \in \partial_j^{/i}} \dd J_{jk} \dd J_{kj} \pi\left( J_{jk}, J_{kj} \right) \nonumber \\
			&\hspace{2cm}\times \left[ 1-{\rm i}\frac{\hat\Phi_2}{\Phi} J_{ij}J_{ji} \sum_{k \in \partial_j^{/i}} J_{jk}J_{kj} - \frac{\hat\Phi_2^2}{2\Phi^2} \left( J_{ij}J_{ji} \sum_{k \in \partial_j^{/i}} J_{jk}J_{kj}\right)^2 \right]	\Bigg\} \nonumber \\
			&= \int \frac{\dd\hat\Phi_2}{2\pi} e^{{\rm i}\hat\Phi_2 \Phi_2} \prod_{j\in \partial_i} \Bigg\{\sum_{c_j}\frac{c_j}{p} P_\mathrm{deg}(c_j)  \Bigg[ 1-{\rm i} \frac{\hat\Phi_2}{\Phi} J_{ij}J_{ji} (c_j -1) \left\langle J_{12}J_{21}\right\rangle_\pi \nonumber \\
			&\hspace{4cm}- \frac{\hat\Phi_2^2}{2\Phi^2} \left( J_{ij}J_{ji} \right)^2 (c_j -1)\left\langle\left(J_{12}J_{21}\right)^2\right\rangle_\pi \Bigg]	\Bigg\} \nonumber \\
			&= \int \frac{\dd\hat\Phi_2}{2\pi} e^{{\rm i}\hat\Phi_2 \Phi_2} \prod_{j\in \partial_i}\Bigg\{ 1 - {\rm i} \frac{(p_2-p)}{p}\frac{\hat \Phi_2}{\Phi} J_{ij}J_{ji} \left\langle J_{12}J_{21}\right\rangle_\pi\nonumber \\
			&\hspace{4cm}- \frac{\hat\Phi_2^2}{2\Phi^2} \left( J_{ij}J_{ji} \right)^2 \frac{(p_2-p)}{p}\left\langle\left(J_{12}J_{21}\right)^2\right\rangle_\pi \Bigg\} \nonumber \\
			&\approx \int \frac{\dd\hat\Phi_2}{2\pi}  \exp\Bigg(   {\rm i} \hat \Phi_2 \left[\Phi_2 - \frac{(p_2-p)}{p}  \left\langle J_{12}J_{21}\right\rangle_\pi\nonumber \right]\\
			&\hspace{4cm} - \frac{\hat\Phi_2^2}{2}  \frac{(p_2-p)}{p} \left\langle\left(J_{12}J_{21}\right)^2\right\rangle_\pi \frac{\sum_{j \in \partial_i}\left( J_{ij}J_{ji} \right)^2}{\Phi^2}\Bigg), \label{eq:longSeries}
		\end{align}
		where we recall that  $p_2 = \sum_c c^2 P_\mathrm{deg}(c)$.     Hence,  if $\Phi^2$ is large relative to $\sum_{j\in \partial_i} (J_{ij}J_{ji})^2$, then we recover a Dirac distribution.    Thus for large enough $\Phi$,  $\Phi_2$  takes  the deterministic value  given by Eq.~(\ref{eq:Phi2Det})
    independent of $\Phi$, as we were meant to show. Notice that the approximation signs in  (\ref{eq:longSeries}) are due to higher order terms in $\Phi$ having been neglected.

   Note that in cases when the network defect comprises of a single heavily-weighted edge, then $\Phi^2$ is not large in comparison to $\sum_{j\in\partial_i} (J_{ij}J_{ji})^2$.   Hence, the present line of  reasoning  applies  to  hubs that have a  large degree . 
		
		\section{Asymptotic analysis for random graphs with geometric and power-law degree distributions and with edge weights of fixed absolute value (corresponding with Sec.~\ref{section:example:constantweight})}\label{app:geomPower}
		
		In this appendix, we  discuss and motivate in more detail some of the results in Section \ref{section:example:constantweight}, where we consider the tail eigenvalue density and disorder-averaged IPR for signed graphs with uniform edge weight magnitudes. 
		
		In Sec.~\ref{appendix:configmodelconditional}, we derive an approximate expression for the conditional probability $P(c \vert \Delta)$ [see the discussion after Eq.~(\ref{configdensitysum})], which is valid for large $\Delta$ and $c$. Using  this approximation, we  derive the expressions for the tail region eigenvalue density in  geometric random graphs Eq.~(\ref{eq:geometric_spectral_density_fixed_weights}), and in power-law random graphs, Eqs.~(\ref{epslhalf}) and (\ref{epsghalf}) in Section \ref{appendix:eigenvaluedensityconfig}. Furthermore,  in Sec.~\ref{appendix:configurationipr}  we use the aforementioned approximation to  derive the expressions for the disorder-averaged IPR given in Eqs.~(\ref{iprgeo}) and (\ref{iprpow}) of the main text, for geometric and power-law degree distributions, respectively. Finally, in Sec.~\ref{appendix:crossover}, we discuss briefly the crossover regime that occurs for $\epsilon \approx 1/2$ in the case of the power-law degree distribution.

		\subsection{ Gaussian approximation of \texorpdfstring{$P(\Delta , c)$}{P(Delta,c)} for large \texorpdfstring{$\Delta$}{Delta}}\label{appendix:configmodelconditional}
		
		In this appendix, we  derive simple approximate expressions for $P(c,\Delta)$, which will help us  evaluate the sum over $c$ in Eq.~(\ref{configdensitysum}).   We use these expressions in Sec.~\ref{appendix:eigenvaluedensityconfig} to extract the asymptotic eigenvalue density given in Eqs.~(\ref{eq:geometric_spectral_density_fixed_weights}), (\ref{epslhalf}) and (\ref{epsghalf}), and in Sec.~\ref{appendix:configurationipr} we use these expression for  $P(c,\Delta)$ to derive the expected IPR in Eqs.~(\ref{iprgeo}) and (\ref{iprpow}).
		
		Let us begin with the expression for $P(\Delta, c)$ given in  Eq.~(\ref{eq:PDeltaC}).  Using Stirling's approximation for the binomial coefficient (valid for large $c$ and $c\pm \Delta$, which we will find to be the case \textit{a posteriori}), we obtain
		\begin{align}
			\ln P(\Delta, c) \approx& \ln P_\mathrm{deg}(c) + c \ln c - \frac{(c-\Delta) }{2} \ln \left(\frac{c-\Delta }{2}\right) - \frac{(c+\Delta) }{2} \ln \left( \frac{c+\Delta }{2} \right) \nonumber \\
			&+  \frac{(c+\Delta)}{2} \ln (1-\epsilon)  + \frac{(c-\Delta)}{2}  \ln \epsilon + \frac{1}{2} \ln\left(\frac{4c}{(c+\Delta)(c-\Delta)} \right) . \label{lnpdeltac}
		\end{align}
		Let us now find the value $c = c_0$ for which $\ln P(\Delta, c)$ is greatest for fixed $\Delta$.  Equating the derivative of the right-hand side of (\ref{lnpdeltac}) to zero, we find the equation
		\begin{align}
			\frac{P_\mathrm{deg}'(c_0)}{P_\mathrm{deg}(c_0)} + \ln\left(\frac{2c_0 \sqrt{\epsilon(1-\epsilon)}}{\sqrt{(c_0+\Delta)(c_0-\Delta)}} \right) - \frac{c_0^2 + \Delta^2}{2 c_0(c_0+\Delta)(c_0-\Delta)}= 0 \label{eq:c0Max}
		\end{align}
        for $c_0$.

        Next we examine (\ref{eq:c0Max}) for the two examples for $P_\mathrm{deg}(c)$ that we give in the main text: the geometric distribution and the power-law.  In both cases, we make the ansatz 
        \begin{equation}
        c_0 = c_0^{(0)} \: \Delta  + c_0^{(1)} + \mathcal{O}(1/\Delta) \label{eq:ansatz}
        \end{equation}
        in Eq.~(\ref{eq:c0Max}) and solve towards $c_0^{(0)}$ and $c_0^{(1)}$.

        In the case of the geometric distribution, it holds that $\ln P_\mathrm{deg}(c) = -c \ln\alpha + \mathrm{const.}$, with $\alpha = 1 +p^{-1}$. Hence from  (\ref{eq:c0Max}) it follows that
		\begin{equation}
			\frac{4 c_0^2 \epsilon(1-\epsilon)}{\alpha^2 (c_0+\Delta)(c_0-\Delta)} = \exp\left(-\frac{1}{c_0} + \frac{1}{c_0+\Delta} +\frac{1}{c_0-\Delta} \right).  \label{eq:asymptoC0}
		\end{equation}
Using the ansatz (\ref{eq:ansatz}) in (\ref{eq:asymptoC0}),  expanding the resultant Eq.~(\ref{eq:asymptoC0}) as a series in $1/\Delta$, and solving towards  the coefficients $c_0^{(0)}$ and $c_0^{(1)}$ we obtain
\begin{align}
			c^{(0)}_0 &= \frac{\alpha }{\sqrt{\alpha^2 - 4 \epsilon (1-\epsilon)}} \label{c0geomgeom}
            \end{align}
            and 
            \begin{align}
			c^{(1)}_0 &= -\frac{\alpha^2 - 4 \epsilon (1-\epsilon)}{8\epsilon(1-\epsilon)}\left( \frac{\alpha^4}{[\alpha^2 - 4 \epsilon (1-\epsilon)]^2} -1\right).   \label{c0geom}
		\end{align}  
		Note that the denominator in the expression for $c_0^{(0)}$ is real for all choices of $p>0$. 
		
		In the case of the power-law distribution, we instead have $\ln P_\mathrm{deg}(c) = -\gamma \ln c + \mathrm{const.}$, for which  (\ref{eq:c0Max})  yields the equation
		\begin{align}
			\frac{4 c_0^2 \epsilon(1-\epsilon)}{ (c_0+\Delta)(c_0-\Delta)} = \exp\left(\frac{2\gamma}{c_0} -\frac{1}{c_0} + \frac{1}{c_0+\Delta} +\frac{1}{c_0-\Delta} \right). 
		\end{align}
		If we once again use the ansatz we obtain in this case 
		\begin{align}
			c^{(0)}_0 &=  \frac{1 }{ \vert 1 - 2 \epsilon \vert } \label{eq:c00pow}
        \end{align}    
        and 
          		\begin{align}
			c^{(1)}_0 &= -\frac{  1+ 2 \epsilon (1-\epsilon) (2\gamma-1)}{(1-2\epsilon)^2} . \label{c0pow}
		\end{align}
        We thus find that in both geometric and power law degree distributions, $c_0 \sim \Delta$. As we demonstrate in the following appendices, the subleading $\mathcal{O}(\Delta^0)$ terms in the expansion of $c_0$ contribute at the same order to the eigenvalue density and the IPR  as the contributions from the wider network.   This approach does not apply for Erd\H{o}s-R\'{e}nyi graphs, as we discuss in Sec.~\ref{appendix:poissonian}.
	
		Expanding the combined probability distribution $P(\Delta, c)$  in Eq.~(\ref{lnpdeltac}) about the point $c = c_0$, we find the Gaussian approximation 
		\begin{align}
			\ln P(\Delta, c) =& \ln P(\Delta, c_0) - \frac{1}{2} \frac{\Delta^2}{c_0 (c_0 + \Delta)(c_0-\Delta)} (c - c_0)^2  \nonumber\\  
            & + \mathcal{O}(1/\Delta) + \mathcal{O}((c-c_0)^3).   \label{lnpgauss}
		\end{align}
		Hence $c$ fluctuates within roughly $\sim \pm\sqrt{\Delta}$ of its mean value.

		\subsection{Asymptotic analysis for the eigenvalue density}\label{appendix:eigenvaluedensityconfig}
		We use the approximation in Eq.~(\ref{lnpgauss}) to extract the asymptotic eigenvalue density in Eqs.~(\ref{eq:geometric_spectral_density_fixed_weights}), (\ref{epslhalf}) and (\ref{epsghalf}) from Eq.~(\ref{configdensitysum}).  
		
		We write for the quantity $P(\Delta)$ in Eq.~(\ref{configdensitysum})  as 
		\begin{align}
			P(\Delta) &= \sum_{c \in S(\Delta)}^\infty P(\Delta, c) \nonumber \\
			&= \frac{1}{2}\int_\Delta^\infty \dd c  P(\Delta, c) + \frac{P(\Delta, \infty) + P(\Delta, \Delta)}{2} + \frac{\partial_c P \vert_{c \to \infty} -\partial_c P\vert_{c = \Delta} }{12} + \cdots, 
		\end{align}
		where we  have used the Euler-Maclaurin formula to approximate the sum as an integral. Approximating now the integral using Eq.~(\ref{lnpgauss}), we find 
		\begin{align}
			P(\Delta) &= \frac{1}{2}P(\Delta, c_0)\int_\Delta^\infty \dd c e^{ - \frac{1}{2} \frac{\Delta^2}{c_0(c_0+\Delta)(c_0-\Delta)}(c-c_0)^2 + \frac{1}{3!} \frac{B_1}{\Delta^2} (c-c_0)^3 + \frac{1}{4!} \frac{B_2}{\Delta^2} (c-c_0)^4 + \cdots}  \nonumber \\
			&+ \frac{P(\Delta, \infty) + P(\Delta, \Delta)}{2} + \frac{\partial_c P \vert_{c \to \infty} -\partial_c P\vert_{c = \Delta} }{12} + \cdots,\nonumber \\
			&= \frac{1}{2}P(\Delta, c_0)\Bigg[  \sqrt{\frac{\pi c_0 (c_0+\Delta)(c_0-\Delta)}{2 \Delta^2}}\mathrm{Erfc}\left(\sqrt{ \frac{\Delta^2(c_0-\Delta)}{c_0 (c_0+\Delta)}}\right)  \nonumber \\
			&+ \int_{-\infty}^\infty \dd c e^{ - \frac{1}{2} \frac{\Delta^2}{c_0(c_0+\Delta)(c_0-\Delta)}(c-c_0)^2 }  + \mathcal{O}\left( \frac{1}{\sqrt{\Delta}}\right)\Bigg] \nonumber \\ 
			&+ \frac{P(\Delta, \infty) + P(\Delta, \Delta)}{2} + \frac{\partial_c P \vert_{c \to \infty} -\partial_c P\vert_{c = \Delta} }{12} + \cdots.\label{approximationspdelta}
		\end{align}
		From Eq.~(\ref{lnpdeltac}), we see that $P(\Delta, \infty) = 0$, and so are the derivatives evaluated at $c \to \infty$. On the other hand, we have from (\ref{eq:PDeltaC}) that $P(\Delta, \Delta) = P_\mathrm{deg}(\Delta) (1-\epsilon)^\Delta $. Considering the relative size, we see that $P(\Delta,\Delta)$ is exponentially smaller than $P(\Delta, c_0)$ in $\Delta$, and similarly for the derivative terms that arise from the Euler-Maclaurin formula. We can therefore safely neglect these terms. With regards to the complementary error function that arises from the lower limit of the integral, we use the fact that $\mathrm{Erfc}(x) e^{x^2} x \sqrt{\pi} \to 1$ as $x \to \infty$. From this we see that the term containing the error function goes like $e^{-\alpha \Delta}$, where $\alpha$ is a positive constant, and hence we can also neglect this term. Ultimately, we obtain
		\begin{align} \label{eq:pdelta}
			P(\Delta) &= \frac{1}{2}P(\Delta, c_0)\Bigg[ \sqrt{2 \pi \frac{c_0(c_0+\Delta)(c_0-\Delta)}{\Delta^2}} + \mathcal{O}\left( \frac{1}{\sqrt{\Delta}}\right)\Bigg]  .
		\end{align}
        One notes that the last term in Eq.~(\ref{lnpdeltac}) is $\sim 1/\sqrt{\Delta}$. Combining it with the prefactor of $P(\Delta,c_0)$ in Eq.~\eqref{eq:pdelta},  recalling that $c_0 \sim \Delta$, and neglecting subleading order contributions,  we find 
		\begin{align}
			\ln P(\Delta) \approx& \ln P_\mathrm{deg}(c_0) + c_0 \ln c_0 - \frac{(c_0-\Delta) }{2} \ln \left(\frac{c_0-\Delta }{2}\right) - \frac{(c_0+\Delta) }{2} \ln \left( \frac{c_0+\Delta }{2} \right) \nonumber \\
			&+  \frac{(c_0+\Delta)}{2} \ln (1-\epsilon)  + \frac{(c_0-\Delta)}{2}  \ln \epsilon . \label{lnpdelta}
		\end{align}		
        Therefore, we obtain (making the substitution $\Delta = \kappa \omega^2$) 
		\begin{align}
			\rho(\omega) &\sim \sum_{\Delta= 0}^\infty P(\Delta) \delta\left(\omega-\sqrt{a^2[\Delta + (p_2-p)  (1-2\epsilon)/p ] }\right), \nonumber \\
			&\approx \omega^2 \int \dd\kappa P(\omega^2 \kappa) \delta\left(\omega-\sqrt{a^2[\omega^2 \kappa + (p_2-p)  (1-2\epsilon)/p ] }\right) \nonumber \\
			&\sim \omega P\left(\Delta(\omega) \right), \label{rhoomegapm}
		\end{align}
		where $\Delta(\omega) = \frac{\omega^2}{a^2} - (p_2-p)  (1-2\epsilon)/p$. 
        
     In conclusion, the expressions for the eigenvalue density in Eqs.~(\ref{eq:geometric_spectral_density_fixed_weights}), (\ref{epslhalf}) and (\ref{epsghalf})   of the main text are obtained by evaluating $\rho(\omega) \approx \omega P(\Delta)$ at $\Delta = \omega^2/a^2 - (p_2-p) (1-2 \epsilon)/p$, where $P(\Delta)$ is given by Eq.~(\ref{lnpdelta}) and $c_0(\Delta)$ is given by Eq.~(\ref{c0geom}) for the geometric degree distribution and Eq.~(\ref{c0pow}) for the power-law degree distribution. 
		
		One notes that in Eqs.~(\ref{eq:geometric_spectral_density_fixed_weights}), (\ref{epslhalf}) and (\ref{epsghalf})   of the main text, we ignore the `shift' $- (p_2-p)  (1-2\epsilon)/p$ from the wider network in $\Delta(\omega)$. This shift only contributes to $\ln P(\Delta)$ at the order $\mathcal{O}(1/\Delta)$, which we neglect in our approximation scheme. That is, we see that the wider network has little influence on the asymptotic eigenvalue density for the geometric and power-law networks. We show that the wider-network contribution to the eigenvalue density in the case of the ER network is more significant, but still small, in Sec.~\ref{appendix:poissonian}. In contrast, the influence of the wider network is far more important in the calculation of the IPR, as we will show in the next subsection.

		\subsection{Asymptotic analysis for the disordered averaged IPR}\label{appendix:configurationipr}
		We now use the approximation (\ref{lnpgauss}) for $P(\Delta,c)$, which also yields a corresponding approximation of  $P(c\vert \Delta)$, to evaluate the disorder average of the IPR (for fixed $\omega$) in Eq.~(\ref{iprgeneral}), ultimately deriving Eqs.~(\ref{iprgeo}) and (\ref{iprpow}).   Furthermore, we obtain the expressions (\ref{eq:AGeo}) and (\ref{eq:APow}) for the coefficients $A_{\rm geo}$ and $A_{\rm pow}$ in Eqs.~(\ref{iprgeo}) and (\ref{iprpow}), respectively. 
		
		Beginning with the expression in Eq.~(\ref{iprasymgeneral}), we use the following approximations, which are valid for large $\Delta$ [where we neglect relative fluctuations that are $\mathcal{O}(1/\Delta)$, as justified in Sec.~\ref{appendix:selfaveraging}, using the law of large numbers]   
		\begin{align}
			\sum_{j\in\partial_i} J_{ij} J_{ji} \sum_{k\in\partial_j^{/i}} J_{jk} J_{kj} &\approx  a^2 (1-2\epsilon) \sum_{j\in\partial_i} J_{ij} J_{ji} (c_j-1)  \nonumber \\
			&= a^4 (1-2\epsilon) \Delta \sum_{c_j} P_\mathrm{deg}(c_j) \frac{c_j^2 - c_j}{p} \nonumber \\
			&=  a^4 (1-2\epsilon) \Delta \frac{(p_2-p)}{p},\label{sum1}
		\end{align}
		and, similarly, we also have 
		\begin{align}
			\sum_{j\in\partial_i} J_{ji}^2 \sum_{k\in\partial_j^{/i}}J_{kj}^2  &\approx \frac{a^4 c (p_2-p)}{p} ,  \nonumber \\
			\sum_{j\in\partial_i} J_{ji}^2 \sum_{k\in\partial_j^{/i}} J_{jk} J_{kj}   &\approx  \frac{a^4 c (p_2-p) (1-2\epsilon)}{p} .\label{sum2}
		\end{align}
        
		We note that we have explicitly used the distribution in Eq.~(\ref{piconstantmag}) in evaluating these sums. Inserting these expressions into Eq.~(\ref{iprasymgeneral}), we obtain the IPR of a hub with given values of  $\Delta$ and $c$, 
		\begin{align}
			q^{-1}(\Delta, c) \approx \left( \frac{\Delta}{\Delta + c}\right)^2 \left[ 1  - \frac{4\Delta}{(\Delta + c)} \frac{ c (p_2-p) (1-\epsilon)}{p\Delta^2} + \frac{c}{\Delta^2}\right]. \label{confignetworkiprbeforeav}
		\end{align}    
		We now wish to perform the average over $c$ for fixed $\Delta$, keeping only terms that contribute to $\mathcal{O}(1/\Delta)$. We note that fixing $\Delta$ is equivalent to fixing $\omega$ when the edge-weight distribution obeys Eq.~(\ref{piconstantmag}). This is because the eigenvalue associated with a hub is given by $\omega^2 \approx a^2 [\Delta - (p_2-p)/p(1-2\epsilon)]$   [see Sec.~\ref{appendix:selfaveraging}]. That is, we wish to evaluate
		\begin{align}
			\overline{q^{-1}} = \frac{\sum_{c} q^{-1}(\Delta, c) P(c, \Delta)}{\sum_c P(c, \Delta)},\label{eq:sumsums}
		\end{align}
		where $q^{-1}(\Delta, c)$ takes the form of Eq.~(\ref{confignetworkiprbeforeav}). 

        To evaluate the right-hand side of (\ref{eq:sumsums}), we approximate the  sums in the numerator and the denominator  as integrals using the Euler-Maclaurin formula, just as in  Eq.~(\ref{approximationspdelta}). The resulting integrals can be evaluated as a series in $1/\Delta$ for large $\Delta$. Just as we saw in Eq.~(\ref{approximationspdelta}), all of the terms apart from the Gaussian integral do not contribute to leading order.  This procedure yields the formula
		\begin{align}
			\overline{q^{-1}} = \frac{ P(c_0', \Delta)}{P(c_0, \Delta)}  \sqrt{ \frac{(\sigma'_c)^2}{(\sigma_c)^2}}\left( \frac{\Delta}{\Delta + c_0'}\right)^2 \left[ 1  - \frac{4\Delta}{(\Delta + c_0')} \frac{ c_0' (p_2-p) (1-\epsilon)}{p\Delta^2} + \frac{c_0'}{\Delta^2}\right].  \label{iprafterlapaceapprox}
		\end{align}
		Here  
      \begin{equation}  
        c_0' =  {c'}_0^{(0)} \Delta + {c'}_0^{(1)},
        \end{equation}
        where
		\begin{align}
			{c'}_0^{(0)} &= c_0^{(0)} ,
		\end{align}
        the same coefficient that appears in Eq.~(\ref{eq:ansatz}), and takes the forms (\ref{c0geomgeom}) and (\ref{eq:c00pow}) for  geometric and power-law  degree distributions, respectively.  For geometric degree distributions,     
   	\begin{align}
			{c'}_0^{(1)} &= \frac{1}{[\alpha^2 - 4 \epsilon (1-\epsilon) ]^{3/2}} \left(2 \alpha^3 - 8 \alpha\epsilon (1-\epsilon)+ [2 (1-\epsilon)\epsilon - 3 \alpha^2 ]\sqrt{\alpha^2- 4 \epsilon (1-\epsilon)} \right),\label{c1dashgeom}
		\end{align}
        and for power-law degree distributions, 
        \begin{align}
			{c'}_0^{(1)} &= \frac{-3+2\vert1-2\epsilon\vert+2\epsilon (1-\epsilon) (1-2 \gamma)}{(1-2\epsilon)^2} .\label{c1dashpow}
		\end{align}
        Furthermore, in Eq.~(\ref{iprafterlapaceapprox}) we used the symbols 
		\begin{align}
			(\sigma'_c)^2 &= -\frac{1}{c_0'} + \frac{1}{2 (c_0'+\Delta)}+ \frac{1}{2 (c_0'-\Delta)} + \frac{2}{(\Delta+ c_0')^2} \nonumber \\ &- \frac{1}{2}\left[ \frac{1}{(c_0')^2} - \frac{1}{(c_0' + \Delta)^2}- \frac{1}{(c_0' - \Delta)^2}\right]
            		\end{align}
                    and 
		\begin{align}
			(\sigma_c)^2 &= -\frac{1}{c_0} + \frac{1}{2 (c_0+\Delta)}+ \frac{1}{2 (c_0-\Delta)} + \frac{2}{(\Delta+ c_0)^2} \nonumber \\ &- \frac{1}{2}\left[ \frac{1}{(c_0)^2} - \frac{1}{(c_0 + \Delta)^2}- \frac{1}{(c_0 - \Delta)^2}\right]. 
		\end{align}

	Expanding Eq.~(\ref{iprafterlapaceapprox}) up to next to leading order in $1/\Delta$, we obtain  for geometric degree distributions the asymptotic form 
		\begin{align}
			\overline{q^{-1}} &= \left( \frac{1}{1 + c_0^{(0)}}\right)^2 \Bigg[ 1  - \frac{1}{\Delta} \Bigg(\frac{4}{(1 + c_0^{(0)})} \frac{ c^{(0)}_0 (p_2-p) (1-\epsilon)}{p} - \frac{2(c_0^{(0)})^2}{(1+c_0^{(0)})} +  \frac{2{c'}_0^{(1)}}{(1+c_0^{(0)})} \nonumber \\
			&- \frac{{c'}_0^{(1)}- {c}_0^{(1)}}{c_0^{(0)}} + \frac{\left({c'}_0^{(1)}- {c}_0^{(1)}\right)\left({c'}_0^{(1)}+ {c}_0^{(1)}\right)}{2 c^{(0)}_0 \left(1 + c^{(0)}_0 \right) \left( c^{(0)}_0 -1\right)}\Bigg)\Bigg], \label{eq:geomFinal}
		\end{align}
		and in the power-law case we get
		\begin{align}
			\overline{q^{-1}} &= \left( \frac{1}{1 + c_0^{(0)}}\right)^2 \Bigg[ 1  - \frac{1}{\Delta} \Bigg(\frac{4}{(1 + c_0^{(0)})} \frac{ c^{(0)}_0 (p_2-p) (1-\epsilon)}{p} - \frac{2(c_0^{(0)})^2}{(1+c_0^{(0)})} +  \frac{2{c'}_0^{(1)}}{(1+c_0^{(0)})} \nonumber \\
			&+ \frac{(\gamma-1)[{c'}_0^{(1)}- {c}_0^{(1)}]}{c_0^{(0)}} + \frac{\left({c'}_0^{(1)}- {c}_0^{(1)}\right)\left({c'}_0^{(1)}+ {c}_0^{(1)}\right)}{2 c^{(0)}_0 \left(1 + c^{(0)}_0 \right) \left( c^{(0)}_0 -1\right)}\Bigg)\Bigg], \label{eq:powFinal}
		\end{align}
        which are of the same form as the Eqs.~(\ref{iprgeo}) and (\ref{iprpow})  that we were meant to derive.   Note that comparing Eqs.~(\ref{eq:geomFinal}) and (\ref{eq:powFinal}) with    Eqs.~(\ref{iprgeo}) and (\ref{iprpow}), respectively, we find that 
\begin{align}
			A_\mathrm{geo}(\epsilon, p) &= \frac{4}{(1 + c_0^{(0)})} \frac{ c^{(0)}_0 (p_2-p) (1-\epsilon)}{p} - \frac{2(c_0^{(0)})^2}{(1+c_0^{(0)})} +  \frac{2{c'}_0^{(1)}}{(1+c_0^{(0)})} \nonumber \\
			&- \frac{{c'}_0^{(1)}- {c}_0^{(1)}}{c_0^{(0)}} + \frac{\left({c'}_0^{(1)}- {c}_0^{(1)}\right)\left({c'}_0^{(1)}+ {c}_0^{(1)}\right)}{2 c^{(0)}_0 \left(1 + c^{(0)}_0 \right) \left( c^{(0)}_0 -1\right)}\label{eq:AGeo}
		\end{align}
        and
     \begin{align}
			A_\mathrm{pow}(\epsilon, \gamma) &= \frac{4}{(1 + c_0^{(0)})} \frac{ c^{(0)}_0 (p_2-p) (1-\epsilon)}{p} - \frac{2(c_0^{(0)})^2}{(1+c_0^{(0)})} +  \frac{2{c'}_0^{(1)}}{(1+c_0^{(0)})} \nonumber \\
			&+ \frac{(\gamma-1)[{c'}_0^{(1)}- {c}_0^{(1)}]}{c_0^{(0)}} + \frac{\left({c'}_0^{(1)}- {c}_0^{(1)}\right)\left({c'}_0^{(1)}+ {c}_0^{(1)}\right)}{2 c^{(0)}_0 \left(1 + c^{(0)}_0 \right) \left( c^{(0)}_0 -1\right)}.\label{eq:APow}
		\end{align}
Recall that in  these formulae, for geometric degree distributions $c^{(0)}_0$ is given by (\ref{c0geomgeom}), $c^{(1)}_0$ by (\ref{c0geom}), and ${c'}_0^{(1)}$ by (\ref{c1dashgeom}).   Analogously, for power-law degree distributions we use the formulae (\ref{eq:c00pow}), (\ref{c0pow}), and (\ref{c1dashpow}).

	Note that in finding the asymptotic  $\omega$ dependence of the IPR, it was crucial to take into account the effect of the wider network, and not just the hub itself.

		\subsection{Random graphs with  power-law degree distribution and equal probability of reinforcing and antagonistic links(\texorpdfstring{$\epsilon=1/2$}{epsilon=0.5})}\label{appendix:crossover}
      
		As we highlight in Eq.~(\ref{iprpow}) of the main text, the eigenvectors are extended when $\epsilon = 1/2$ (i.e. $\overline{q^{-1}} \to 0$) for $\omega \to \infty$. This indicates that  the `single-defect' approximation breaks down in this limit -- one can no longer consider the network hubs with the same arrangements of links as effectively isolated from one another.
        Indeed,  for $\epsilon = 1/2$,  the ansatz solution (\ref{eq:ansatz}), $c_0 = \Delta c_0^{(0)} + c_0^{(1)} + \dots$, breaks down. Let us instead use the more general ansatz $c_0 = \Delta^\beta c_0^{(0)} + \cdots$ in (\ref{eq:c0Max}), from which we find that in this case $\beta = 2$ and $c_0^{(0)} = 1/(1+2 \gamma)$ for large $\Delta$.   Indeed, substituting $c=\Delta^2 c^{(0)}_0$ in Eq.~(\ref{confignetworkiprbeforeav}), we see that the factor $(\Delta/(c+\Delta))^2$ vanishes as $\Delta\to \infty$. However, we note that the subleading term is of the same order in $\Delta$ as the leading order term, and hence the asymptotic  expansion in $1/\Delta$ breaks down. This is to be expected since the single defect approximation relies on non-overlapping eigenvectors.

        	\section{Asymptotic analysis for Erd\H{o}s-R\'{e}nyi  graphs with edge weights of fixed absolute value (corresponding with Sec.~\ref{section:example:constantweight})} \label{appendix:poissonian}

		We now discuss the case of   Erd\H{o}s-R\'{e}nyi  graphs with edge weights of fixed absolute values.   In Sec.~\ref{app:asymptERSpec} we derive the expression Eq.~(\ref{rhoer})   for the  asymptotic eigenvalue density, and in Sec.~\ref{app:asymptERIPR} we derive the   Eq.~(\ref{iprer}) for the disordered averaged IPR.

        \subsection{Asymptotic analysis for the eigenvalue density}\label{app:asymptERSpec}
		The  Erd\H{o}s-R\'{e}nyi graph must be handled slightly differently to the power-law and geometric degree distributions considered in the previous appendix. This is due to the fact that the degree distribution $P_\mathrm{deg} \approx e^{-p}p^c/c!$ is much more quickly decaying than the aforementioned cases. This means that, typically, the most likely value of $c$ for a hub with a given value of $\Delta$ to have is $c \approx \Delta$, i.e. with comparatively few antagonistic links.

        For this reason, Stirling's approximation for Eq.~(\ref{eq:PDeltaC}), as we applied it in (\ref{lnpdeltac}), does not work, as we show next.  We use in what follows a different parameterisation of the sums in comparison to the geometric and power-law cases of Sec.~\ref{app:geomPower}, in anticipation of the fact that the typical connectivity $c$ will be very close to $\Delta$. We denote $m = \sum_{j \in \partial_i} \Theta(-J_{ij}J_{ji})$ as the number of antagonistic links belonging to the hub at node $i$. Let us start by identifying the typical value of $m$. In this case, one has (assuming large $m$ and $\Delta$ and using Stirling's approximation)
		\begin{align}
			\ln P(\Delta, m) \approx (2m + \Delta) \ln p - m \ln (m/\epsilon) - (m + \Delta )\ln [(m + \Delta )/(1-\epsilon)] .
		\end{align}
		For fixed $\Delta$, one finds the peak value $m_\mathrm{max} \approx  p^2 \epsilon (1-\epsilon) /\Delta$, which contradicts our assumption of $m$ being large, confirming that this approach breaks down.

    If we instead approximate $m$ as being much smaller than $\Delta$, then we can use
		\begin{align}
			(\Delta+m)! \sim \sqrt{2\pi \Delta} \left(\frac{\Delta}{e}\right)^\Delta \Delta^m,
		\end{align}
		and we obtain that the approximate marginal distribution for $m$ for large values of $\Delta$ is the following Poisson distribution
		\begin{align}
			P(m\vert \Delta) \approx \frac{1}{m!}\left[\frac{ p^2\epsilon (1-\epsilon)}{ \Delta} \right]^m \exp\left[ - \frac{ p^2 \epsilon (1-\epsilon)}{ \Delta} \right].  \label{mgivenk}
		\end{align}
		Extract the conditional mean from the Poisson distribution we find 
		\begin{align}
			\mathrm{E}(m\vert \Delta) \approx \frac{p^2 \epsilon (1-\epsilon)}{ \Delta} ,\label{condmer}
		\end{align}
		confirming that on an Erd\H{o}s-R\'{e}nyi graph the typical number of antagonistic links is  small for hubs of a given large value of $\Delta$.
		
		We are now in a position to find the asymptotic eigenvalue density. We begin with Eq.~(\ref{configdensitysum}) and note that $p_2 = p^2 +p$ for ER graph. We thus obtain
		\begin{align}
			\rho(\omega) &\approx \frac{1}{2}\sum_{l= 0}^\infty\sum_{r= 0}^\infty \frac{e^{-p} p^{r+l}}{r!l!} \left(1-\epsilon\right)^{r} \epsilon^{l} \delta\left[\omega-\sqrt{a^2[\Delta + p  (1-2\epsilon) ] }\right] \nonumber \\
			&=\frac{1}{2}\sum_{\Delta= 0}^\infty\sum_{m= \mathrm{max}(0, -\Delta)}^\infty \frac{e^{-p} p^{2m +\Delta}}{m!(m+\Delta)!} \left(1-\epsilon\right)^{m+\Delta} \epsilon^{m} \delta\left[\omega-\sqrt{a^2[\Delta + p  (1-2\epsilon) ] }\right], \label{ersum}
		\end{align}
		where from the first to the second lines we have made the substitution $\Delta = r - l$ and $m = l$. We see that once again large $\omega$ corresponds to large $\Delta$ (i.e. many more reinforcing links than antagonistic links). 
		
		We can thus now carry out the sum over $m$ in Eq.~(\ref{ersum}) using Eq.~(\ref{mgivenk}) to obtain
		\begin{align}
			\rho(\omega) &\approx \frac{1}{2}\sum_{\Delta= 0}^\infty \frac{1 }{\sqrt{2 \pi \Delta}} \left(\frac{ep(1-\epsilon)}{\Delta}\right)^{\Delta} \exp\left[-p\left( 1- \frac{p\epsilon(1-\epsilon)}{\Delta}\right) \right] \delta\left[\omega-\sqrt{a^2[\Delta + p  (1-2\epsilon) ] }\right].
		\end{align}
		Making the same coarse-graining as for the cases in Section \ref{appendix:eigenvaluedensityconfig} by writing $\sum_k \to \omega^2 \int \dd\kappa$ with $\Delta = \omega^2 \kappa$, and carrying out the integral over $\kappa$, we finally obtain the expression in Eq.~(\ref{rhoer}) of the main text. One notes that we have neglected terms that contribute at the order $O(1/\omega^2)$ to $\ln\rho(\omega)$ as in Section~\ref{appendix:eigenvaluedensityconfig}. 

        \subsection{Asymptotic analysis for the disordered averaged IPR}\label{app:asymptERIPR}
        We derive the asymptotic expression (\ref{iprer}) for the disordered average IPR  of the Erd\H{o}s-R\'{e}nyi ensemble  with edge weights of fixed absolute value.    Note that Eq.~(\ref{iprer})  is more accurate than the expressions for geometric and power-law degree distributions as it considers terms up to $\mathcal{O}(\omega^{-6})$.    This implies that we need to consider one order of accuracy higher than the calculation in Section~\ref{appendix:hermexpansion}. 

Analogously to the procedure in Sec.~\ref{appendix:hermexpansion}, we use  that the right eigenvectors associated with tail eigenvalues are  localised on   hubs $i$ of large effective degree $|\Phi_i|$. Therefore, evaluating the right-hand side of Eq.~(\ref{iprfromres}) at a tail eigenvalue $\omega=\lambda_{i}$, we can  truncate the sum as 
        \begin{equation}
        \overline{q^{-1}(\omega)} \approx \lim_{\eta\to 0}\eta^2 \left\langle |\mathcal{H}^{22}_{ii}|^2 + \sum_{j\in\partial_i} |\mathcal{H}^{22}_{jj}|^2 \right\rangle, \label{eq:DisorderAVP}
        \end{equation}
        where we use Eq.~(\ref{eq:B5}) to write the IPR in terms of the Hermitised resolvent. We note that the next-nearest neighbours would contribute to the IPR at the order $\mathcal{O}(1/\omega^6)$, and we neglect these terms. One can see this following the arguments  in Section \ref{appendix:exploc} on exponential localisation of the right-eigenvector components associated with the hub $i$; the squared eigenvector component of the next-nearest neighbours goes as $\vert r_k \vert^2 \sim  1/\omega^4$. Since there are roughly $c p$ next-nearest neighbours, and $c\sim \omega^2$ the contribution to the IPR is $\sum_{k\in\partial^{/i}_j }\vert r_k\vert^4 \sim c p /\omega^8 \sim 1/\omega^6$, as claimed.

       \subsubsection{Asymptotic expansion of the Hermitised cavity equations.}
    To determine $\mathcal{H}^{22}_{ii}$ and  $\mathcal{H}^{22}_{jj}$, we first perform an asymptotic expansion in $1/\omega$ of the Hermitised cavity equations [Eqs.~(\ref{cavityhermP}) and (\ref{cavityherm})], similarly to the expansion in  Sec.~\ref{app:HermitisedCav}.   As the final expression  for the IPR has one additional order of accuracy then the one obtained in Sec.~\ref{app:HermitisedCav},  we need to consider one extra layer of nodes (the ``next-next-nearest'' neighbours $\ell\in \partial^{\setminus j}_k$).    Thus,  we get for the Hermitised cavity resolvents  the following  asymptotic expressions  for large and real $\omega$,
		\begin{align}
			&D_{ll}^{(k)} \approx A_{ll}^{(k)}  \approx -\frac{\eta}{\omega^2},\nonumber \\
			&G_{ll}^{(k)} \approx \frac{1}{\omega}, \nonumber \\
			&D_{kk}^{(j)}   \approx -\frac{\eta}{\omega^2}\left(1 + \frac{2}{\omega^2}\sum_{l\in\partial_k^{/j}} J_{lk}J_{kl}  + \frac{1}{\omega^2} \sum_{l\in\partial_k^{/j}} J_{lk}^2 \right), \nonumber \\
			&G_{kk}^{(j)} \approx \frac{1}{\omega}\left(1 +\frac{1}{\omega^2}\sum_{l\in\partial_k^{/j}} J_{kl}  J_{lk} \right), \nonumber \\
			&A_{jj}^{(i)}  \approx -\frac{\eta}{\omega^2}\Bigg(1 + \frac{1}{\omega^2}\left[2\sum_{k\in\partial_j^{/i}} J_{jk}J_{kj}  +  \sum_{k\in\partial_j^{/i}} J_{jk}^2 \right]\nonumber \\
			&+\frac{1}{\omega^4} \Bigg[2 \left(\sum_{k \in \partial_j^{/i}} J_{jk} J_{kj}\right)  \left(\sum_{k \in \partial_j^{/i}} J_{jk}^2 \right) + 3 \left(  \sum_{k \in \partial_j^{/i}} J_{jk} J_{kj}  \right)^2 + \sum_{k \in \partial_j^{/i}} J_{jk}^2  \sum_{l \in \partial_k^{/l}}J_{lk}^2 \nonumber \\
			&+ 2 \sum_{k \in \partial_j^{/i}} J_{jk}^2 \sum_{l \in \partial_k^{/l}}J_{kl}J_{lk} + 2 \sum_{k \in \partial_j^{/i}} J_{jk} J_{kj} \sum_{l \in \partial_k^{/l}}J_{kl} J_{lk} \Bigg]\Bigg), \nonumber \\
			&G_{jj}^{(i)} \approx \frac{1}{\omega} +\frac{1}{\omega^3}\sum_{k\in\partial_j^{/i}} J_{jk}  J_{kj} \nonumber \\
			& + \frac{1}{\omega^5}\left[\left(\sum_{k\in\partial_j^{/i}} J_{jk}  J_{kj} \right)^2 + \sum_{k\in\partial_j^{/i}} J_{jk}  J_{kj} \sum_{l \in \partial_k^{/j}} J_{kl}  J_{lk} \right], \label{truncatedhermcavres2}
		\end{align}
        where $i$ is the hub, $j\in \partial_i$, $k\in \partial^{/i}_j$, and $\ell\in \partial^{/j}_k$.  Just as in Sec.~\ref{app:HermitisedCav},  approximation signs in the above equations signify that higher order terms in $1/\omega$ and $\eta$ have been neglected.

		Using the fact that the pair of weights $(J_{ij}, J_{ji})$  are drawn from the joint distribution in Eq.~(\ref{piconstantmag}),  so that they have constant magnitudes, we can write the above quantities in terms of the degrees  
		\begin{eqnarray}
			 c_j^{(i)} = \sum_{k \in \partial_j^{/i}} 1,  \quad {\rm and} \quad 
			c_i = \sum_{j \in \partial_i}  1,     
		\end{eqnarray}
        and the quantities 
        		\begin{eqnarray}
        \Delta_i = \sum_{j \in \partial_i}  \mathrm{sign}\left(J_{ij}J_{ji}\right), \quad  {\rm and}    \quad  \Delta_j^{(i)} &= \sum_{k \in \partial_j^{/i}} \mathrm{sign}\left(J_{jk}J_{kj}\right),
        		\end{eqnarray}
		so that we obtain the simpler formulae 
		\begin{align}
			&A_{jj}^{(i)}  \approx -\frac{\eta}{\omega^2}\Bigg(1 + \frac{a^2}{\omega^2}\left[2\Delta_j^{(i)}  +  c_j^{(i)} \right] \nonumber \\
			&+\frac{a^4}{\omega^4} \Bigg[2 \Delta_j^{(i)}c_j^{(i)} + 3 \left(  \Delta_j^{(i)} \right)^2 + \sum_{k \in \partial_j^{/i}} \left[ c_k^{(j)} + 2 \Delta_k^{(j)} + 2 \frac{J_{jk}J_{kj}}{a^2} \Delta_{k}^{(j)} \right] \Bigg]\Bigg),
            \end{align}
            and 
            \begin{align}
			&G_{jj}^{(i)} \approx \frac{1}{\omega} +\frac{a^2}{\omega^3}\Delta_j^{(i)}  + \frac{a^4}{\omega^5}\left[\left(\Delta_j^{(i)} \right)^2 + \sum_{k\in\partial_j^{/i}} \frac{J_{jk}  J_{kj}}{a^2} \Delta_k^{(j)} \right] .
		\end{align}

      \subsubsection{Evaluating the Hermitised resolvent at the tail eigenvalues.}
		We wish to calculate the Hermitised resolvent elements $D_{ii} = \mathcal{H}^{22}_{ii}(\omega)$ and $D_{jj} = \mathcal{H}^{22}_{jj}(\omega)$ with $j\in\partial_i$ at the point  $\omega=\lambda_i$ where $G_{ii}$ has a pole (corresponding to the location of an eigenvalue).

        Since we assume that the connectivity $c_i$ of the hub $i$ is large, we can replace the sums over $j$ by their averages [we have checked that the fluctuations contribute only to $\mathcal{O}(1/\omega^6)$ in the final result].  Furthermore, using the law of large numbers we find the following useful equalities, which are valid for the distribution in Eq.~(\ref{piconstantmag}) and the Erd\H{o}s-R\'{e}nyi graph (i.e. when the existence of each link is an independent binary event, and whether or not a link is reinforcing or antagonistic is also a binary event)  
		\begin{align}
			\frac{1}{c_i}\sum_{j \in \partial_i}\left\langle \Delta_{j}^{(i)}\right\rangle &= p (1- 2\epsilon), \nonumber \\
			\frac{1}{c_i}\sum_{j \in \partial_i}\left\langle c_{j}^{(i)}\right\rangle &= p , \nonumber \\
			\frac{1}{c_i}\sum_{j \in \partial_i} \left\langle \Delta_j^{(i)}c_j^{(i)}\right\rangle &= (1 -2\epsilon) p_2 , \nonumber \\
			\frac{1}{c_i}\sum_{j \in \partial_i}\left\langle (\Delta_j^{(i)})^2\right\rangle &= p_2 (1-2\epsilon)^2 + 4 \epsilon (1-\epsilon)p , \nonumber \\
			\frac{1}{c_i}\sum_{j \in \partial_i}\left\langle \sum_{k \in \partial_j^{/i}} c_k^{(j)}\right\rangle &= p^2, \nonumber \\
			\frac{1}{c_i}\sum_{j \in \partial_i}\left\langle \sum_{k \in \partial_j^{/i}} \Delta_k^{(j)}\right\rangle &= p^2 (1-2\epsilon) , \nonumber \\	
			\frac{1}{c_i}\sum_{j \in \partial_i}\left\langle \sum_{k \in \partial_j^{/i}} J_{jk}J_{kj}\Delta_k^{(j)}\right\rangle &= p^2 (1-2\epsilon)^2 .
		\end{align}
     Recall hat $p_2 = p^2 + p$ for Poisson degree distributions.

          Using the above formula in Eq.~(\ref{implicitomi}), we find that the  pole of $G_{ii}$  solves 
		\begin{align}
			\frac{\lambda^2_i}{a^2} \approx& \Delta_i + \frac{a^2\Delta_i  p (1-2\epsilon) }{\lambda^2_i} + \frac{a^4\Delta_i\left[p^2 (1-2\epsilon)^2  + p_2 (1-2\epsilon)^2 + 4 p \epsilon (1-\epsilon) \right]}{\lambda^4_i}. \label{omer2}
		\end{align}
		Rearranging this, we obtain the following relationship between the pole location $\lambda_i$ and $\Delta_i$ 
		\begin{align}
			\Delta_i \approx \frac{\lambda^2_{i}}{a^2} - p(1-2\epsilon) + \frac{a^2[4 \epsilon(1-\epsilon)p^2 - p(1+p)]}{\lambda^2_{i}}, \label{delter}
		\end{align}
        where we have neglected $O(\lambda^{-4}_i)$ contributions.

		At the pole, we have that [see the discussion around Eq.~(\ref{dii})]
		\begin{align}
			&D_{ii} = \left( \eta - \sum_{j \in \partial_i} A_{jj}^{(i)}\right)^{-1} 
		\end{align}
        and [see Eq.~(\ref{messagepassingasym1})]
        \begin{equation}
        D_{jj} = J_{ij} J_{ji} \left( G_{jj}^{(i)}\right)^2 D_{ii}.  
        \end{equation}
		Again exploiting the large connectivity of the pole, we obtain  (neglecting higher order terms in $1/\omega$)
		\begin{align}
			&D_{ii}  \approx \frac{1}{\eta} \Bigg\{1+\frac{c_i}{\omega^2}\Bigg(1 + \frac{a^2 p}{\omega^2}\left[2(1-2\epsilon)  +  1 \right] \nonumber \\
			&+\frac{a^4}{\omega^4} \Bigg[2 p_2(1-2\epsilon) + 3 \left[p_2 (1-2\epsilon)^2 + 4 p \epsilon (1-\epsilon)  \right] + p^2\left[1 + 2 (1-2\epsilon) + 2 (1-2\epsilon)^2 \right]\Bigg]\Bigg)\Bigg\}^{-1}. \label{diier2}
		\end{align}
		A major simplifying observation is that $c_i = \Delta_i + 2 m_i$ is very close to $\Delta_i$ in the case of the  Erd\H{o}s-R\'{e}nyi  graph (i.e. there are typically few antagonistic links $m_i$). This can be seen from Eq.~(\ref{condmer}). This means that we can write (again neglecting higher order terms in $1/\omega$)
		\begin{align}
			&D_{ii}  \approx \frac{1}{\eta} \Bigg\{1+\frac{a^2\Delta_i}{\omega^2}\Bigg(1 + \frac{a^2 p}{\omega^2}\left[2(1-2\epsilon)  +  1 \right] \nonumber \\
			&+\frac{a^4}{\omega^4} \Bigg[2  p^2 \epsilon(1-\epsilon)+2 p_2(1-2\epsilon)\nonumber \\
			&+ 3 \left[p_2 (1-2\epsilon)^2 + 4 p \epsilon (1-\epsilon)  \right] + p^2\left[1 + 2 (1-2\epsilon) + 2 (1-2\epsilon)^2 \right]\Bigg]\Bigg)\Bigg\}^{-1}
            		\end{align}
                    and 
\begin{align}
			\sum_{j \in \partial_i} D_{jj}^2 &=  D_{ii}^2\sum_{j \in \partial_i} \left(J_{ij}J_{ji}\right)^2 \left[G_{jj}^{(i)}\right]^4 = D_{ii}^2 c_i \frac{a^4}{\omega^4}\left[1 + \frac{4a^2}{\omega^2}p(1-2\epsilon)\right] \nonumber \\
			&\approx D_{ii}^2 \Delta_i \frac{a^4}{\omega^4}\left[1 + \frac{4a^2}{\omega^2}p(1-2\epsilon)\right] .
		\end{align}

		All that remains is to insert the expression for $\Delta_i$ in Eq.~(\ref{delter}) into the above equations, and insert these expressions into Eq.~(\ref{eq:DisorderAVP}). After expanding in $1/\lambda^2_{i}$, this then yields Eq.~(\ref{iprer}).

		\section{Saddle point computation of the disorder-averaged IPR in Eqs.(\ref{iprpsiphi}) and (\ref{saddlepsi})}\label{appendix:saddleipr}
        We now derive the Eqs.(\ref{iprpsiphi}) and (\ref{saddlepsi}) from Eq.~(\ref{generaliprintegral}). Using (\ref{densitygeneral}) in (\ref{generaliprintegral}), we can rewrite the latter equation as 
		\begin{align}
			\overline{q^{-1}(\omega)} \approx \frac{ \int \dd \Psi \int \dd\Phi \sum_c \delta(\omega^2 - \Phi ) \left(\frac{\Phi}{\Phi + \Psi} \right)^2P(\Psi, \Phi \vert c)P_\mathrm{deg}(c) }{  \int \dd\Phi \sum_c \delta(\omega^2 - \Phi )P( \Phi \vert c)P_\mathrm{deg}(c) }, \label{iprsaddlegeneral}
		\end{align}
		where using the exponential representation of the Dirac distribution we have defined
		\begin{align}
			P(\Phi, \Psi \vert c ) &=  \int \prod_{j}(\dd J_{ij} \dd J_{ji}) \delta\left(\Phi - \sum_{j= 1}^c J_{ij}J_{ji}\right)\delta\left(\Psi - \sum_{j= 1}^c J_{ji}^2\right) \prod_{j= 1}^c \pi(J_{ij}, J_{j,i}) \nonumber \\
			&= \int \frac{\dd \hat\Phi}{2\pi} \int \frac{\dd \hat\Psi}{2\pi}  \exp\left({\rm i} \hat \Phi \Phi + {\rm i} \hat \Psi \Psi + c \ln \left\langle e^{-{\rm i}\hat\Phi uv -{\rm i}\hat\Psi v^2}\right\rangle_{\pi(u,v)} \right) , \label{ppsiphigivenc}
		\end{align}
        and where $P(\Phi|c)$ is defined in Eq.~(\ref{pdeltagivenk}).
		Using that both $\Phi$ and $\Psi$ are large in the tails of the spectrum, we perform a saddle-point approximation of the integrals over $\hat\Psi$ and $\hat \Phi$ and obtain
		\begin{align}
			\ln P(\Phi, \Psi\vert c) \sim -\Phi\xi - \Psi\chi+ c \beta ,\label{logpphipsi}
		\end{align}
		where we have  defined 
		\begin{align}
			\beta = \ln \left\langle e^{uv\xi + v^2 \chi } \right\rangle_{\pi(u,v)}, 
		\end{align}
		and where the functions $\xi$ and $\chi$ are the solutions of the equations
		\begin{equation}
			\begin{split}
				\frac{\left\langle uv e^{uv\xi + v^2 \chi } \right\rangle_{\pi(u,v)}}{\left\langle e^{uv\xi + v^2 \chi } \right\rangle_{\pi(u,v)}} &= \frac{\Phi}{c} \, ,  \\
				\frac{\left\langle v^2 e^{uv\xi + v^2 \chi } \right\rangle_{\pi(u,v)}}{\left\langle e^{uv\xi + v^2 \chi } \right\rangle_{\pi(u,v)}} &= \frac{\Psi}{c} .
			\end{split}
		\end{equation}
        Using again that $\Phi$ and $\Psi$ are large, we now approximate the integral over $\Psi$ in Eq.~(\ref{iprsaddlegeneral}) using the saddle-point method.   Differentiating the exponent with respect to $\Psi$, we obtain simply that
		\begin{align}
			\frac{\partial \ln P(\Phi, \Psi\vert c)}{\partial \Psi} = \chi = 0.
		\end{align}
        Using that at the saddle point $\chi=0$, it follows from Eqs.~(\ref{logpphipsi}) and (\ref{lnpdgivek}) that at the saddle point $\ln P(\Phi,\Psi|c) = \ln P(\Phi|c)$.  
		We therefore find that the exponential factors in the saddle point integrals in the numerator and the denominator of Eq.~(\ref{iprsaddlegeneral})  cancel, and we arrive at the expressions  Eqs.~(\ref{iprpsiphi}) and (\ref{saddlepsi}).
		
		\section{Derivation of the asymptotic expressions for \texorpdfstring{$\rho(\omega)$}{rho(omega)} and \texorpdfstring{$\overline{q^{-1}(\omega)}$}{IPR(omega)} in Sec.~\ref{section:saddlepoint}  } \label{appendix:expproduct}
We illustrate with a specific example how the saddle-point procedure developed in Sec.~\ref{section:saddlepoint} works.  
 		In particular, taking the distribution of edge weights to be the one defined in Sec.~\ref{section:example:general}, we determine the asymptotic eigenvalue density and IPR for the e canonical  degree distributions, geometric, power-law and Poisson.   

In the present case of exponentially distributed weights, the equation (\ref{logrho}) for the spectral distribution reads 
\begin{align}
			\ln \rho(\omega) \sim -\omega^2 \xi-   f(\omega)\kappa \:  \ln\left( \frac{1+(1-2\epsilon)d\xi}{1 - d^2 \xi^2} \right) + \ln P_\mathrm{deg}\left(f(\omega) \kappa\right), \label{logrho2}
		\end{align} 
where $\xi$ and $\kappa$ solve the equations [Eqs.~(\ref{saddlekappa}) and (\ref{saddlexi2})]
\begin{align}
			\ln\left( \frac{1+(1-2\epsilon)d\xi}{1 - d^2 \xi^2} \right)= \frac{P_\mathrm{deg}'\left(f(\omega) \kappa\right)}{P_\mathrm{deg}\left(f(\omega) \kappa\right)} , \label{saddlekappa2}
		\end{align}
and 
   \begin{align}
			\; d \frac{2d\xi+(1-2\epsilon)\left(1+d^2\xi^2\right)}{\left(1 - d^2 \xi^2\right)(1+(1-2\epsilon)d\xi)}  = \frac{\omega^2}{\kappa f(\omega)}. \label{saddlexi3}
		\end{align}
   The IPR  follows from Eqs.~(\ref{iprpsiphi}) and (\ref{saddlepsi}), which in the present case gives 
		\begin{equation} \label{eq:ipr_example_exp_weights}
			\begin{split}
				&\lim_{\omega\to \infty}\overline{q^{-1}(\omega ) }= \lim_{\omega\to \infty} \left(\frac{\langle uv e^{\xi u v} \rangle_\pi}{\langle uv e^{\xi u v} \rangle_\pi + \langle v^2 e^{\xi u v} \rangle_\pi } \right)^2 \\
				&= \lim_{\omega \to \infty} \left( \frac{2d\xi+(1-2\epsilon)\left(1+d^2\xi^2\right)}{2d\xi\left( 1 +\frac{(1-2\epsilon)}{2} \left(\nu+\frac{1}{\nu}\right)\right) + \left( 1+d^2\xi^2 \right) \left( (1-2\epsilon) +\frac{1}{2} \left(\nu+\frac{1}{\nu}\right)\right)} \right)^2 \, .
			\end{split}
		\end{equation}

        In what follows we solve the   Eqs.~(\ref{saddlekappa2}) and (\ref{saddlexi3}) towards $\xi$, $\kappa$, and $f$, with $\omega$ an arbitrary but large value.
        The functional form of $f(\omega)$ is set so that the function $\kappa$ for large values of $\omega$ is independent of $\omega$.   Thus, from  Eq.~(\ref{saddlexi3}) it follows that if $\xi$ is independent of $\omega$, then $f(\omega)= \omega^2$.  As we show in Secs.~\ref{app:geom5} and \ref{app:pow5}, this is the case for random graphs with geometric and power law degree distributions with $\epsilon\neq 1/2$.   On the other hand, 
        for Poisson degree distributions we find $f(\omega)=\omega$  (see Sec.~\ref{app5:ER}).   For power-law degree distributions with $\epsilon=1/2$ we find that  $f(\omega)=\omega^4$ (see  Sec.~\ref{app:pow5}), but in this limiting case we also find that the single-defect approximation, which relies on localisation of eigenvectors, does not apply. We end this Appendix with a discussion of the behaviour of the distribution of the effective degree $P(\Phi|c)$  conditioned on the hub degree in Sec.~\ref{app:PHiC}.

		\subsection{Geometric degree distribution}\label{app:geom5}
		First we consider the case of the geometric degree distribution with  
		\begin{equation}
			\ln P_\mathrm{deg}(c) \sim -c\ln\left[(p+1)/p\right] \; .  \label{eq:degGeom}
		\end{equation}
		Using the form (\ref{eq:degGeom}) of the degree distribution  in Eq.~(\ref{saddlekappa2}), we get
		\begin{equation}
			\frac{1+(1-2\epsilon)d\xi}{1 - d^2 \xi^2} = \frac{p+1}{p} \; ,
		\end{equation}
		that is solved by
		\begin{equation} \label{eq:xi_saddle_point_exp_weights_geometric_app}
			\xi = -\frac{(1-2\epsilon)p}{2d(p+1)} \pm \sqrt{\left[\frac{(1-2\epsilon)p}{2d(p+1)}\right]^2 + \frac{1}{d^2(p+1)}} \; .
		\end{equation}
		Notice that the two solutions are always one negative and one positive. Since we are interested in the tails on the real axis, we focus on the positive solution $\xi$ (while the negative one is relevant for the tails on the imaginary axis).
Using (\ref{eq:xi_saddle_point_exp_weights_geometric_app}) in (\ref{saddlexi3}), we find that for $f(\omega) = \omega^2$  the quantity $\kappa$ is independent of $\omega$ and given by 
		\begin{align}
			\kappa =  \frac{p+1}{pd} \frac{\left(1 - d^2 \xi^2\right)^2}{2d\xi+(1-2\epsilon)\left(1+d^2\xi^2\right)} \; .
		\end{align}
		Plugging these expressions for $\kappa$, $\xi$,and $f$  into Eq.~\eqref{logrho}, we obtain a simple gaussian decay
		\begin{align}
			\rho(\omega) \sim \omega e^{-\xi \omega^2},
		\end{align}
		Note that  differently from the case of bounded weights discussed in Section \ref{section:example:constantweight}, the scale $\xi$ converges to a finite value as we approach the fully sign-antisymmetric limit, \ie, $\lim_{\epsilon\to1} \xi = 1/d$.
		
		The IPR follows from plugging the expression Eq.~\eqref{eq:xi_saddle_point_exp_weights_geometric_app}  for $\xi$ into Eq.~(\ref{eq:ipr_example_exp_weights}). We examine some limiting cases. First, we take the case $\epsilon=1/2$. In this case, the solution of Eq.~\eqref{eq:xi_saddle_point_exp_weights_geometric_app} is simply $\xi = \left(d\sqrt{p+1}\right)^{-1}$, from which we obtain
		\begin{align}
			\lim_{\omega\to \infty}\overline{q^{-1}(\omega) } = \left[\frac{1}{1+\frac{1}{4}\left(\nu+\frac{1}{\nu}\right)\frac{2+p}{\sqrt{1+p}}}\right]^2 \, . \label{eq:ipr_example_exp_weights_eps0.5_geom}
		\end{align}
		On the other hand, in the case $\epsilon=0$, we have $\xi = \left(d(p+1)\right)^{-1}$. Upon plugging this expression into \eqref{eq:ipr_example_exp_weights}, we obtain
		\begin{align}
			\lim_{\omega\to \infty}\overline{q^{-1}(\omega ) } = \left[\frac{2\nu}{(1+\nu)^2}\right]^2. \label{eq:ipr_example_exp_weights_eps0or1_geom}
		\end{align}
		Finally, in the case $\epsilon=1$ (fully sign-antisymmetric), we get $\xi = d^{-1}$, and we once again obtain the result in Eq.~\eqref{eq:ipr_example_exp_weights_eps0or1_geom}. 
        
        Interestingly, we have found that the scale of the Gaussian decay $\xi$ decreases as we move from $\epsilon=0$ (fully sign-symmetric) to $\epsilon=1$ (fully sign-antisymmetric). In contrast, the inverse participation ratio increases, symmetrically, the further we are from $\epsilon=0.5$, and it decreases as we increase the weights asymmetry $\nu$.

		\subsection{Power-law degree distribution}\label{app:pow5}
		Let us now move on to the case of a power-law degree distribution, where 
		\begin{equation}
			\ln P_\mathrm{deg}(c) \sim -\gamma \ln c \; .
		\end{equation}

        Using this form in  Eq.~(\ref{saddlekappa2}), we find
		\begin{equation} \label{eq:saddle_point_exp_weights_geometric_powerlaw}
			f(\omega) \kappa\ln\left( \frac{1+(1-2\epsilon)d\xi}{1 - d^2 \xi^2} \right)  = \gamma  \; .
		\end{equation}
		Accordingly, the second term in Eq.~\eqref{logrho2} is not having any role for a power-law degree distribution. Moreover, from Eq.~\eqref{eq:saddle_point_exp_weights_geometric_powerlaw} we get
		\begin{equation} \label{eq:saddle_point_exp_weights_geometric_powerlaw_step1}
			\ec^{\frac{\gamma}{f(\omega) \kappa}} (d^2 \xi^2-1) + d(1-2\epsilon) \xi + 1 = 0 \; .
		\end{equation}
		Taking the limit of $\omega\gg 1$ and assuming that  $f(\omega) \sim \omega^a$ with $a\ge1$ so that   $\ec^{\frac{\gamma}{f(\omega) \kappa}} \approx 1$, we find that   Eq.~\eqref{eq:saddle_point_exp_weights_geometric_powerlaw_step1} becomes
		\begin{equation} \label{eq:saddle_point_exp_weights_geometric_powerlaw_step2}
			d \xi \left( d \xi + (1-2\epsilon) \right) = 0 \; ,
		\end{equation}
		which admits the two  solutions
		\begin{equation} \label{eq:xi_exp_weights_geometric_powerlaw_eps<0.5}
			\xi = 0 \; ,
		\end{equation}
		and
		\begin{equation} \label{eq:xi_exp_weights_geometric_powerlaw_eps>0.5}
			\xi = \frac{2\epsilon-1}{d} \; .
		\end{equation}

       If $\epsilon<1/2$, then  only the first solution $\xi=0$ is feasible.    Indeed, plugging the second solution into Eq.~\eqref{saddlexi3}, we get
		\begin{equation}
			\frac{\omega^2}{f(\omega) \kappa} = -d \frac{1-2\epsilon}{4\epsilon(1-\epsilon)} \; , 
		\end{equation}
          which yields a negative value of the  degree parameter $\kappa$ when  $\omega$ is real.    Therefore whenever $\epsilon < 0.5$, we are left with the first solution $\xi=0$. Plugging it into Eq.~\eqref{saddlexi3} we get 
		\begin{equation}
			\frac{\omega^2}{f(\omega) \kappa} = d (1-2\epsilon), 
		\end{equation}
        from which we obtain again the scaling $f(\omega)=\omega^2$.   
		Accordingly the eigenvalue density in Eq.~\eqref{logrho2} displays a power-law behaviour
		\begin{equation} \label{eq:spectral_density_saddle_point_exp_weights_powerlaw_eps>0.5}
			\rho(\omega) \sim 2 \omega^{-2\gamma+1} \; .
		\end{equation} 
		
		On the other hand, if $\epsilon > 0.5$, then the   feasible solution is $\xi = (2\epsilon-1)/d$, and we get from  Eq.~\eqref{saddlexi3} the saddle poin equation   
		\begin{equation}
			\frac{\omega^2}{f(\omega) \kappa} = d \frac{2\epsilon-1}{4\epsilon(1-\epsilon)} \; ,
		\end{equation}
        which yields the same scaling $f(\omega)=\omega^2$ for the degrees of the hubs.    Hence, the spectral density in Eq.~\eqref{logrho2} becomes
		\begin{equation} \label{eq:spectral_density_saddle_point_exp_weights_powerlaw_eps<0.5}
			\rho(\omega) \sim 2 \omega^{-2\gamma+1} \exp\left[-\frac{2\epsilon-1}{d} \omega^2 \right] \; ,
		\end{equation}
		where, next to the power-law contribution, it is present a leading Gaussian decay.
		
		Hence, just as we have found in Sec.~\ref{sec:eigenavlue41} for edge weights of fixed magnitude,  for exponentially distributed   edge weights the  tails on the real axis for power-law graphs behave differently depending on whether $\epsilon < 0.5$ or $\epsilon > 0.5$. More specifically, when the weights are more probably sign-symmetric ($\epsilon<0.5$) the tails are going to zero as a power-law. On the other hand, when the weights are more probably sign-antisymmetric ($\epsilon>0.5$) the tails are exponentially suppressed. The behaviour is mirrored for the tails on the imaginary axis ($\omega^2 < 0$).
		
		Concerning the IPR, we can plug the solutions of Eq.~\eqref{eq:saddle_point_exp_weights_geometric_powerlaw_step1} into Eq.~\eqref{eq:ipr_example_exp_weights}, finding
		\begin{equation} \label{eq:ipr_saddle_point_exp_weights_powerlaw_app}
			\begin{split}
				&\lim_{\omega \to \infty} \ipr(\omega)  = \left[\frac{1}{1+\frac{\left(\nu + \frac{1}{\nu}\right)}{2\lvert 1-2\epsilon \rvert}}\right]^2 \; ,
			\end{split}
		\end{equation}
		which interestingly resembles the asymptotic part of Eq.~\eqref{iprpow}. Similarly to the case discussed in Section~\ref{section:example:constantweight}, the asymptotic IPR reaches its maximum value at $\epsilon=0$ and $\epsilon=1.0$, and it decreases in a mirrored fashion up to $\epsilon=0.5$, where it is null. 
		
		Finally, let us briefly discuss the special case $\epsilon=0.5$. From Eq.~\eqref{eq:saddle_point_exp_weights_geometric_powerlaw_step1} we get
		\begin{equation}
			\xi = \frac{1}{d} \; \sqrt{\frac{\gamma}{f(\omega)\kappa}}  + \mathcal{O}\left(1/f^{3/2}(\omega)\right)\; ,
		\end{equation} 
		and, plugging it into Eq.~\eqref{saddlexi3} we find
		\begin{equation}
			\frac{\omega^2}{f(\omega)\kappa} = 2d \frac{d\xi}{\left(1 - d^2 \xi^2\right)} = 2d \sqrt{\frac{\gamma}{f(\omega)\kappa}}+ \mathcal{O}\left(1/f^{3/2}(\omega)\right) .
		\end{equation}
        Thus, in this case the scaling
        \begin{equation}
        f(\omega)=\omega^4
        \end{equation}
        provides a degree $\kappa$ that is asymptotically independent of $\omega$.   Moreover, since
		\begin{equation} \label{eq:xi_exp_weights_geometric_powerlaw_eps=0.5}
			\xi \omega^2 \approx \sqrt{\frac{\omega^4 \gamma}{\omega^4 \kappa}} = \mathcal{O}(1) \; ,
		\end{equation}
		we don't have any Gaussian decay and, also in this case, the spectral density in Eq.~\eqref{logrho} has a power-law behaviour but with a larger exponent
		\begin{equation} \label{eq:spectral_density_saddle_point_exp_weights_powerlaw_eps=0.5}
			\rho(\omega) \sim 2 \omega^{-4\gamma+1} \; .
		\end{equation}
		However, if $\epsilon=0.5$, the eigenvectors are extended, and thus the single-defect approximation, which is based upon localisation, fails.   Indeed, we can plug Eq.~\eqref{eq:xi_exp_weights_geometric_powerlaw_eps=0.5} into Eq.~\eqref{eq:ipr_example_exp_weights} obtaining
		\begin{equation}
			\begin{split}
				&\lim_{\omega \to \infty} \ipr(\omega) \approx \lim_{\omega \to \infty} \left[\frac{1}{1 + \frac{1}{4}\left(\nu + \frac{1}{\nu}\right)\sqrt{\frac{f(\omega)\kappa}{\gamma}}} \right]^2 \longrightarrow 0 \; .
			\end{split}
		\end{equation}
        Consequently, the assumption that the spectral tails form a pure point spectrum of eigenvalues breaks down in this limit.   
		Notice that this result is also consistent with Eq.~\eqref{eq:ipr_saddle_point_exp_weights_powerlaw_app} in the case $\epsilon=0.5$. 
		
		\subsection{Erd\H{o}s-R\'{e}nyi graph}\label{app5:ER}
		
		Finally, we consider the Erd\H{o}s-R\'{e}nyi graph, for which the degree distribution decays faster than exponentially,
		\begin{equation} \label{eq:example_exp_weights_er}
			\ln P_\mathrm{deg}(c) \sim c - c\ln( c/p) \; .
		\end{equation} 
		Combining Eq.~\eqref{eq:example_exp_weights_er} with Eq.~(\ref{saddlekappa2})  we get
		\begin{equation}
			\frac{1+(1-2\epsilon)d\xi}{1 - d^2 \xi^2} = \frac{c}{p} \; ,
		\end{equation}
		which is solved by 
		\begin{equation} \label{eq:xi_saddle_point_exp_weights_poisson_complete}
			d\xi(c) = -\frac{(1-2\epsilon)p}{2c} \pm \sqrt{\left(1-\frac{p}{2c}\right)^2 + 4\epsilon(1-\epsilon)\left(\frac{p}{2c}\right)^2}  \; .
		\end{equation}
		These two solutions are always one negative and one positive. As before, we are going to focus on the positive one, which is relevant for the tails on the real axis.
		 Moreover, under the assumption $c\gg1$, if $\epsilon\neq 1/2$, we can neglect the terms $4\epsilon(1-\epsilon)\left(p/2c\right)^2$ in the square root obtaining
		\begin{equation}
			\xi = \frac{1}{d} \left(1 - \frac{(1-\epsilon) p}{c} \right) + \mathcal{O}\left(\left(\frac{p}{c}\right)^2\right) \; . \label{eq:xiPoisson1}
		\end{equation}
		If $\epsilon=1/2$, we expand the square root in Eq.~\eqref{eq:xi_saddle_point_exp_weights_poisson_complete}, finding the corresponding result
		\begin{equation}
			\xi  =  \frac{1}{d} \left(1 - \frac{p}{2c}\right) + \mathcal{O}\left(\left(\frac{p}{c}\right)^2\right)  \; .\label{eq:xiPoisson2}
		\end{equation}
		
		Plugging the positive solution into of $\xi$ into Eq.~\eqref{saddlexi3} we get
		\begin{equation}
			\begin{split}
				\frac{\omega^2}{f(\omega) \kappa} = d \left( \frac{f(\omega) \kappa}{(1-\epsilon) p} - 1 \right) + \mathcal{O}\left(\left(\frac{p}{\kappa f(\omega)}\right)^2\right)  \; .  
			\end{split}
		\end{equation}
	Requiring that $\lim_{\omega\rightarrow \infty}\kappa$ is a constant,  we find the linear scaling
		\begin{equation}
			f(\omega)=\omega \; ,
		\end{equation}
		and thus 
		\begin{equation}
			\kappa = \sqrt{\frac{(1-\epsilon) p}{d}} + \mathcal{O}(1/\omega) \; .  
		\end{equation}
		Using that $c=f(\omega)\kappa = \omega \kappa$ in Eqs.~(\ref{eq:xiPoisson1}) and (\ref{eq:xiPoisson2}) we find 
		\begin{equation} \label{eq:xi_saddle_point_exp_weights_poisson}
			\xi(\omega) = \frac{1}{d} - \frac{1}{\omega} \sqrt{\frac{(1-\epsilon) p}{d}} + \mathcal{O}\left(1/\omega^2\right) \; .
		\end{equation}
		
		At the level of the spectral density, plugging these expressions in Eq.~\eqref{logrho} we find that the leading term $\omega \kappa \log\left( \frac{\omega \kappa}{c} \right)$ actually elides and the remaining terms recombine into
		\begin{equation} \label{eq:spectral_density_exp_weights_poisson_app}
			\begin{split} 
				\rho(\omega) \sim \omega \exp\left[ -\left(\frac{\omega}{\sqrt{d}} - \sqrt{(1-\epsilon) p} \; \right)^2 \right] = \omega \exp\left[ -\left(\omega - \sqrt{(1-\epsilon) pd} \; \right)^2 \Big/ d\, \right] \; .
			\end{split}
		\end{equation}
		This is in contrast to the case where the magnitude of the edge weights was fixed, as discussed in Section~\ref{section:example:constantweight}, where we found $\rho \sim (\omega^2)^{-\omega^2}$. We thus see how the edge-weight distribution can drastically affect the asymptotic decay of the eigenvalue density.
		
		Finally, plugging Eq.~\eqref{eq:xi_saddle_point_exp_weights_poisson} into Eq.~\eqref{eq:ipr_example_exp_weights} we get that the disordered-averaged IPR is 
		\begin{equation} \label{eq:IPR_exp_weights_prod_poisson_app}
			\begin{split}
				&\lim_{\omega \to \infty} \ipr(\omega) = \left[ \frac{2\nu}{\left( 1+\nu \right)^2} \right]^2 \; ,
			\end{split}
		\end{equation}
		which interestingly does not depend on the sign pattern (i.e., the $\epsilon$ parameter). 
		   \subsection{The function \texorpdfstring{$P(\Phi|c)$}{P(Phi|c)}}\label{app:PHiC}
        We analyse the behaviour of the function $P(\Phi|c)$, as defined in Eq.~\eqref{lnpdgivek}, as a function of $c$ for the case of weights that are drawn from an exponential distribution as defined in Sec.~\ref{section:example:general}; we use these results in  Sec.~\ref{sec:RhoQ} to rationalise the  behaviour of $\rho(\omega)$.    
        
        Computing the derivative of $\ln P(\Phi|c)$ with respect to the degree $c$, and combining it with the condition in Eq.~\eqref{saddlexi}, we find that its stationary points  are identified by         $\beta = 0$.  In the  present example of an exponential weight distribution, as defined in Sec.~\ref{section:example:general}, we get 
        \begin{equation}
       \beta = \ln \left\langle e^{uv\xi} \right\rangle_{\pi(u,v)}  = \ln  \frac{1+(1-2\epsilon)d\xi}{1 - d^2 \xi^2}, 
        \end{equation}
        and thus using that $\beta=0$,  we get  the equation 
 \begin{equation}
d\xi \left( d\xi + (1-2\epsilon)\right) = 0. \label{eq:xiSol}
\end{equation} 
        Equation (\ref{eq:xiSol})
         admits two simple solutions,  viz., 
         \begin{equation}
         \xi = 0 \quad {\rm and} \quad \xi=(2\epsilon-1)/d. \label{eq:solutions}
         \end{equation}
        Interestingly, only one of the two solutions  in (\ref{eq:solutions}) is feasible, depending on the sign-antisymmetric probability $\epsilon$.   
        
        If $\epsilon<1/2$, then only the first solution, $\xi=0$,  is feasible (as the second solution leads to $\xi<0$).  Plugging $\xi=0$ into  Eq.~(\ref{saddlexi}), we find that  for $\epsilon<1/2$, 
        \begin{equation}
       c^{(0)}_\mathrm{max}= \frac{\Phi}{\left\langle uv \right\rangle_{\pi(u,v)}} =  \frac{\Phi}{d (1-2\epsilon)},  \label{eq:c0Max_phi1} 
        \end{equation}
        where the last equality  uses again the specific distribution $\pi$ as defined in Sec.~\ref{section:example:general}.        
        
         On the other hand, if  $\epsilon>1/2$, then  Eq.~(\ref{eq:c0Max_phi1}) is negative, which is incompatible with $c^{(0)}_{\rm max}>0$.  Thus, for $\epsilon>1/2$ only the second solution of (\ref{eq:solutions}), $\xi=(2\epsilon-1)/d$, is feasible. 
          From    Eq.~\eqref{saddlexi},  we then find 
           \begin{equation}
       c^{(0)}_\mathrm{max}=\Phi  \frac{\langle e^{uv (2\epsilon-1)/d}  \rangle_{\pi(u,v)}}{\left\langle uv \:  e^{uv (2\epsilon-1)/d}    \right\rangle_{\pi(u,v)}} .  \label{eq:c0Max_phi2} 
        \end{equation}
        
        In conclusion,  the function $P(\Phi|c)$ for an exponential weight distribution (see  Sec.~\ref{section:example:general}) has a unique maximum, whose expression qualitatively depends on the sign-antisymmetric probability $\epsilon$ (whether $\epsilon<1/2$ or $\epsilon>1/2$). 

        Note that from Eq.~\eqref{saddlexi2}  we observe that, since $\xi$ is independent of $\omega$, the stationary point is such that the degree scales quadratically, \ie, $c \sim \omega^2$. This explains, as we see below, why the geometric and power-law degree distributions yield $f(\omega) \sim \omega^2$, where as the more quickly varying Erd\H{o}s-R\'{e}nyi graph has a different scaling.
        
		\section{Breakdown of the saddle-point procedure}\label{appendix:breakdowngeneral}
		Let us now explore an example where the method of Section \ref{section:saddlepoint} of the main text breaks down. We saw that in the examples involving the power-law degree distribution, when $\epsilon = 1/2$, the states in the tail region are no longer localised, undermining our single defect assumption. 
		
		There are also circumstances in which the sates may still be localised, but where the procedure that we outline is not valid, due to $P(\Phi \vert c)$ not being a peaked function of $c$. This means that there are circumstances under which the eigenvalues in the tail region are no longer dominated by well-connected hubs, but instead by small collections of heavily weighted edges. We show that in these cases, it can still be possible to compute the asymptotic behaviour of the tails.
		
		Let us examine the case where $P(y) = (\gamma - 1)/[2(1 + \vert y\vert)^{\gamma}]$ [instead of the exponential distribution in Eq.~(\ref{expedgedist})]. In this case, we see that the averages in Eq.~(\ref{saddlexi}) do not converge for any real $\xi$ for $\gamma >1$. Let us then examine the functions $P(\Phi \vert c)$ by evaluating the first line of Eq.~(\ref{pdeltagivenk}) directly. First, one has trivially
		\begin{align}
			P(\Phi \vert 1) = (\gamma-1)/[2(1 +\Phi)^{\gamma}] \sim \Phi^{-\gamma}. 
		\end{align}
		Let us now consider
		\begin{align}
			P(\Phi \vert 2) = \frac{(\gamma-1)^2}{2^2}\int \dd x \frac{1}{(1 + \vert x\vert)^{\gamma}} \frac{1}{(1 + \vert \Phi - x\vert)^{\gamma}} .
		\end{align}
		We see that this integral is dominated by values of $x$ around $x \approx 0$ and $x \approx \Phi$, corresponding to one link being the far more dominant contribution. We therefore also have
		\begin{align}
			P(\Phi\vert 2) \sim \Phi^{-\gamma}. 
		\end{align} 
		This pattern continues for higher $c$, and we see that $P(\Phi \vert c)$ scales the same way with $\Delta$ for all values of $c$. There is thus no trade-off between $P(\Phi\vert c)$ and $P_\mathrm{deg}(c)$, invalidating the approach of Section \ref{section:saddlepoint}. One therefore finds simply from Eq.~(\ref{densitygeneral}) that
		\begin{align}
			\rho(\omega) \sim \frac{1}{\omega^{2\gamma - 1}}. 
		\end{align}
		Given the reasoning above, we therefore expect states to be localised on single edges in this case, rather than well-connected hubs. 
		
		\section{Finite size effects for the disorder-averaged Inverse Participation Ratio (IPR)} \label{appendix:finite_size_effects_ipr}

\begin{figure}[ht]
			\centering
			\begin{subfigure}[b]{0.95\textwidth}
				\centering
				\includegraphics[width=\textwidth]{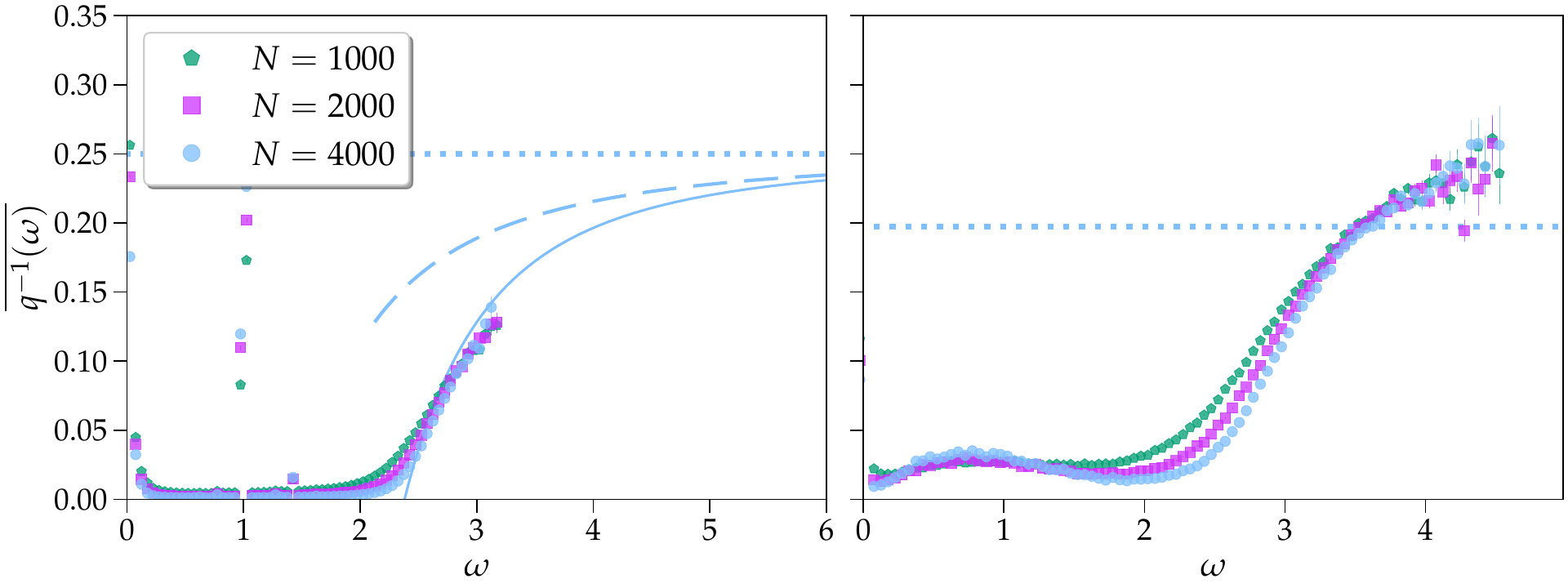}
			\end{subfigure}
			\begin{subfigure}[b]{0.95\textwidth}
				\centering
				\includegraphics[width=\textwidth]{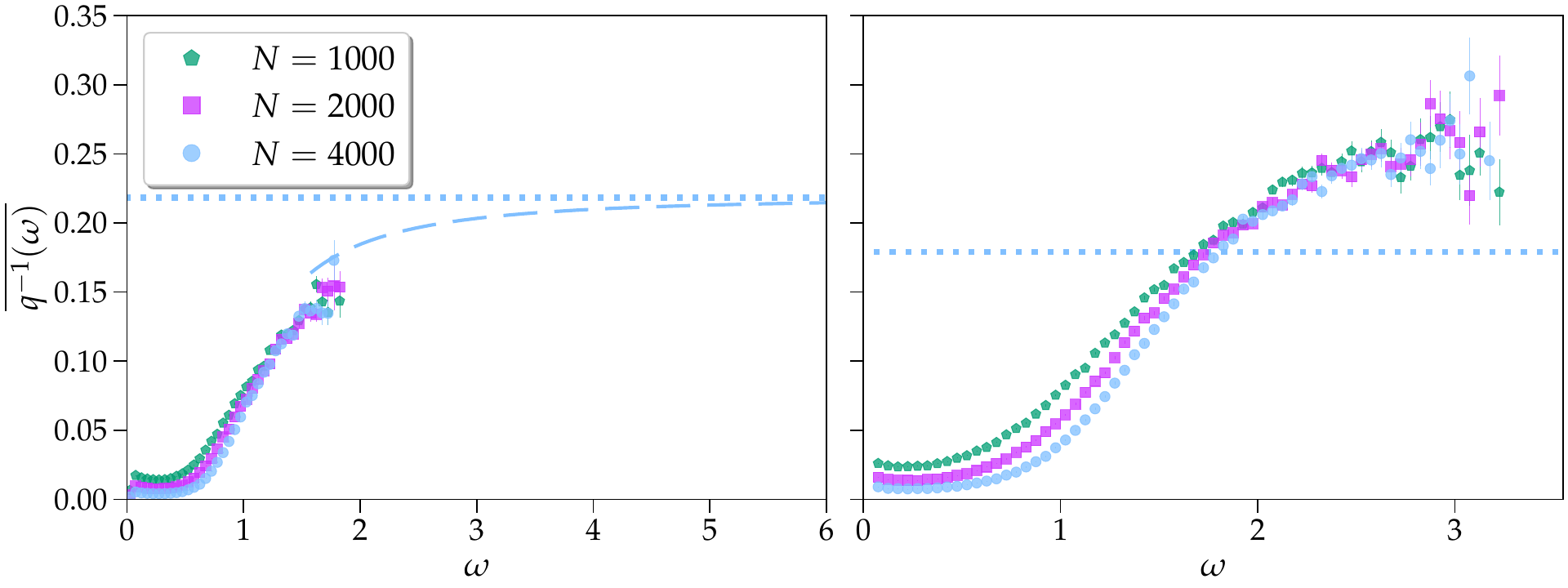}
			\end{subfigure}
    			\begin{subfigure}[b]{0.95\textwidth}
				\centering
				\includegraphics[width=\textwidth]{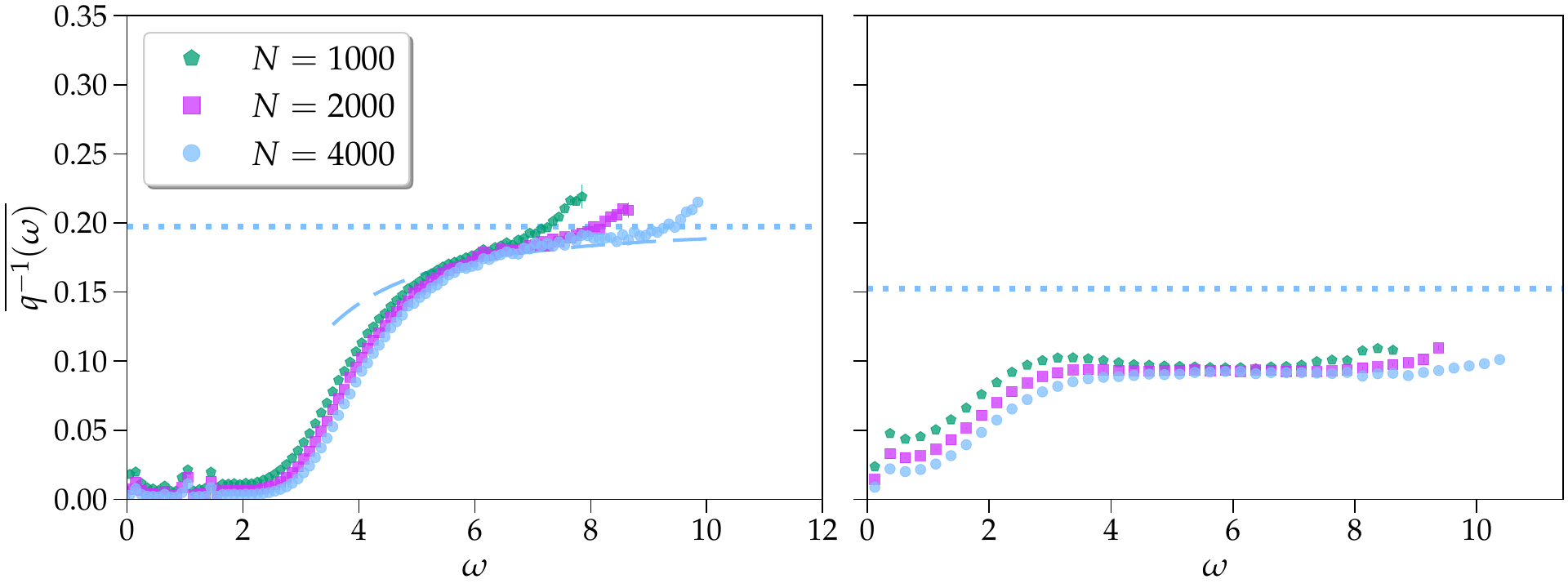}
			\end{subfigure}
    	\caption[]{\textit{Finite size effects for the disorder-averaged IPR.} Finite size effects for the disorder-averaged IPR for Erd\H{o}s-R\'{e}nyi (top panels, $\epsilon=0.60$), geometric (middle panels, $\epsilon=0.90$), and power-law (bottom panels, $\epsilon=0.10$) graphs with $\pm1$ weights (panels on the left) and weights that are exponentially distributed (panels on the right). Different markers and colours are associated to different sizes, as indicated in the legend: $N=1000$ is depicted by green pentagons, $N=2000$ by purple squares and $N=4000$ by light blue circles. The numbers of samples are $40000$, $20000$ and $10000$, respectively, ensuring that the total number of eigenvalues considered is the same across the three sizes. The other parameters are the same as in Fig.~\ref{fig:ipr_grid_fixed_edge} and Fig.~\ref{fig:ipr_grid_expprod_weights}.}
		\label{fig:finite_size_effects_ipr}
		\end{figure}

        In order to provide further evidence of eigenvector localisation, in this Supplementary Section we analyse finite size effects in the disorder-averaged IPR for some of the ensembles presented in Fig.~\ref{fig:ipr_grid_fixed_edge} and Fig.~\ref{fig:ipr_grid_expprod_weights}. Indeed, localised eigenvectors are characterised by an IPR which, in the large $N$ limit, is independent of the system size. This is in contrast to extended or multifractal eigenvectors, which display a dependence on $N$ \cite{evers2008anderson, cugliandolo2024multifractal}. In particular, in Fig.~\ref{fig:finite_size_effects_ipr} we compare the disorder-averaged IPR for three values of the system size $N$, for each combination of the three degree distributions (Erd\H{o}s-R\'{e}nyi, top panels, geometric, middle panels, and power-law, bottom panels) and the two weight distributions discussed in the main text (fixed edge weights, on the left, and unbounded continuous edge weights, on the right). For all the ensembles considered, it can be observed that the IPR in the tails remains finite and exhibits little dependence on the system size. These results are compatible with the theory presented in this paper and the explicit verification of exponential localisation in Section~\ref{appendix:exploc}, corroborating that the eigenvectors associated with eigenvalues in the tail regions are indeed localised.

		\section{Numerical methods}\label{appendix:numerics}
		\subsection{Large simple graphs with a prescribed degree sequence} \label{app:generate_graphs_numerics}
		In this Appendix we briefly illustrate the algorithm used in this paper to extract a simple graph with a prescribed degree sequence, which is a slight modification of the one described in \cite{delgenio2010efficient}. In particular, this algorithm has been used to sample simple graphs with geometric and power-law degree distribution and obtain the numerical results in Fig.~\ref{fig:spectral_denisty_grid_fixed_edge}, \ref{fig:ipr_grid_fixed_edge}, \ref{fig:spectral_denisty_grid_expprod_weights} and~\ref{fig:ipr_grid_expprod_weights}.
		
		The purpose of this algorithm is to extract simple graphs, in which self loops and multiple edges between the same pair of nodes are not allowed, such that the degrees of its nodes are prescribed by the non-increasing sequence
		\begin{equation}
			\mathcal{D} = \{ c_0 \; , c_1 \;, \; \cdots \; , c_{N-1} \}
		\end{equation}
		A degree sequence is said a graphical sequence if there is at least one simple graph with that sequence. The graphicality of a degree sequence can be assessed using the Erd\H{o}s–Gallai test, stating that a non-increasing sequence of degrees $\mathcal{D} = \{ c_0 \; , c_1 \;, \; \cdots \; , c_{N-1} \}$ is graphical if and only if their sum is even and the following inequality holds for all $0 \le m < N-1$:
		\begin{equation} \label{eq:Erdos-Galli_cond} 
			\sum_{i=0}^m c_i \le m(m+1) + \sum_{i=m+1}^{N-1} {\rm min}\left(m+1, c_i\right)
		\end{equation}
		In the following we assume to deal with graphical sequences. Notice that in general, even if a sequence is graphical, placing at random the edges between the nodes may result in a graph which is not simple.
        
		One possible approach, called the configuration model \cite{newman2018networksbook, latora2017complex, van2017random}, consists in starting from a node $i$ (usually, but not necessarily, the one with largest degree) which we call the hub, randomly pick a node $j$ and connect them. At this point we consider the reduced degree sequence in which we replace $c_i \to c_i-1$ and $c_j \to c_j-1$ and we proceed in this way up to the point in which all the reduced degrees are zero. However, as anticipated, this approach may result in graphs which are not simple. In this eventuality, since we are interested in generating simple graphs, we have to reject the graph realisation and start again from the beginning, inducing an implicit bias in the sampling.
		
		On the other hand, in this paper we employ a different approach: we identify the set $\mathcal{A}$ of nodes which are allowed to be connected to the hub, we select uniformly at random one of them, and connect it to the hub. Repeating this procedure we are always able to generate a simple graph from a graphical sequence. The crucial point is that the allowed set should be built each time we place a new edge, and therefore should be identified in a very efficient way. Starting from theoretical results in graph theory,  Ref.~\cite{delgenio2010efficient} provide an efficient method to build the allowed set by testing the sequence $\mathcal{D}'$, which is obtained from the original degree sequence $\mathcal{D}$ assuming that the hub $i$ is connected to all the nodes in its leftmost adjacency set, defined as the set of the $k_i$ nodes with the largest degrees that have not been already connected to the hub, except for the last one. Therefore we can choose a hub, build the sequence $\mathcal{D}'$, test it, identify the set $\mathcal{A}$ and pick one allowed node at random and keep repeating this procedure. In order to make the algorithm even more efficient, we have identified rules that allow us to update the sequence $\mathcal{D}'$ instead of constructing it anew each time. The code in \texttt{C++} is publicly available on the GitHub page of one of the authors at the link \href{https://github.com/kingofbroccoli/PowerLaw_Graph_Extraction_and_Direct_Diagonalisation}{\nolinkurl{https://github.com/kingofbroccoli}} \cite{valigigithub_powerlawextraction}.
		
		\subsection{Single-defect samples} \label{app:single_defect_numerics}        

		For the purposes of verifying our expressions for the asymptotic values of the IPR and the eigenvalues corresponding to rare network events, it is computationally costly to build networks where such rare events occur. However, we can go some way to testing our results by building a network of a moderate size (in which very few such rare events occur) and `forcing' the occurrence of a single rare event. We then weight our samples according to the probability that such a rare event would occur `naturally' in a large network. This numerical approach does not verify the validity of the single-defect approximation itself, but it does help us to understand how the wider network affects the states and eigenvalues associated to defects, and whether indeed such states are exponentially localised. The results of this numerical approach are compared with the results of exact diagonalisation in Fig. \ref{fig:ipr_grid_fixed_edge}.
		
		More precisely, suppose we construct a network with $N-1$ nodes. Now we add a single `defect'. We draw the connectivity and the edge weights of the defect from a distribution of our choice. For each realisation of the network with an artificially imposed defect, we can then calculate whatever observable we are interested in (the IPR or leading eigenvalue) that is associated with the defect. The important point is to weight each sample correctly, so that when we average the results over many realisations, we obtain statistically correct results.

        We consider a set of $n_{\mathrm{real}}$  matrices  $\{ (M_{ij}, M_{ji})\}_r$, with $r=1,2,\ldots,n_{\mathrm{real}}$, constructed as follows.   For each $r$,  the entries  $(M_{ij}, M_{ji})$  for $1 \leq i,j\leq N-1$ are drawn from the model of interest.   We then add to each of the realisations of this "natural ensemble" a particular realisation of the  hub.  We do this  by adding, for each $r$ independently,   the matrix entries $(M_{Nj}, M_{jN})$ with  $1 \leq j\leq N-1$.   Let us denote the probability that we would see these entries occur `naturally'  as $P_\mathrm{nat}(\{ (M_{ij}, M_{ji})\}_r)$ and the probability of occurring in our simulations with a "planted" hub as $P_\mathrm{sim}(\{ (M_{ij}, M_{ji})\}_r)$. We emphasise that $P_\mathrm{nat}(\{ (M_{ij}, M_{ji})\}_r)$ is determined by the model of interest, while  $P_\mathrm{sim}(\{ (M_{ij}, M_{ji})\}_r)$ is obtained from the model of interest by planting a hub into it. We then compute the ensemble average of the observable $O_r$ over $n_\mathrm{real}$ realisations as  
		\begin{align}
			\langle O \rangle = \frac{1}{n_\mathrm{real}}\sum_{r = 1}^{n_\mathrm{real}} \frac{O_r P_\mathrm{nat}(\{ (M_{ij}, M_{ji})\}_r)}{P_{sim}(\{ (M_{ij}, M_{ji})\}_r)}, \label{weightedobservable}
		\end{align}
        where   $O_r$ is a function of  $\{ (M_{ij}, M_{ji})\}_r$, and the  $\{ (M_{ij}, M_{ji})\}_r$ on the right-hand side of Eq.~(\ref{weightedobservable}) are realisations  drawn from the planted ensemble with the hub .  
		One notes that the normalisation constraint $\langle 1\rangle = 1$ is satisfied as $n_{\mathrm{real}} \to \infty$. 
        Using this approach, we can effectively turn rare events into frequently occurring events. Depending on what areas of the spectrum we are interested in particularly, we can can tune $P_\mathrm{sim}(\{ (M_{ij}, M_{ji})\}_r)$ to sample these areas.
		
		Let us take the example of the configuration model networks used to produce the results in Fig.~\ref{fig:ipr_grid_fixed_edge} (which is the case of fixed edge weights).   In these figures, we find the eigenvalue density and IPR (respectively) associated with defects with given values of $\omega$. We accomplish this by producing hubs with large values of $\Delta = r- l$, where $r$ is number of reinforcing links connected to the defect and $l$ is number of antagonistic links. These hubs are guaranteed to produce eigenvalues in the tail region in which we are interested.  We then average the observable in which we are interested over realisations of the wider network, weighting the realisation according to the hub arrangement as in Eq.~(\ref{weightedobservable}).
		
		More precisely, for a given value of $\Delta$, we now wish to average the observables over all possible local configurations of the defect, and also various realisations of the wider network. First, one constructs the configuration model network of $N-1$ nodes in the usual manner, with link weights being drawn from the distribution in Eq.~(\ref{piconstantmag}). One then adds the defect, which has the chosen value of $\Delta$. We then draw the value of the connectivity $c$ of the defect uniformly in the range $c_\mathrm{min} = \min(\Delta, \bar c-\delta c/2)$ to $c_\mathrm{max} =  \bar c+\delta c/2$, with $\delta c$ chosen so as to include a broad range of likely connectivities (a low value of $\delta c$ increases the speed of the computation at the cost of the accuracy of the results), and $\bar c$ is chosen to coincide with the most likely value of $c$ as predicted by the theory. The edges of the defect are connected to $c$ nodes in the existing network. Out of these $c$ connections, $(c+\Delta)/2$ are chosen at random to be reinforcing, and the remaining $(c-\Delta)/2$ are chosen to be antagonistic. The set of nodes $\{j\}$ to which we connect the defect are chosen from the existing $N-1$ nodes with probabilities proportional to $c_j+1$.  This preserves the neighbour degree distribution of the configuration model, namely $P_\mathrm{neighbour}(c) = \frac{c}{p} P_\mathrm{deg}(c)$. One performs this for many values of $\Delta$ using the following for the weightings in Eq.~(\ref{weightedobservable})
		\begin{align}
			P_\mathrm{nat}(\{ (M_{Nj}, M_{jN})\}_r) &= P_\mathrm{deg}(c) \binom{c}{\frac{c+\Delta}{2}} \epsilon ^{(c-\Delta)/2}\left(1-\epsilon \right)^{(c+\Delta)/2} \, , 		\end{align}
            and
\begin{align}
			P_\mathrm{sim}(\{ (M_{Nj}, M_{jN})\}_r) &= \sum_{c = c_\mathrm{min}}^{c_\mathrm{max}} P_\mathrm{deg}(c) \binom{c}{\frac{c+\Delta}{2}} \epsilon ^{(c-\Delta)/2}\left(1-\epsilon \right)^{(c+\Delta)/2} \, .
		\end{align}

\printbibliography[heading=subbibliography, title={References for Supplemental Material}]
\end{refsection}
		
\end{document}